%% file: main.tex
\documentclass[acmsmall]{acmart}

\input{macro}

\acmJournal{TOSEM}

\begin{document}

\title{Program Repair}

\author{Xiang Gao}\affiliation{\institution{Beihang University}\country{China}}\email{xiang_gao@buaa.edu.cn}
\author{Yannic Noller}\affiliation{\institution{National University of Singapore}\country{Singapore}}\email{yannic.noller@acm.org}

\author{Abhik Roychoudhury}\authornote{Corresponding author}\affiliation{\institution{National University of Singapore}\country{Singapore}}\email{abhik@comp.nus.edu.sg}

\begin{abstract}
Automated program repair is an emerging technology which consists of a suite of techniques to automatically fix bugs or vulnerabilities in programs. In this paper, we present a comprehensive survey of the state of the art in program repair. We first study the different suite of techniques used including search based repair, constraint based repair and learning based repair. We then discuss one of the main challenges in program repair namely patch overfitting, by distilling a class of techniques which can alleviate patch overfitting. We then discuss classes of program repair tools, applications of program repair as well as uses of program repair in industry. We conclude the survey with a forward looking outlook on future usages of program repair, as well as research opportunities arising from work on code from large language models. 
\end{abstract}

\maketitle

\input{chapters/1_intro}

\input{chapters/2_search}

\input{chapters/3_semantics}
\input{chapters/4_learning}
\input{chapters/5_overffint_in_synthesis}

\input{chapters/7_tools}

\input{chapters/6_application}

\input{chapters/8_future}

\section*{Acknowledgments}
This work was partially supported by a Singapore Ministry of Education (MoE) Tier 3 grant "Automated Program Repair", MOE-000332-01, and National Natural Science Foundation of China under Grant No (62202026).

 The authors would like to acknowledge all our collaborators in the area of program repair. Discussions with the collaborators have helped us gain perspectives on the research area. Abhik Roychoudhury would like to thank Sergey Mechtaev, Shin Hwei Tan and Jooyong Yi for past collaborations in the area of program repair. 
 
Abhik Roychoudhury would like to thank participants at the following meetings for valuable discussions which have over the years shaped the reflections on the field --- (a) Dagstuhl seminar 17022 on Automated Program Repair (January 2017), and (b) Shonan meeting 160 on Fuzzing and Symbolic Execution (September 2019).

Several colleagues read and commented on parts of the draft to improve its readability including Luciano Baresi and Shin Hwei Tan. The authors thank them for their valuable suggestions.

\bibliographystyle{ACM-Reference-Format}
\bibliography{biblio}

\end{document}

%% file: macro.tex
\usepackage{xspace}
\newcommand{\extractfix}{\textsc{ExtractFix}\xspace}
\newcommand{\genprog}{\textsc{GenProg}\xspace}
\newcommand{\angelix}{\textsc{Angelix}\xspace}
\newcommand{\prophet}{\textsc{Prophet}\xspace}
\newcommand{\semfix}{\textsc{SemFix}\xspace}

\newcommand{\footpatch}{\textsc{Footpatch}\xspace}
\newcommand{\saver}{\textsc{Saver}\xspace}
\newcommand{\senx}{\textsc{Senx}\xspace}
\newcommand{\infer}{\textsc{Infer}\xspace}
\newcommand{\sapfix}{\textsc{SapFix}\xspace}
\newcommand{\sapienz}{\textsc{Sapienz}\xspace}
\newcommand{\memfix}{\textsc{MemFix}\xspace}
\newcommand{\spr}{\textsc{SPR}\xspace}
\newcommand{\cpr}{\textsc{CPR}\xspace}
\newcommand{\directfix}{\textsc{DirectFix}\xspace}
\newcommand{\rsrepair}{\textsc{RSRepair}\xspace}
\newcommand{\trpautorepair}{\textsc{TrpAutoRepair}\xspace}
\newcommand{\prapr}{\textsc{PraPR}\xspace}
\newcommand{\uniapr}{\textsc{UniAPR}\xspace}
\newcommand{\vulnfix}{\textsc{VulnFix}\xspace}
\newcommand{\fixie}{\textsc{Fixie}\xspace}

\usepackage{listings}
\usepackage{multicol}
\usepackage{enumerate}
\usepackage{graphicx}
\usepackage{textcomp}
\usepackage{xcolor}
\usepackage{subcaption}
\usepackage{multirow}
\usepackage{color, soul}
\usepackage{pifont}
\usepackage{framed}
\usepackage{tcolorbox}
\usepackage{makecell}
\usepackage{balance}
\usepackage{mathtools}
\usepackage{arydshln}
\newcommand{\highlight}[1]{%
  \colorbox{gray}{$\displaystyle#1$}}
\usepackage{fancyvrb}
\usepackage{algorithm,algorithmic}

\usepackage{xcolor}
\definecolor{main-color}{rgb}{0.6627, 0.7176, 0.7764}
\definecolor{string-color}{rgb}{0.3333, 0.5254, 0.345}
\definecolor{key-color}{rgb}{0.8, 0.47, 0.196}
\lstdefinestyle{mystyle}
{
    language = Java,
    basicstyle = {\ttfamily \color{main-color}},
    stringstyle = {\color{string-color}},
    keywordstyle = {\color{key-color}},
    keywordstyle = [2]{\color{lime}},
    keywordstyle = [3]{\color{yellow}},
    keywordstyle = [4]{\color{teal}},
    morekeywords = [3]{<<, >>},
    morekeywords = [4]{++},
    basicstyle=\ttfamily\footnotesize,
    commentstyle=\color{blue}\ttfamily,
    morecomment=[f][\lstbg{red!20}]-,
    morecomment=[f][\lstbg{green!20}]+,
    morecomment=[f][\lstbg{yellow!20}]++,
    morecomment=[f][\lstbg{yellow!20}]--,
    morecomment=[f][\textit]{@@},
    texcl=false
}

%% file: chapters/1_intro.tex
\section{Introduction}

Programming is one of the most challenging activities undertaken by humans, because of the stringent demands it places on human creativity {\em and} precision. A casual comparison of programming with other significant activities of scientific or artistic nature - will convince of the significant challenges which a computer science professional seldom stops to think of. As a comparison, consider the task of painting a landscape or, or composing a novel. While there may be some demands on precision, the demands on precision of exactly how the story will unfold or how the painting will look like -- is not too high. In such a human activity, the main demand placed is on human creativity and imagination.  In contrast, consider the task of solving a partial differential equation, or conducting structural load testing in civil engineering for the purpose of building a bridge. Even though there may be some demands on creativity in such activities --- the focus is on precision. The main stringent demand is on precision and correctness - we want the structural load testing, or the differential equation solving to be very precise.

Let us now examine the task of programming. Given a requirement specification, it involves significant creativity - coming up with the right sub-tasks of the problem, suitable data structures, and suitable algorithms for solving the sub-tasks. At the same time, there are significant demands on precision when we conduct a programming activity --- the program needs to pass all the tests at the very least! This dual demands of creativity and precision, makes programming inherently difficult, and causes programmers to make mistakes!  As a result, the debugging and fixing of errors takes up significant time and resources in a software project, sometimes taking up 75---90\% of the resources. Part of the reason for such mistakes is also how a software project evolves over time, the challenge in obtaining and maintaining specifications for a software system, and the shared authorship of code which also evolves over time. This has prompted software practitioners to call the difficult process of software evolution, and the errors that creep in as a result, as the {\em legacy crisis} \citep{Seacord}. This brings us to the prospect of automated program repair, where a program can heal itself from errors and vulnerabilities!

\begin{center}
\fbox{
{\em Automatically repairing program errors goes beyond productivity enhancement. }
}
\end{center}

\subsection{Program Repair in a Nutshell}

Imagine that a test case for a large, established program has failed. 
What’s a novice developer to do?
In many cases, developers use the failing test to analyze the root cause of this bug and manually fix it. Automated program repair represents a suite of technologies that attempt to automatically fix errors or vulnerabilities in software systems. 
As a problem statement, automated program repair takes in a buggy program and a correctness criterion. The correctness criterion is usually in the form of a test-suite, a set of tests where for each test we provide the test input and the expected output(s).
Specifically, given a buggy program $P$ and a set of passing tests $pT$ and failing tests $fT$, test-driven program repair tries to find a (minimal) change to $P$ and make it pass all tests from $pT$ and $fT$.

There exist a suite of techniques that can generate (minimal) program edits that allow the transformed program to pass a given set of tests. Meta-heuristic search-based approaches traverse an explicitly defined search space of program edits (e.g. GenProg~\cite{GenProg}). Such repair methods can scale to large programs but not large search spaces as the search space is explicitly represented. Implicitly represented search spaces can be supported via symbolic execution where a specification of the desired patch can be inferred as a repair constraint, followed by patch generation via program synthesis \citep{semfix}. Such an approach can also be augmented to produce the minimal fix \citep{directfix}.
Due to the quick improvement of deep/machine learning techniques, learning repair strategies from human patches has gained attention recently.
The learning-based repair techniques (e.g., GetaFix~\cite{getafix}) first mine human patches that fix defects in existing software repositories, train a general repair rule and apply the model to buggy programs to produce patches.

Since a test-suite is an incomplete specification of the intended behavior of a program --- repairs which meet the expected outputs of a given set of tests can be {\em over-fitting}. In other words, these may pass the given tests but may fail other tests. This provides a significant technical challenge - since we have not witnessed or encountered tests outside the given test-suite, how do we generate patches which generalize over those unseen tests? This article studies a plethora of methods to ameliorate or address this key challenge. Since there are many {\em plausible} patch candidates, there are opportunities in terms of using machine learning techniques to rank patch candidates \citep{prophet}.

\subsection{Supporting Software Evolution}

Automated program repair goes to the heart of how software evolution can proceed "correctly", where the notion of correctness, or the formal specification, is not documented. 
Software systems, by their very nature, are flexible to change. At the same time, repeated changes to improve software or add features, makes software construction extremely error-prone. This situation cannot simply be improved by allocating additional personnel as explained by Turing award winner Fred Brooks in his now legendary book \emph{``The mythical man month''} \citep{Brooks}. This is due to inherent complexities in comprehending software systems. Automated program repair provides the necessary support for such correct software evolution. It also supports a different view of correct software development - as opposed to the age-old established techniques which develop correct software by formally proving its correctness. Instead of viewing the software as an inanimate piece of code about which theorems of correctness are proved, automated program repair views the software as a living entity --- which seeks to protect itself by gleaning specifications of intended behavior, and seeking to heal itself to meet these intended specifications.

The vision of automated program repair is to retain the workflows of software construction as they exist today in the form of Continuous Integration (CI) systems, and yet support \textit{correct software evolution}. In today's Continuous Integration systems used in companies, a developer will make changes to a code base and will seek to deposit them in the form of code-base commits. Even though the developer typically conducts unit testing prior to submitting the commits, the commits may fail system-level tests when submitted. The developer is then tasked with manually \textit{\textbf{fixing the errors}} so that at least all given tests pass, by informally following a "chain" of reasoning starting from an observable test failure. Our goal in automated program repair is to retain this software development workflow and yet alleviate the developer burden by \emph{automating} the program repair. 

\subsection{Challenges}

Most of the current work in program repair is based on user-provided tests, where a test suite is given as the correctness criterion, and repairs (to the given buggy program) are generated to pass the given tests. 
As a result, one of the key concerns with the current works on repair is that the patches produced are \emph{overfitting} to the given test-suite, and might result in a program that is not correct in general.

\begin{center}
    \fbox{
    Overfitting is a key challenge in automated program repair.}
\end{center}

Specifically, \textit{search-based program repair} techniques attempt to represent program repair as a navigation of a search space of program edits. These techniques are also called generate and validate techniques, since they generate patch candidates and validate them against the given tests.  Search-based repair techniques often rely on the ``plastic surgery hypothesis'': the ``correct'' code is present elsewhere in the codebase, from where it can be copied and moved via search operations.
In contrast to search-based repair, \textit{semantic program repair}, represents the repair process as an explicit specification inference. Given a correctness criterion, a constraint capturing the edits that need to be made is inferred, which is followed by the synthesis of a (minimal) patch meeting the constraint. Semantic repair techniques are less dependent on plastic surgery hypothesis based assumptions and can synthesize the needed code to pass a given set of tests. Both classes of repair techniques (search-based and semantic) can be enhanced by employing machine learning to prioritize fixes that better resemble human patch patterns.

A key challenge in correct program repair lies in the \textit{absence of detailed formal specifications} of intended behavior. Suppose, the correctness criterion guiding the repair is given in the form of tests (as is often the case in practice), which form an incomplete description of intended program behavior. The patches generated may be ``overfitting''; the patched program passes the tests guiding the repair process, but it can fail other tests. Combating overfitting in automated program repair and automated code generation is a key technical challenge. In this article, we discuss some of the approaches in this direction, so that developer trust in automatically generated code can be enhanced. 

\begin{center}
    \fbox{Combating patch overfitting enhances developer trust in automatically generated code.}
\end{center}

\subsection{Applications}

\begin{figure}
    \centering
    \includegraphics[width=0.6\textwidth]{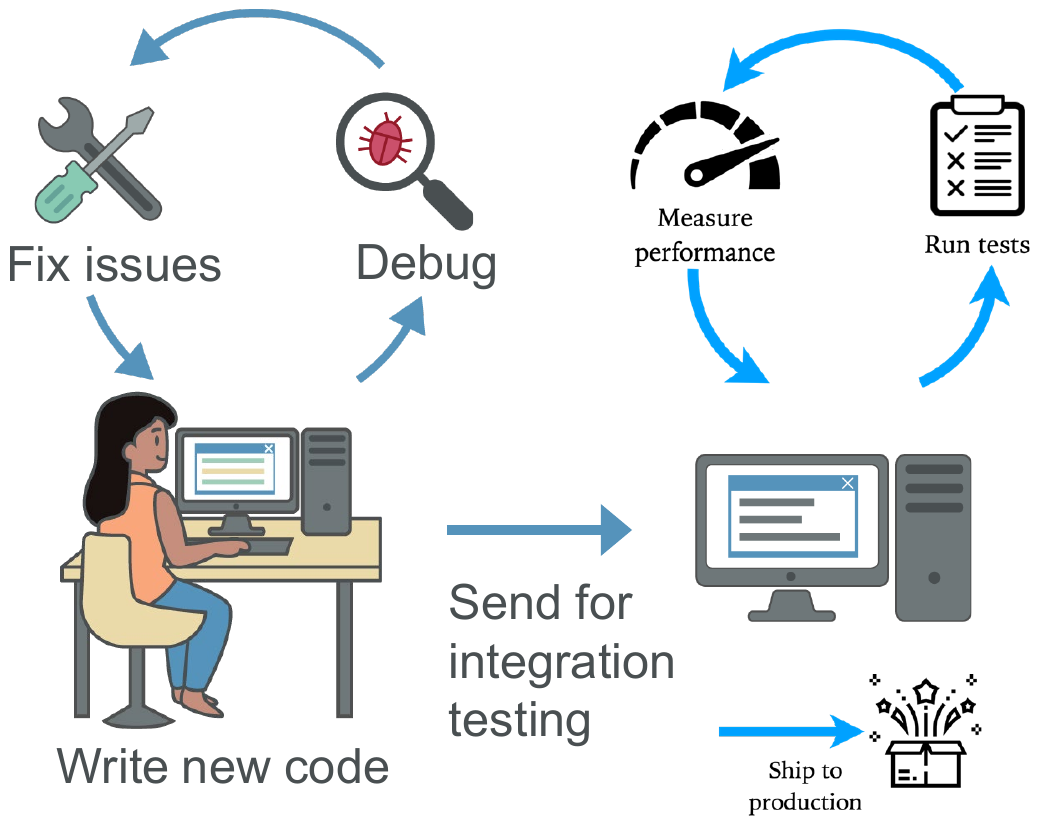}
    \caption{Software Development Life-Cycle}
    \vspace{-4pt}
    \label{fig:dev-lifecycle}
    \vspace{-4pt}
\end{figure}

Automated program repair, where a program morphs and protects itself of its own flaws, thus remains an enticing possibility. Given the huge cost incurred in software projects in debugging and fixing errors, software productivity enhancement is one of the most immediate applications of automated program repair. This is partly shown by where debugging and fixing sit in the software development life-cycle, as shown in Figure~\ref{fig:dev-lifecycle}.

As mentioned in the now-famous book "The mythical man month" by Fred Brooks \citep{Brooks}, "adding manpower to a late software project makes it later". This is partly because of the difficulties in program comprehension for large software systems by humans, particularly when the code is composed by other programmers. 
It has been mentioned that 50-75\% of the time in software projects are consumed by debugging and fixing errors \citep{ACMqueue17}. Automated program repair provides a convenient mechanism for retaining the software development life-cycle, and yet achieving correct (or at least more trustworthy) software construction.

Apart from applications in software productivity enhancement to improve software correctness, automated program repair has important applications in security vulnerability repair, with the goal of reducing exposure of software to security vulnerabilities. Currently, most security vulnerabilities are found by detection tools called {\em greybox fuzz testing tools} \citep{fuzzing21}. Fuzz testing tools proceed by a biased random search over the domain of inputs and search over the domain of inputs via (random) mutations. The mutations and the search are guided by objective functions such as coverage guided fuzzing. One could think of automated program repair as a follow-up back-end to fuzz testing tools, where vulnerabilities found are submitted for automated fixing. We note that both fuzz testing, and automated fixing can be viewed as biased random searches, with the fuzz testing being a biased random search over the domain of program inputs, and program repair being viewed as a biased random search over the domain of program edits. The two searches can re-enforce each other. For example, the additional tests generated by fuzz testing can be used to reduce over-fitting in selected patches so that the patches are correct with respect to a larger population of tests. Similarly, the selected patches can help in the test generation via directed fuzz testing campaigns with the goal of reaching the patch location and exercising the patches. In this way, one can envision a combined program toolkit which combines fuzz testing and automated patching together as process, with testing and patching re-enforcing and helping each other. 

A final application of automated program repair is in programming education, or building intelligent tutoring tools to teach programming. A feasibility study of using automated program repair tools for enhancing programming education appears in \citep{Yi17}. The main usage of automated program repair in such settings can be in providing hints or feedback to students. However, there are several reasons why automated program repair methods cannot be used straightforwardly to find out how the student's submission differs from a reference or model programming solution for an assignment. One of the reasons for this lies in the "competent programmer hypothesis" which essentially assumes that the programs written by professional programmers almost are correct. However, the programs written by novice programmers are not almost correct, hence the search space for edits is high. Furthermore, if we want to use automated repair for teaching, its goal will not be to give the full solution to the struggling student, but to give hints on the next step to the student. For this reason, when we use automated program repair for education - there would be value in partial repairs, such as repairs which may increase the number of passing tests, instead of repairs which make the patched program pass all tests in a test-suite. In general, the use of automated repair for programming education \citep{Yi17} is an emerging area with many tools being proposed \citep{Clara,Sarfgen,Refactory}, and we hope to see increased activity in the future in this important area.

\subsection{Organization}
The organization of the article is as follows. 
The next three chapters are devoted to the three prominent classes of repair techniques: search-based (Chapter~\ref{cha:search}), semantics-based (Chapter~\ref{cha:semantic_repair}), and learning-based repair (Chapter~\ref{sec:learning}). We then devote one chapter to study a variety of ideas to combat and alleviate overfitting in program repair (Chapter~\ref{sec:overfitting}), since this is the key technical challenge. 
We review the existing program repair technology in Chapter~\ref{sec:technology}. We then devote one chapter (Chapter~\ref{sec:applications}) to studying applications of program repair where we discuss advancements made in application domains such as developer productivity enhancement, software security and programming education. 
The article concludes with some perspectives on the area in Chapter~\ref{sec:perspectives}, specifically focusing on (a) methods to enhance developer trust, and (b) the futuristic possibility of combining program repair with recently proposed language model-based AI pair programmers.

\subsection{Existing overview articles}

There exist some other works which provide an overview of the field of program repair.

We are aware that Monperrus, et al. \cite{monperrus2018automatic} maintains a bibliography of APR papers.

The work of Le Goues, Pradel and Roychoudhury 
 \cite{LPR19} provides a summary of the field in the form of an introduction to the field. Thus it is not a comprehensive survey of program repair.

One other survey on automated program repair \cite{Mariani} exists. It presents results from techniques on program repair published up to January 2017. Our article presents only the repair technology as a whole but also its challenges (e.g. patch over-fitting), applications (e.g. intelligent tutoring) and future outlook (e.g. large language models). Moreover, we cover the advances in the field all the way until 2022.

%% file: chapters/2_search.tex
\section{Search-Based Program Repair}\label{cha:search}
Automated program repair is a process of automatically finding fixes to observable errors or vulnerabilities in a program.
This process can be conducted via search, where meta-heuristic search frameworks are used to look for candidate fixes in the space of program edits.


\subsection{Basic Search-Based Repair Workflow}
Search-based repair takes a buggy program and a correctness criterion as inputs.
The correctness criterion is usually demonstrated as a test suite consisting of both passing and failing tests, where failing tests demonstrate the bug, and passing tests represent the functionality that needs to be preserved.
Most techniques first identifies the code locations that are likely to be buggy.
The fault localization procedure provides a set of code locations ranked based on their potential buggy-ness.
The exact process of fix localization is not shown here. 
It may involve finding locations that appear with significantly greater frequency in failing tests than in passing tests.

Once fix locations are determined, search-based APR employs a generate-and-validate methodology. It constructs a search space of syntactic program modifications at the fix locations and iterates over the patch space to find patches such that the patched program satisfies the given correctness criterion.
These techniques can be explained as follows:
\begin{figure}[ht]
\vspace{-6pt}
\centering
\begin{minipage}{0.5\linewidth}
\begin{algorithmic}
\FOR{$\mathit{cand} \in \mathit{SearchSpace}$}
    \STATE validate $cand$ using criterion
    \STATE break if successful
\ENDFOR
\end{algorithmic}
\end{minipage}
\vspace{-6pt}
\end{figure}

where \textit{SearchSpace} denotes the set of considered modifications of the program, i.e., program patches.
Validation involves checking whether the patched program satisfies the correctness criterion when a suggested patch has been applied.
For instance, given the criterion is provided via a set of test cases, validation checks whether the patched program passes all the given tests. 

\subsection{Search Space Exploration}
Due to the combinatorial explosion of possible mutations and fix locations, the number of possible candidate patches that can be generated is usually very large.
Different exploration strategies have been proposed to find the correct patches among huge search space.

\subsubsection{Genetic Programming}\label{sec:gp}
Inspired by biological evolution, Genetic Programming (GP) is a stochastic search method that is used to discover computer programs for a particular task.
The GP algorithm has been proven to be faster and more efficient when compared to the traditional brute-force search methods.
In the context of program repair, the GP algorithm is applied to search for program variants, i.e., patched programs after applying particular patches, that retain the required functionality and fix the given bug.
\genprog~\citep{GenProg} is one of the most well-known APR tools that rely on the GA algorithm.
\genprog has been applied to fix real-world programs and showed that automatically fixing one real defect only requires 8 US dollars \citep{genprog-icse12}.
Because of the huge cost spent on fixing software bugs, \genprog has gained a lot of attention from both academia and industries.
Algorithm~\ref{alg:gp_algo} shows the overall GP algorithm in more detail.
The algorithm maintains a population of program variants in a multiple generation process.
In each generation, it modifies source code to produce a population of candidate variants.
It then uses a user-defined fitness function to guide the evolution of each variant.
Specifically, the variants with large fitness are selected and passed to the next generation (see line \ref{line:select}).
At the same time, it generates new variants using mutation and crossover operations and adds them to the new population.
This process continues until a variant that satisfies the given criterion is obtained or a timeout is reached.


\begin{algorithm}
\caption{Genetic Programming Algorithm}
\label{alg:gp_algo}
\begin{algorithmic}[1]
\STATE $\mathit{population} \gets$ init\_population($P$, $\mathit{pop\_size}$)
\REPEAT
    \STATE $\mathit{new\_pop} \gets$ select($\mathit{ranked\_pop}$, fitness\_function) \label{line:select}
    \FORALL{<$p_1, p_2$> $\in$ $\mathit{new\_pop}$}
        \STATE $c_1, c_2 \gets$ crossover($p_1, p_2$)
        \STATE $\mathit{new\_pop} \gets$ $\mathit{new\_pop}\cup \{c_1, c_2\}$
    \ENDFOR
    \FORALL{$p_i$ $\in$ $\mathit{new\_pop}$}
        \STATE $\mathit{population} \gets$ $\mathit{population}\cup$\{mutate($p_i$)\}
    \ENDFOR
\UNTIL{$\exists p_i \in \mathit{population}: fitness\_function(p_i) = \texttt{max\_fitness}$}
\RETURN $p$ with largest fitness where $p \in \mathit{population}$
\end{algorithmic}
\end{algorithm}

\paragraph{Mutation}
The mutation operator changes a particular statement with some probability.
This probability can be the same probability of being the fix location, which is computed during the fix localization step.
A statement $\mathit{stmt}$ is mutated by one of the three operators: \texttt{delete}, \texttt{insert}, and \texttt{replace}.
For instance, expression $a+b$ coule be mutated as $a-b$ or $a*b$.
The delete operator directly deletes statement $\mathit{stmt}$ from the original program.
The insert operator copies a statement $\mathit{stmt_2}$ from somewhere else in the program and inserts it after $\mathit{stmt}$, while the replace operator replaces $\mathit{stmt}$ with another statement $\mathit{stmt_2}$.

\paragraph{Crossover}
The crossover operator generates new variants by combining the ``first part'' of one program variant with the ``second part'' of another program variant, i.e., creating an offspring variant by combining two parent variants.
Given two variants $p_1$ and $p_2$, and a \texttt{cutoff} point in the program, crossover operator creates $c_1$ by combining $p_1$[0:\texttt{cutoff}] with $p_2$[\texttt{cutoff}:\texttt{END}], and creates $c_2$ by combining $p_2$[0:\texttt{cutoff}] with $p_1$[\texttt{cutoff}:\texttt{END}].
For instance, two programs 
\[prog_1: \{res1 = a + b; res2 = c - b;\}\ \ \ \ prog_2: \{res1 = a - b; res2 = c + b;\} \]
could produce two new programs:
\[prog'_1: \{res1 = a + b; res2 = c + b;\} \ \ \ \ prog'_2: \{res1 = a - b; res2 = c - b;\} \]
via crossover operation.

\paragraph{Fitness}
The fitness of a program variant evaluates its acceptability with regard to the provided correctness criterion.
The fitness is used to (1) guide the selection of variants passed to the next generation and (2) provide the termination condition for the search.
In the test-driven program repair, the fitness of a variant is measured by evaluating its ability to pass the given test cases.
First, if a variant cannot be compiled, it will be given $0$ fitness.
If a compilable variant passes all the given tests, it is assigned the maximal fitness value.
Otherwise, the fitness of a variant is defined as follows:
\begin{align*}
\mathtt{fitness\_function}(p) &= W_p * |\{t\in Pass\_Test\ |\ p\ \texttt{pass}\ t\}|\\
&+\ W_n * |\{t\in Fail\_Test\ |\ p\ \texttt{pass}\ t\}|
\end{align*}
where $W_p$ and $W_n$ are the weight applied to passing and failing tests, respectively.
Basically, this function calculates the fitness based on the number of passing and failing tests of a program variant.
The variant that passes more test cases will be given higher priority to be passed to the next generation.
The fitness function drives the search direction until we find a variant that passes all the given test cases.

\begin{table}[t!]
\centering
\begin{tabular}{>{\raggedright\arraybackslash}p{3.1cm}|p{10cm}}
\hline
Pattern & Description \\
\hline\hline
Parameter Replacer &
For a method call, this pattern replaces a selected parameter with a compatible variable or expression within the same scope.\\
\hline
Method Replacer &
For a method call, replace it with another compatible method.\\
\hline
Parameter Adder and Remover &
For a method call, this pattern adds or removes parameters if the method has overloaded methods.\\
\hline
Expression Replacer &
For a conditional branch, this pattern replaces its predicate by another expression collected in the same scope.\\
\hline
Expression Adder and Remover &
For a conditional branch, this pattern inserts or removes more term(s) to its predicate. \\
\hline
Null Pointer Checker &
For object reference, this pattern adds \texttt{if} statements to check whether an object is null. \\
\hline
Object Initializer &
For a variable in a method call, this pattern inserts an initialization statement before the call. \\
\hline
Range Checker &
For an array references, this pattern adds \texttt{if} statements that check whether an array access index exceeds upper and lower bounds.\\
\hline
Collection Size Checker &
For collection access, this pattern adds \texttt{if} statements to check whether the access index exceeds the size of the collection object.\\
\hline
Class Cast Checker &
For a class-casting statement, this pattern inserts an \texttt{if} statement checking that the castee is an object of the casting type.\\
\hline
\end{tabular}
\caption{Fix pattern adapted from PAR~\citep{par} for generating search space.}
\label{tab:patterns}
\end{table}

\subsubsection{Pattern-Based Search}
Instead of relying on random program mutations such as statement addition, replacement, and removal, which may generate nonsensical patches, another line of repair technique generate search space according to pre-defined patterns~\citep{par}.
The main intuition behind these approaches is that human-written patches have common fix patterns.
Therefore, the fix templates can be inferred from the human-written patches and apply them to fix other similar bugs.
The main advantage of these approaches is that the type and variety of edits considered for a repair candidate can be easily controlled, keeping great flexibility while limiting search space size.
Table~\ref{tab:patterns} shows some of patch patterns adapted from \cite{par}.
For instance, to fix a null point dereference, a typical fix pattern is to insert an \textit{Null Pointer Checker} ``\texttt{if ($var$ == null) return;}''.
PAR~\cite{par} showed that such a pattern-based repair technique can generate more acceptable patches than genetic programming, and the generated patches tend to be more comparable to human-written patches.

\subsubsection{Heuristic Search}\label{sec:heuristics}

Out of huge candidate patch space, searching for the correct patch is usually a time-consuming task.
Heuristics are defined to guide the search process.
The heuristics evaluate the quality of the patch candidate and then determine the search direction or terminate the search when a certain goal is satisfied.
The typical and most apparent heuristic would be to measure the number of passing and failing test cases. Similar to the fitness function defined in Chapter~\ref{sec:gp}, a variant of the heuristic is illustrated as follows:
\begin{equation}\label{eq:passing}
\begin{split}
f_{1a}(x) = w_p * |\{t \in T_p \;|\;x \text{ passes } t\}| + w_f * |\{t \in T_f\;|\;x \text{ passes } t\}|
\end{split}
\end{equation}
where $x$ represents the patch candidate, $T_p$ is the set of positive tests, $T_f$ is the set of failing tests, and $w_p, w_f \in (0, 1]$ are the weights applied to passing and failing tests, respectively. The objective would be to maximize the number of passing test cases. Similarly, one can define the weighted failure rate \citep{arja}:
\begin{equation}\label{eq:weighted-failure-rate}
\begin{split}
f_{1b}(x) = w * \dfrac{|\{t \in T_p \;|\;x \text{ fails } t\}|}{|T_p|} + \dfrac{|\{t \in T_f\;|\;x \text{ fails } t\}|}{|T_f|}
\end{split}
\end{equation}
$w \in (0, 1]$ is a bias towards negative tests. Here the objective is to minimize the weighted failure rate.
Furthermore, ARJA~\cite{arja} propose to combine minimizing the weighted failure rate $f_{1b}$ (Equation \ref{eq:weighted-failure-rate}) with minimizing the patch size captured in $f_2$:
\begin{equation}\label{eq:patch-size}
\begin{split}
f_{2}(x) = \sum^{n}_{i=1}{b_i}
\end{split}
\end{equation}
where $b_i$ is a binary value that indicates whether patch $x$ is modifying the $i$th modification point. In sum, this is a proxy for the number of edit operations.

In addition to selecting patches, heuristics are also applied to rank patches when multiple plausible patches are available. Such ranking heuristics can look similar to the selection heuristics.
Typically patches would be ranked by preferring patches that only make a minimal change to the program. Change can mean minimal \textit{syntactical} change but also \textit{semantic} change, e.g., with regard to the altered control flow.
Concolic program repair (CPR)~\citep{cpr} ranks patches based on their dynamic behavior during its patch refinement phase. The more often a patch exercises the buggy location without triggering any violation, the higher it is ranked relative to the other patches. Furthermore, CPR also deprioritizes patches that change the control flow for inputs that are actually correctly handled.
Instead of defining ranking policies upfront, Prophet~\cite{prophet} attempts to \textit{learn} heuristics in form of a statistical model. They train a model on a labeled data set of incorrect and correct patches. The training goal is that the learned model assigns a high probability to correct patches so that the model becomes an estimator for \textit{correct} patches. Their tool \prophet explores the search space by \spr~\citep{spr} and uses the trained model to sort the generated patch candidates. In the given order, \prophet validates the patches with the available test suite and returns an ordered and validated list of patches (see more details in Chapter \ref{sec:learn-ranking}).

\subsubsection{Test-Equivalent Analysis}
As we mentioned above, the search space size could be very large, and searching for correct patches from a huge search space could be inefficient.
To solve this problem, one idea is to optimize the patch evaluation via test-equivalence analysis according to patch semantic behaviors.
The patch candidates can be divided into equivalence groups based on the test-equivalence analysis.
When exploring patch space, instead of validating candidate patches one by one, the patches in the same group can be evaluated together, hence increasing patch exploration efficiency.

Determining whether two programs A and B are semantically equivalent is challenging.
AE~\cite{genprogae} proposes to approximate semantic equivalent of A and B, i.e., A $\approx$ B implies that A and B are semantically equivalent.
To do so, three heuristics are applied to determine semantic equivalence: \textit{syntactic equality}, \textit{dead code elimination}, and \textit{instruction scheduling}.
Specifically, syntactic equality means that if two patches are syntactically the same, they must be semantically equivalent.
This heuristic applies to search-based APRs, e.g., GenProg, which uses existing programs (at the other locations) as the source to form a patch.
For instance, if the statement \texttt{if(x!=NULL)} appears three times in the program, GenProg might consider three different patches, i.e., inserting each instance of \texttt{if(x!=NULL)} to the fault location.
All those three patches are syntactically the same and hence semantically equivalent.
Second, dead code elimination means that if the value of variable \textit{var} does not affect program execution, all the patches that change \textit{var} will be regarded as semantically equivalent patches.
For instance, statement \textit{var}=0 and \textit{var}=1 are inserted at fault location as patch, but \textit{var} is never used after the fault location, patch \textit{var}=0 and \textit{var}=1 will have the same semantic effect on the program.
Third, instruction scheduling means that if two instructions $S1$ and $S2$ do not have write-write or read-write dependencies, $S2; S1;$ and $S1; S2;$ are semantically equivalent.
For instance, consider the program snippet L1: x=$i_1$; L2: y=$x_2$; L3: z=$x_3$; and the candidate patches are to insert ``a=0;'' at L1, L2 or L3.
Since ``a=0;'' is not dependent on the statements at L1, L2, and L3, so all three patches are semantically equivalent.
When evaluating semantically equivalent patches, evaluating only one of them is sufficient to validate the whole equivalence partition.

Besides the approximated approaches, \cite{TOSEM18Sergey} introduces an approach to determine semantic behaviors of patch expressions based on value-based test-equivalence analysis.
Specifically, the patch's test-equivalence relation is defined as follows:
\begin{definition}[Test-equivalence]\label{def:test-equivalence}
Let $prog_1$ and $prog_2$ be two program variants, and $t$ is a test. $prog_1$ is test-equivalent to $prog_2$ w.r.t. $t$ if $prog_1$ and $prog_2$ produce same output when executing test $t$.
\end{definition}

If two patched programs using patch $p_1$ and $p_2$ are test-equivalent w.r.t. a given test, then $p_1$ and $p_2$ will be put in the same group.
To detect the test-equivalence of any two program variants w.r.t. $t$, executing both of them on test $t$ is expensive.
Actually, detecting the test-equivalence of two variants does not necessarily require executing each variant individually.
Instead, while executing only one of program variants, dynamic analysis can be performed to determine whether they are test-equivalence, which can help reduce the number of test executions required for evaluation.
Mechtaev, et al. \cite{TOSEM18Sergey} consider one such analysis referred to as \emph{value-based test-equivalence}.
\begin{definition}[Value-based test-equivalence]
Let $e$ and $e'$ be expressions, $prog_1$ and $prog_2$ be programs such that $prog_1 = prog_2[e\!\mapsto\!e']$, meaning that the only difference between $prog_1$ and $prog_2$ is expression $e$.
We say that $prog_1$ is value-based test-equivalent to $prog_2$ w.r.t. test $t$ if $e$ is evaluated into the same sequence of values during the execution of $prog_1$ with $t$, as the value sequence of $e'$ during the execution of $prog_2$ with $t$.
\end{definition}

Using a value-based test-equivalence relation, the patch space can be then represented as a set of patch partitions.
The patches within a patch partition will be evaluated together, hence increasing the efficiency of patch evaluation.

\begin{figure}[!t]
\begin{subfigure}[c]{0.45\linewidth}
\begin{lstlisting}
void cap_first_letter(char s[]) { 
  int len = strlen(s);
  for (int i = 0; i < len; i++) {
    if (s[i]==`p') {
      s[i] = toupper(s[i]);
    }
  }
}
\end{lstlisting}
\end{subfigure}
\hfill
\begin{subfigure}[c]{0.4\linewidth}
\begin{tabular*}{\textwidth}{ l @{\extracolsep{\fill}} l l }
\hline
Test   & Input         & Expected output \\
\hline\hline
Test 1 & ``aristotle'' & ``Aristotle''\\
Test 2 & ``plato''     & ``Plato''\\
\hline
\end{tabular*}
\end{subfigure}
    \caption{The left part is a simple example that capitalizes the first letter of a given string, where the bug happens on the if-condition. The correct condition is i == 0. The right part shows two tests and their corresponding expected outputs.}
    \label{fig:f1x_example}
\end{figure}

Figure~\ref{fig:f1x_example} shows an example that capitalizes the first letter of a given string. A bug exists at line 4, where the if-condition should be $\mathtt{if} (i==0)$. The right part of Figure~\ref{fig:f1x_example} presents two test cases, where Test 1 is a failing test and Test 2 is a passing test. To fix this bug, suppose the fix location is determined as the if-condition and the patch candidates include s[i]==`a', s[i]==`b', s[i]==`b',\dots, i==0, i==1, i==2\dots.
Evaluating the patch candidates on the given two tests one by one is time-consuming.
To solve this problem, the patch candidates can be grouped according to value-based test-equivalence relations.
Specifically, over Test 1, patch s[i]==`a' and i==0 will be divided into the same group, since the produced values on ``aristotle" is <T, F, F, F, F, F, F, F, F>, while patches s[i]==`r' and i==1 are in the same group, since their produced values are <F, T, F, F, F, F, F, F, F>.
Similarly, patches s[i]==`b', s[i]==`c', s[i]==`d', s[i]==`e' and \dots will be put in the same group.
The patches in the same group will produce the same output. Therefore, evaluating one patch of a group is enough to determine whether they can pass the given tests.
For instance, all the patches in the group consisting of s[i]==`a' and i==0 can pass Test 1, while all patches in the group s[i]==`r' and i==1 fail Test 1.
The grouping mechanism can significantly accelerate the evaluation of patch candidates.

Besides, CPR~\cite{cpr} use \textit{Concolic Program Repair} (CPR) to maintain a set of \textit{abstract} patches. Each abstract patch holds multiple concrete patches by using a patch template and corresponding constraints on the parameters of this template. By refining the constraint, multiple concrete patches can be eliminated simultaneously (see subsection \ref{sec:cpr} for more details).

\subsection{Search Space Validation}\label{sec:validation}
The goal of the second part of "generate-and-validate" methodology is to \textit{validate} the generated patches and produce a subset with the high-quality patches.
In general, there are three aspects to consider:
\begin{enumerate}
    \item a patch should fix the provided failing test case(s),
    \item a patch should not harm existing functionality, and hence, still pass the other test cases in the test suite, and
    \item a patch should generalize beyond the provided test suite.
\end{enumerate}
Patches that satisfy aspects (1) and (2) are called \textit{plausible patches}. They can be identified by test execution. However, since the test cases only represent a partial specification of the intended behavior, such plausible patches do not need to represent so-called \textit{correct} patches, i.e., patches that are semantically equivalent to an actual human developer fix. High-quality patches account for aspect (3) and go beyond satisfying existing test cases. Identifying these patches, and thereby ruling out \textit{overfitting} patches, is still an open research challenge, which we discuss separately in Chapter \ref{sec:overfitting}.
In reality, the validation phase does not need to be strictly separated from the generating phase. For example, in the search exploration based on genetic programming (see subsection~\ref{sec:gp}), the fitness evaluation of the population is integrated in the overall workflow, and helps to guide the remaining search space exploration.
Disregarding of how and where validation is integrated in the repair workflow, a common challenge is to perform it \textit{efficiently}. Otherwise, it can become a bottleneck for the overall time to repair, and hence, limits the scalability of the APR technique.
In fact, it was empirically shown that the runtime of SPR~\citep{spr} is dominated by patch compilation and test execution \citep{AcceleratingSBPR}. Part of the problem is the repeated execution of developer tests \citep{UniAPR}.
The literature on APR has proposed a few solutions to tackle this problem, which are discussed in the following paragraphs.

\paragraph{Test Case Prioritization}
The problem of expensive test execution is not new to automated program repair but is a known challenge in software testing, in particular regression testing \citep{TestPrio, RegTest}.
Given a test suite $T$, the set of permutation of this test suite $T$ denoted as $PT$, and a function $f$ from $PT$ to real numbers; TestPrio~\cite{TestPrio} defines the test case prioritization problem as follows: Find $T' \in PT$ such that $(\forall T'') (T'' \in PT) (T'' \neq T') [f(T') \geq f(T'')])$.
Classic prioritization techniques choose $f$ to represent information like \textit{code coverage} and \textit{fault-exposing potential} (FEP) of test cases to decide on an ordering.
Many research has been conducted to find techniques for the selection and prioritization of test cases that have a high chance of revealing defects (i.e., maximizing their value) and avoiding redundant or overlapping execution traces (i.e., minimizing their execution effort).
RegTest~\cite{RegTest} provides an overview of existing minimization, selection, and prioritization techniques. 

In program repair, similar techniques can be applied to reduce the effort of patch validation. For example, \rsrepair~\citep{RSRepair} applies classic test case prioritization to maximize the invalid-patch detection rate of the leveraged test suite, and hence, early identify insufficient patch candidates.
A downside of classic prioritization techniques is that they require previous test execution information (like code coverage) prior to the actual repair process. Such effort causes additional costs. \trpautorepair~\citep{TrpAutoRepair} tries to avoid that by incrementally extracting the required information from the repair process itself.

\paragraph{On-the-fly Patch Validation}
Apart from the number of tests or the particular set of tests, the way patches are represented and executed impacts the effort of patch validation. The idea of \textit{on-the-fly} patch validation targets these steps in program repair. It attempts to avoid any re-compilation and restarting of the execution environment. Furthermore, it attempts to generate parameterized patches that represent multiple different patches. By using a parameter, a specific patch variant can be enabled for testing.
For example, Meta-Programming~\cite{Metaprogramming} introduces the concept of a \textit{metaprogram} in the context of repairing null pointer exceptions. Via so-called \textit{modification hooks} the behavior of the original program is equipped with multiple patching strategies. The hooks can be activated by setting specific runtime parameters. Such a metaprogram can be automatically generated using source-to-source transformations. After generation, the metaprogram needs to be compiled only \textit{once} before it can be executed \textit{multiple} times during the patch validation phase.
\prapr~\cite{PraPR} and \uniapr~\cite{UniAPR} are proposed to address this issue in the context of JVM-based APR techniques. 
\prapr performs bytecode-level APR, and hence, does not need to re-compile the generated patch.
\uniapr acts as a patch validation platform around existing source-code level APR tools.
It uses a single JVM session for validating multiple patches, which saves time by skipping costly actions for JVM restart, reload, and warm-up. They leverage JVM's dynamic class redefinition feature to reload only the relevant patched bytecode classes.
UniAPR~\cite{UniAPR} show empirical evidence that such an approach can lead to significant speedups with only a small computation overhead.


\paragraph{Predicting Patch Correctness}
As already mentioned, relying on an in-complete oracle like a test suite may result in incorrect patches. Therefore, as alternative to any test execution, patches can be validated by predicting their correctness with machine learning models \citep{StaticPatchClassifcation, PredictingPatchCorrect}.
For example, Ye et al. \cite{StaticPatchClassifcation} propose to compare the patched program and the buggy program to statically extract relevant code features. Subsequently, they use a supervised learning method to generate a probabilistic model for the assessment of the patch correctness.
%
%
The resulting model can be used in an APR post-processing stage to rule out plausible patches that are likely overfitting.
Finally, machine learning models cannot only be used to detect invalid and overfitting patches, but also to \textit{rank} validated patch candidates \citep{prophet} (see Chapter \ref{sec:heuristics} and \ref{sec:learn-ranking}).


\subsection{Discussion}
Search-based repair generates program patches by 1) generating a set of candidate patch space using pre-defined transformation operators, and 2) searching the correct patch from the patch space according to given heuristics.
The idea of search-based repair is simple and straightforward, and it can be easily extended by implementing new transformation operators or heuristics.
Unfortunately, the repairability of search-based repair is also restricted by pre-defined operators.
The fewer transformation operators are used, the less likely the correct patches can be generated.
On the other hand, the more operators are used, the larger patch space will be produced, resulting in slow searching efficiency.
Furthermore, it is very unlikely that finite pre-defined operators can generate required patches of all types of bugs.

%% file: chapters/3_semantics.tex
\section{Constraint-based Repair}
\label{cha:semantic_repair}

Although the search based approaches towards program repair show promising results in fixing some kinds of bugs, the search process is not efficient for an extremely large search space of patch candidates.
The search based approach does not provide significant opportunities for grouping patch candidates. For a large search space, it is useful to maintain an abstract patch space representation where groups of patch candidates are represented as patch partitions. A more abstract patch space representation could arguably provide leverage in terms of searching for suitable patch candidates, over and above intelligent search heuristics. 

With this thought in mind, we can explore the possibility of representing a collection of patch candidates as a symbolic constraint. If we identify the set of desired patch candidates for the given buggy program as a symbolic constraint, we can then synthesize code meeting these constraints, thereby solving the problem of program repair. This gives us a different viewpoint or outlook towards the program repair problem. Instead of traversing a search space of program edits and deciding which program edits are suitable, we construct a repair constraint capturing the set of program edits which are suitable, and then synthesize a program edit satisfying this repair constraint.

\subsection{Repair Constraints: Idea}

Let us consider a program that takes in three sides of a triangle and determines the kind of triangle constructed out of these three sides. The program may look like the following.

\begin{figure}
    \centering
\begin{verbatim}
1    int tri_detect(int a, int b, int c){
2       if (a <= 0 || b <= 0 || c <= 0)
3           return INVALID;
4       else if (a == b && b == c)
5           return EQUILATERAL;
6       else if (a == b || b == c) 
7           return ISOSCELES;
8       else return SCALENE;
9    }
\end{verbatim}
    \caption{Triangle program adapted from \cite{cacm19}}
    \label{fig:my_label}
\end{figure}

The above program has several bugs. For three sides which violate the triangle inequality - it should return INVALID, but it is not doing so. Similarly, the definition of the isosceles triangle is supposed to check if any two of the three sides are equal. Now, let us show a realistic test-suite consisting one test for invalid triangle, one for equilateral triangle, three tests for isosceles triangle (depending on which two sides are equal), and one test for a scalene triangle. A reasonably constructed test-suite based on the requirements will indeed be of this nature.
\begin{center}
    \begin{tabular}{|c|c|c|c| c|}
    \hline
    a & b & c & Output & Outcome\\
    \hline\hline
    -1 & 1 & 1 & INVALID & Pass\\ \hline
    2 & 2 & 2 & EQUILATERAL & Pass \\\hline
    2 & 2 & 3 & ISOSCELES & Pass \\ \hline
    2 & 3 & 2 & SCALENE & Fail \\ \hline
    3 & 2 & 2 & ISOSCELES & Pass \\ \hline
    2 & 3 & 4 & SCALENE & Pass \\ \hline
    \end{tabular}
\end{center}

Let us assume now that by a control flow analysis of the passing and failing tests, line 6 is inferred as the fix location. 
The fix localization process is the same as the localization of search-based APR techniques.
Once the fix location is identified, the expression in that location is substituted as an unknown or a symbolic variable X.

\begin{verbatim}
1    int tri_detect(int a, int b, int c){
2       if (a <= 0 || b <= 0 || c <= 0)
3           return INVALID;
4       else if (a == b && b == c)
5           return EQUILATERAL;
6       else if (X)
7           return ISOSCELES;
8       else return SCALENE;
9    }
\end{verbatim}

Now, it is required to find out properties about X which would make the program pass the test cases that are given. Note the following constraints. 
\begin{itemize}
    \item the first two tests do not even reach line 6.
    \item among the remaining four tests that reach line 6, X should be true in the third, fourth, and fifth tests. Moreover, X should be false in the sixth test.
\end{itemize}
Getting the above-mentioned requirements, though put intuitively here, is not straightforward. It involves an analysis of the test executions for the given tests. Essentially it amounts to finding the desired value of X (in this case a boolean as it represents a boolean expression) so that it can make the test pass. This amounts to the {\em repair constraint}.

How to formally capture these requirements or constraints on X, which essentially is a placeholder for the code inserted in line 6? A formal way of understanding this repair constraint is that the unknown X is essentially an unknown function on the variables which are live in line 6. Thus essentially 
\[
X = f(a,b,c)
\]
where f is an unknown function that is to synthesize. With this understanding, the fixed program will be of the form
\begin{verbatim}
1    int tri_detect(int a, int b, int c){
2       if (a <= 0 || b <= 0 || c <= 0)
3           return INVALID;
4       else if (a == b && b == c)
5           return EQUILATERAL;
6       else if f(a,b,c)
7           return ISOSCELES;
8       else return SCALENE;
9    }
\end{verbatim}
where the information about the function $f$ is given by the following repair constraint.
\[
f(2,2,3) \wedge f(3,2,2) \wedge f(2,3,2) \wedge \neg f(2,3,4)
\]
This repair constraint can be fed to a program synthesis engine. The synthesis engine can be fed with the ingredients that can appear in the expression: the variables, the constants, and the operators. In this case, the variables are {\tt a}, {\tt b}, {\tt c}, the constants are the integer constants and the operators are the relational operators and logical operators. With these ingredients and the provided repair constraint, a component-based synthesis engine \citep{oracle-guided10} will yield the correct fix 

\begin{verbatim}
    f(a,b,c) = (a == b || b ==c || a == c)
\end{verbatim}

\subsection{Repair Constraint: formal treatment}

Let us now present the formal treatment of repair constraint computation. Statistical fault localization \citep{SBFL-Survey} or other offline analysis techniques are applied to identify potential fix locations. Such offline analysis may involve program dependency analysis, or simply control flow analysis of the passing / failing tests. Let us examine how the control flow analysis of passing / failing tests will proceed under the auspices of statistical fault localization. In such an approach, each statement $s$ in the program is given a suspiciousness score based on the occurrences of $s$ in the passing / failing tests. While many possible scoring mechanisms are possible, one of the earlier scoring mechanisms known as Tarantula \citep{Tarantula} scores the statements as follows 
\[
score(s) = \frac{\frac{fail(s)}{allfail}}{\frac{pass(s)}{allpass} + \frac{fail(s)}{allfail}}
\]
where given a set of tests $T$, allfail is the number of all failing tests in  $T$, allpass is the number of all passing tests in $T$, $fail(s)$ is the number of failing tests where statement $s$ appears, $pass(s)$ is the number of passing tests where statement $s$ appears, and $allfail + allpass = \mid T \mid$.

Given such a scoring mechanism, each statement in the program obtains a suspiciousness score, and the statements can be sorted based on their scores. The statement with the highest score can be considered a candidate fix location, and if a fix meeting the repair constraint cannot be found, the statement with the next highest suspiciousness score is tried. Thus, at any point of time, the repair algorithm is trying to automatically generate a single line fix. Note this does not necessarily mean that the human fix for the same error needs to be a single line fix. Even if the human fix is a multi-line fix, as long as a single line fix is found by the repair algorithm (either equivalent to the human fix, or at least passing all the given tests T), the repair algorithm succeeds.

Once a line to be fixed is decided, we next need to construct a repair constraint. This is a constraint on the expression to be put in the corresponding line as a fix. For the purposes of explanation, let us assume that the fix is either a boolean expression or an arithmetic expression which is the right hand side of an assignment. How to construct the repair constraint? For a boolean expression, the expression can be simply replaced with a new symbolic variable {\tt X} as follows.
\[
{\tt if(e)} \rightarrow {\tt if(X)}
\]
For an arithmetic expression, a new symbolic variable {\tt X} is introduced as follows.
\[
{\tt y = e} \rightarrow {\tt y = X}
\]
Note here that {\tt y} is a program variable and {\tt e} is an expression made out of program variables, while {\tt X} is a symbolic ghost variable which is introduced by us, for the purposes of automated program repair. Note that the symbolic variable {\tt X} is introduced at the deemed fix location, and for now we assume we are generating a one line fix.

\begin{figure}[!t]
    \centering
    \includegraphics[width=0.55\textwidth]{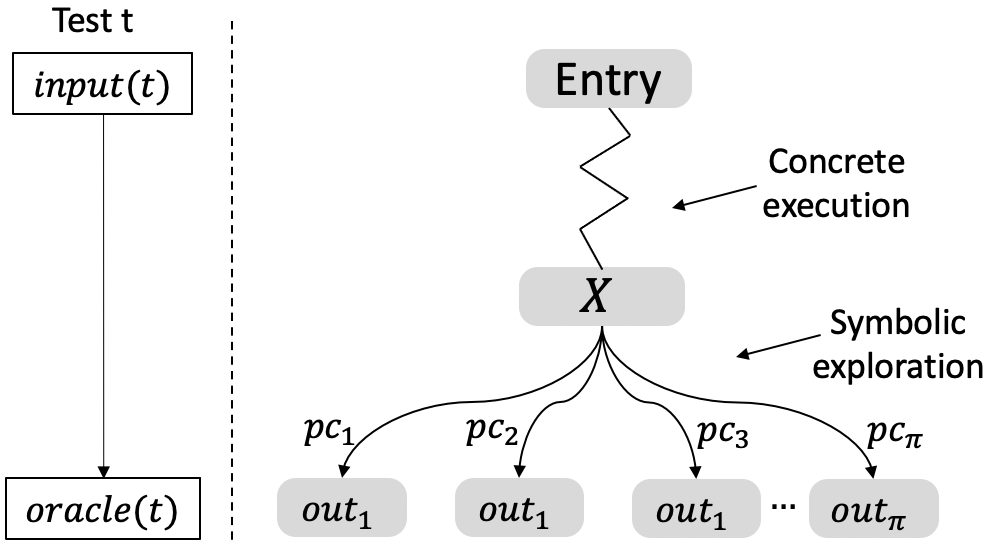}
    \caption{The illustration of the repair constraints.}
    \label{fig:repair_constraint}
\end{figure}

Given such a ghost symbolic variable {\tt X}, the repair constraint is defined in terms of {\tt X} as follows. For a given test $t$, the path up to the fix location $L$ is concrete. From the fix location $L$, there are several possible paths, depending on the value of {\tt X}. Therefore, the path condition of a path $\pi$ from $L$ and the symbolic output along the path in terms of $X$ can be defined. Let these be $pc_\pi$ and $out_\pi$ respectively, as illustrated at Figure~\ref{fig:repair_constraint}. Then a constraint for path $\pi$ can be represented as 
\[
pc_\pi \wedge out_\pi = oracle(t)
\]
where $oracle(t)$ is the expected output for test case $t$. Considering the various paths from $L$ for the execution of test $t$, repair constraint for test $t$ to pass is 
\[
C_t \equiv \bigvee_{\pi} pc_\pi \wedge out_\pi = oracle(t)
\]
The overall repair constraint is the conjunction of the repair constraint collected from all the given tests, since the repaired program is expected to pass all the given tests.
In other words, the repair constraint C is given as follows.
\[
C \equiv \bigwedge_{t} C_t
\]

\smallskip

Once the repair constraint is derived, we do not have the fixed expression immediately. The repair constraint merely describes a property that the fixed expression should satisfy. A fixed expression that satisfies the repair constraint still has to be generated. One mechanism for achieving this - can be via search. The search space of expressions can be represented by 
\begin{itemize}
    \item Variable names
    \item Constants
    \item Arithmetic operators for arithmetic expressions such as +, -, *, ...
    \item Relational operators boolean expressions such as >, <, ==, !=, ...
    \item Other operators that can appear in expressions such as if-then-else, {\em e.g.}, the expression $x > 10? 1: 0$ which returns 1 if $x>10$ and $0$ otherwise.
\end{itemize}

Given such a search space, a layered search in the domain of expressions can be performed as follows. 
\begin{itemize}
    \item Constants appearing in the lowest layer - layer 0
    \item Variables appearing in layer 1
    \item Arithmetic expressions appearing in layer 2 
    \item Boolean expressions which can contain arithmetic expressions as sub-expressions in layer 3
    \item expressions with if-the-else, which can contain arithmetic and boolean expressions as sub-expressions, in layer 4 
    \item and so on, with more and more complex expressions appearing in higher layers.
\end{itemize}
While searching in the domain of expressions, we can also enforce axioms of the operators used in the expressions such as 
\[ 
x + y \equiv y + x 
\]
or 
\[
x + 0 \equiv x 
\]
or 
\[
a \wedge b \equiv b \wedge a 
\]
While we are performing such a (layered) search over the domain of expressions, we are searching for an expression which satisfies the repair constraint. Taking such an approach is different from search-based repair, where we simply search over the domain of expressions, and find an expression which passes all the given tests. The main difference in this approach is that we construct a repair constraint, and then search for expressions which satisfy the repair constraint. The repair constraint amounts to a generalisation of the criterion of passing all given tests, since the repair constraint is derived from a symbolic analysis of the possible test executions. 

\begin{figure}[t!]
\centering
\includegraphics[width=0.6\textwidth]{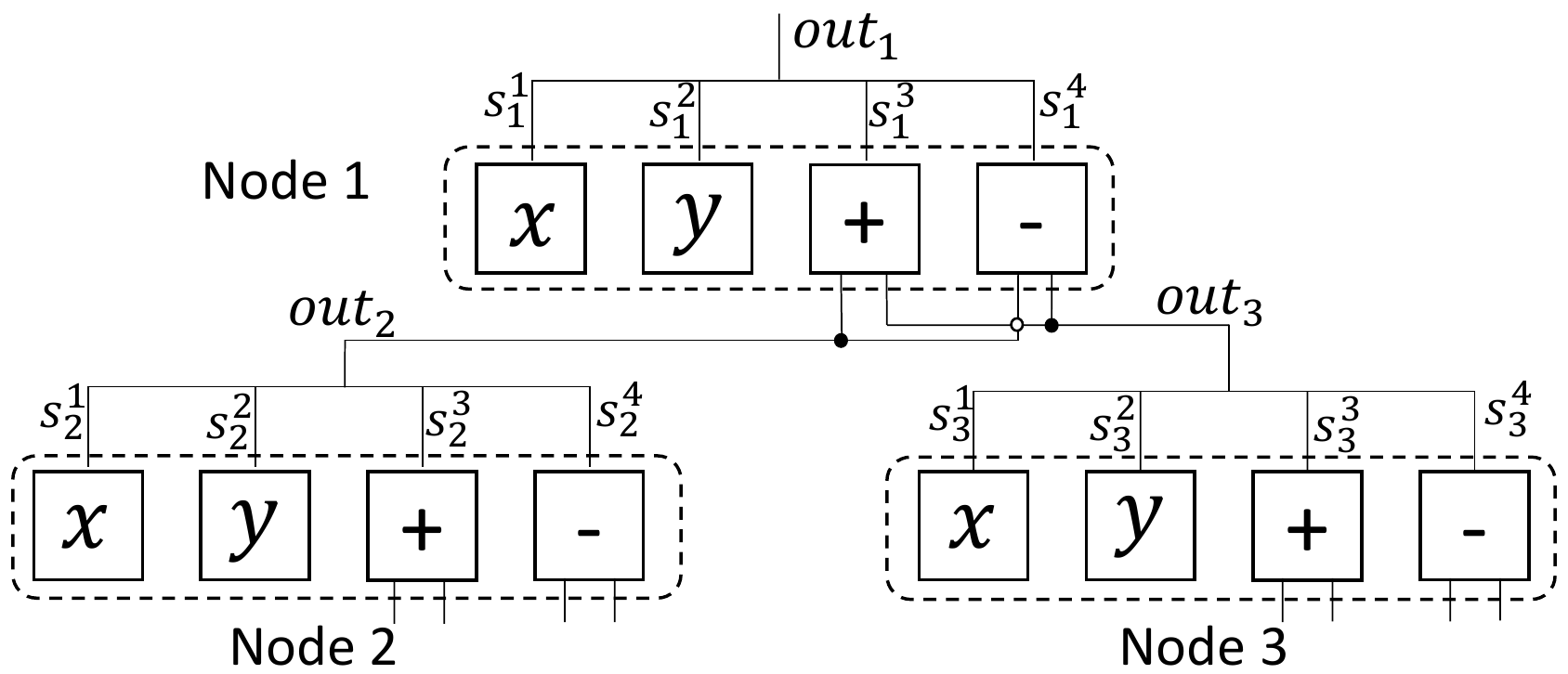}
\caption[SE-ESOC encoding with four components and three nodes.]{SE-ESOC encoding with four components ("x", "y", "+", "-") and three nodes.}
\label{fig:seesoc}
\end{figure}

Apart from conducting a search on the domain of expressions, it is also possible to employ program synthesis techniques to generate an expression meeting the repair constraint. In particular semantic repair techniques \citep{semfix} have advocated the use of component-based program synthesis \citep{oracle-guided10} or second-order synthesis (\emph{SE-ESOC})~\citep{mechtaev2018seesoc}. 
Given a set of pre-defined components $C$, SE-ESOC first constructs a set of candidate programs represented as a tree.
In the constructed tree, each leaf node represents a component without input, while an intermediate node takes the output of its child nodes as inputs.
A tree with three nodes is shown in Figure~\ref{fig:seesoc} shows, where each node is constructed using four components (`x', `y', `+', `-').
Intermediate node 1 (where components `+' and `-' are used) has two child nodes since `+' and `-' take two inputs, while leaf nodes 2 and 3 do not have child node (where components `x' and `y' are used).
Node $i$'s output is donated as $out_i$, and its inputs is donated by \{$out_{i_1}$,$out_{i_2}$,...,$out_{i_k}$\}, meaning the outputs of $i$'s child nodes \{$i_1$,$i_2$,...,$i_k$\} are the inputs of node $i$.
Besides, a boolean variable $s_i^j$ is associated with $j$-th selector of node $i$, representing whether $j$-th component is used in node $i$.
The semantics of $j$-th component is represented as function $F_j$, e.g., $F$ is an addition for component `+'.
For the tree in Figure~\ref{fig:seesoc}, suppose \{$s_1^3$, $s_2^1$, $s_3^2$\} are marked true, the output of the root node is $x+y$.
To ensure the generated program is well-formed, its well-formedness constraint is encoded as $\varphi_{wfp} := \varphi_{node} \wedge \varphi_{choice}$, such that:

\begin{equation}
\begin{aligned}
&\varphi_{node} := \bigwedge_{i=1}^{N}\bigwedge_{j=1}^{|C|}\left({s_i^j \Rightarrow \left(out_i = F_j\left(out_{i_1},out_{i_2},...,out_{i_k}\right)\right)}\right)&
\end{aligned}
\end{equation}
\begin{equation}
\begin{aligned}
&\varphi_{choice} := \bigwedge_{i=1}^{N}exactlyOne\left(s_i^1,s_i^2,...,s_i^C\right)&
\end{aligned}
\end{equation}

Constraint $\varphi_{node}$ demonstrates the semantic relations between each node (the output of child node is the input of parent node). 
To ensure only exactly one component is selected inside each node, constraint $\varphi_{choice}$ restricts is introduced.
With $\varphi_{node}$ and $\varphi_{choice}$, the output of root node is restricted to be a valid function $f$ which connects inputs and outputs of each node (and each node has only one component).
Then, the goal of synthesis is to search for a valid function $f$ by traversing the abstract tree and make sure the generate $f$ satisfies the repair constraint, i.e., the given input-output relations.
With the above formalization, the synthesis problem is transformed into a constraint solving problem.
The generated $f$ meets the repair constraint, and could serve as the patch.

\subsection{Readable and Smallest Fix}

It is often suitable to repair a program by enacting the 'smallest" possible change. The notion of "smallest" that may be a subject of discussion. In general, it is possible to use semantic approaches to find the smallest fix in terms of patch size, as shown by the work of \directfix~\citep{directfix}.

We note that already semantic repair techniques can generate more readable and concise fixes than search based repair techniques like~\genprog. The reason for this is simple, search-based repair techniques as they often insert/delete code may end up inserting/deleting statements, the effect of which could have been accomplished by simple modifications of existing statements in a program. To illustrate this point, let us consider the example in the following, taken from \cite{Mechtaev-thesis}. The statement {\tt x > y} is supposed buggy and needs to be changed to {\tt x >= y}.

\begin{verbatim}
x = E1 ; // E1 represents an expression.
y = E2 ; // E2 represents an expression.
S1 ; // Neither x nor y is redefined by S1.
if (x > y) // FAULT: the conditional should be x >= y
return 0;
else
return 1;
\end{verbatim}

A search-based repair tool like \genprog which works at the statement level could produce a fix like the following by inserting a line of code.

\begin{verbatim}
x = E1 ; y = E2 ;
if (x == y) { S1; return 0; } // This line is one possible repair.
S1 ;
if ( x > y )
return 0;
else
return 1;
\end{verbatim}

A semantic repair tool like \semfix works at the level of expressions and would instead generate the following fix. 

\begin{verbatim}
x = E1 ;
y = E2 ;
S1 ;
if (x >= y) // SIMPLE FIX: >= is substituted for >
return 0;
else
return 1;
\end{verbatim}

The above example shows the utility of semantic program repair tools in generating more maintainable fixes. However, there is not any quality indicator in-built into a tool like \semfix to generate the smallest possible patch. To illustrate this point, we show another example program. We refer the reader to the buggy program in the following – the first two lines are mistakenly swapped,
and the equal signs (=) are omitted.

\begin{verbatim}
if (x > y) // FAULT 1: the conditional should be x >= z
if (x > z) // FAULT 2: the conditional should be x >= y
out = 10;
else out = 20;
else out = 30;
return out ;
\end{verbatim}

\semfix can generate a repair as follows. This also shows the power of \semfix as a single line repair tool to replicate the power of multi-line fixes.
\begin{verbatim}
if ( x > y )
if ( x > z )
out = 10;
else
out = 20;
else out = 30;
return ((x>=z)? ((x>=y)? 10 : 20) : 30); // This line is the fix
\end{verbatim}

Of course, the repair automatically constructed by \semfix is not the smallest one. Consider the following "intuitive" repair which reverses the effect of the "bug" we enunciated earlier
\begin{verbatim}
if (x >= z) // SIMPLE FIX: >= z is substituted for >y
if (x >= y) // SIMPLE FIX: >= y is substituted for >z
out = 10;
else
out = 20;
else out = 30;
return out;
\end{verbatim}

This repair is simpler despite the fact that it modifies two lines
of a program (\semfix cannot modify multiple lines). Generating such concise fixes by an automated repair tool would be valuable. We now briefly discuss some efforts in this direction. More importantly, by showing these examples, we hope to have illustrated that there are often several correct fixes for a bug, and even though they may all be correct - some of the fixes may be more desirable because of conciseness, readability, and maintainability. 
Apart from choosing among correct fixes to get fixes which are more maintainable, there are other benefits of generating smaller fixes. Smaller fixes are unlikely to cater to special cases; they can be more general in nature, and hence less overfitting to a given test-suite. As an example consider the following program with inputs {\tt s, c, k}. The string {\tt s} is of length {\tt k}, and the program is supposed to check if character {\tt c} is in string {\tt s}.

\begin{verbatim}
for ( i =0; i<k-1; i ++)
    if ( s [i] == c ) return TRUE;
return FALSE;
\end{verbatim}

Suppose the test-cases are as follows.
\begin{center}
    \begin{tabular}{|c|c|c|c|c|}
    \hline
    s & c & k & Expected output & Actual Output\\
    \hline
    {\tt "ab?"} & {\tt '?'} & 3 & TRUE & FALSE \\
    {\tt "ab?c"} & {\tt '?'} & 4 & TRUE & TRUE \\
    {\tt "!ab"} & {\tt '!'} &  3 & TRUE & TRUE\\
    \hline
    \end{tabular}
\end{center}
One can have overfitting fixes, based on this test-suite, such as the following fixed program.
\begin{verbatim}
for ( i =0; i < k ; i ++)
  // The following line is one possible overfitting repair.
   if (c == ’?’ || c == ’!’) return TRUE;
return FALSE;
\end{verbatim}
By insisting on the smallest fix, we can generate the following fixed program which not only passes the given tests, but also is not overfitting.

\begin{verbatim}
for ( i =0; i<=k; i ++) // Smallest fix
   if ( s [i] == c ) return TRUE;
return FALSE;
\end{verbatim}

To produce the smallest fix of a program which passes a given test-suite, we can use a partial MaxSMT solver. We can provide the input valuations of the given test-cases $T$, the expected outputs of $T$, as well as the logical formula corresponding to a program $P$ --- all as an SMT formula $\varphi$. Since the program $P$ is buggy and it does not pass all the tests in $T$, the formula $\varphi$ is unsatisfiable. Partial Max-SMT then looks for the smallest change to the formula $\varphi$ that will make it satisfiable. 

The technique for producing the smallest fix is described in the \directfix work \citep{directfix}. The \directfix method treats a program as a circuit. To generate a patch, \directfix changes some of the existing connections in the circuit, adds new components, and add some new connections. To generate the simplest (smallest) repair, \directfix changes as few connections/components as possible.
To achieve this goal, \directfix converts the problem of program repair into an Partial MAX-SMT problem. The reader can obtain more details from \cite{directfix}.

\subsection{Angelic Value based repair}\label{sec:angelix}

Generating a repair constraint to capture the properties of a program satisfying the given tests, is the key characteristic of semantic program repair techniques. However the repair constraint computation can be expensive, the repair constraint can be large depending on the number of tests, and the solving of the repair constraint may involve an implicit search in the domain of expressions (as is accomplished in search-based program synthesis techniques). 

The repair constraint acts as a specification of the desired program. Instead of describing this specification as a constraint, it may also be possible to describe the specification as a collection of values. Essentially if we have a given fix location, we can ask the question, what values if observed at the fix location for a given test $t$, will rescue the execution of test $t$. Such values which will rescue the execution are known as {\em angelic values}. 

The utilisation of value based specifications or angelic values for program repair has been articulated in the work of \angelix~\citep{angelix}. The technique also naturally provides a recipe for multi-line program repair. 
In this work, a controlled customized symbolic execution is conducted by inserting symbols at fix locations. This is where multi-line program repair is naturally supported - since for each fix location a symbolic variable can be substituted. Thus if two symbolic variables 
$\alpha, \beta$ are introduced at two fix locations, along a path the formula to be solved will be 
\[
pc(\alpha, \beta) \wedge out == expected\_out
\]
where \textit{out} is the output variable of symbolic execution, \textit{expected\_out} is the oracle and \textit{pc} is the path condition. Solving this formula will give values of $\alpha, \beta$ which can rescue the execution when driven along a particular path, the one whose path condition $pc$ is computed. The procedure can be repeated for different paths - giving us a {\em collection} of $(\alpha, \beta)$ values --- which we call an angelic forest. This angelic forest represents a value-based specification which is used to drive program synthesis, and produce desired expressions at the fix locations where the symbolic variables $\alpha, \beta$ were inserted.

The concept of using value-based specifications to guide program repair is a general one. The concept of angelic values, or collections of angelic values which rescue a test execution when driven along different paths, represents the crux of the idea of using value-based specifications for program repair. However, there may be different mechanisms for arriving at the value-based specification. In the work of \angelix~\citep{angelix}, the computation of the angelic values is achieved by symbolic execution. For more restricted settings, we can use other techniques to compute angelic values. The work of \spr~\citep{spr} and its extension \prophet~\citep{prophet}, uses enumerative search to search through possible values to rescue test execution. However, the technique only studies repairing conditional expressions which return true or false. So for the execution of a test $t$, if a conditional expression $b$ is being executed $n$ times, the technique could perform an enumerative search over $\{true, false\}^n$ to find value sequences which allow $b$ to be repaired, without altering the number of loop iterations. Given such a value sequence $ V_{desired} \in \{true, false\}^n$, one could look for a "syntactically minimal" modification of $b$, say $b'$ which when replacing $b$, returns the value sequence $V_{desired}$.

\begin{figure}[b]
    \centering
\begin{verbatim}
- if ( max_range_endpoint < eol_range_start )
-   max_range_endpoint = eol_range_start ;
- printable_field = xzalloc ( max_range_endpoint / CHAR_BIT +1);
+ if ( max_range_endpoint )
+   printable_field = xzalloc ( max_range_endpoint / CHAR_BIT +1);
\end{verbatim}
    \caption{An Example adapted from \cite{angelix}}
    \label{fig:angelix}
\end{figure}

We now illustrate the working of the \angelix method with an example taken from \angelix~\cite{angelix} as shown in Figure~\ref{fig:angelix}.
The following shows the code transformations involved in fixing the coreutils bug 13627. 
In this buggy program, the {\tt xzalloc} allocates a block of memory (line 4), however, the call to  {\tt xzalloc} causes segmentation fault. 
To fix this bug, a patch adds a condition before the problematic call to {\tt xzalloc} as shown in line 5.
In the fixed version, {\tt xzalloc} is called only if variable {\tt max\_range\_endpoint} has a non-zero value.
Meanwhile, this patch also removes the original if statement as shown in lines 1–2.
Otherwise, at line 2, {\tt max\_range\_endpoint} will be overwritten with {\tt eol\_range\_start}, which is non-zero.
Hence, the new if condition would not be able to prevent the problematic call to {\tt xzalloc}.

Generating such complex multi-line fixes is beyond the ability of many program repair methods. This is because the changes in one line may impact other lines of code, so it becomes hard to reason locally for the different blocks of code, where changes are needed.  Put more technically, if we see the search space as the space of all fixes, clearly the search space keeps changing as we synthesize changes at each line of the program. 

We can now show, at a high level, how \angelix~\citep{angelix} constructs the multi-line repair, and the technicalities involved in the process. First, we add conditions to each unguarded assignment statement to precisely define the space of fixes. Note that \angelix can synthesize both arithmetic and boolean expressions, it is not restricted to synthesizing boolean expressions. However, adding the trivial conditionals as shown in the following gives the repair technique flexibility to define and work with the fix space precisely.

\begin{verbatim}
if ( max_range_endpoint < eol_range_start )
     max_range_endpoint = eol_range_start ;
if (1)
    printable_field = xzalloc ( max_range_endpoint / CHAR_BIT +1);
\end{verbatim}

Afterward, the repair algorithm finds out the $n$ most suspicious expressions. Such expressions can be boolean expressions, return values, or right-hand sides of assignments. In this case, suppose $n$ is 2, and statistical fault localization of the program points us to the two if conditions. These will then be replaced by symbolic variables, on which the symbolic execution will proceed. So the program can now be seen as conceptually being transformed to 

\begin{tabular}{l l}
{\tt if (} & $\alpha$ {\tt )\{}\\
& {\tt max\_range\_endpoint = eol\_range\_start ;}\\
{\tt if (} & $\beta$ {\tt )}\\
& {\tt printable\_field = xzalloc ( max\_range\_endpoint / CHAR\_BIT +1);}
\end{tabular}

The repair technique then runs symbolic execution on given tests and obtains semantic information relevant to the symbolic variables, in this case, $\alpha$ and $\beta$. 

The repair algorithm then uses a component-based synthesis based on maxSMT reasoning to find the least disruptive fix of the corresponding expressions. This is so that the suspicious expressions are rectified but with the minimal change to the expressions. This finally leads to the fix, as follows

\begin{verbatim}
if (0)
    max_range_endpoint = eol_range_start ;
if (! (max range endpoint == 0))
    printable_field = xzalloc ( max_range_endpoint / CHAR_BIT +1);
\end{verbatim}

We note that techniques like \angelix have significant usage not only in rectifying program errors but also in fixing security vulnerabilities. An example of this capability is the use of the technique to fix the well-known Heartbleed bug, as reported in \angelix~\cite{angelix}. The Heartbleed bug is an exploitable vulnerability which allows attackers to read beyond an intended portion of a buffer, a buffer over-read error. It is a security vulnerability in the popular OpenSSL cryptographic software library, which allows stealing information protected by the SSL/TLS encryption used to secure applications such as web, email, instant messaging and so on. To repair the Heartbleed bug using \angelix we use the publicly available tests as the test-suite and the OpenSSL implementation.  The buggy part of the vulnerable OpenSSL implementation is as follows 

\begin{verbatim}
if ( hbtype == TLS1_HB_REQUEST ) {
    ...
    memcpy ( bp , pl , payload );
    ...
}
\end{verbatim}
and the fix involves checking the {\tt payload} to avoid buffer over-read
\begin{verbatim}
if (1 + 2 + payload + 16 > s->s3->rrec.length)
return 0;
...
if ( hbtype == TLS1_HB_REQUEST ) {
/* receiver side : replies with TLS1_HB_RESPONSE */
     ...
}
else if ( hbtype == TLS1_HB_RESPONSE ) {
/* sender side */
   ...
}
return 0;
\end{verbatim}

In contrast, \angelix produces the following fix which is functionally equivalent to the developer fix.
\begin{verbatim}
if ( hbtype == TLS1_HB_REQUEST && payload + 18 < s->s3->rrec.length) {
/* receiver side : replies with TLS1_HB_RESPONSE */
   ...
   memcpy ( bp , pl , payload );
   ...
}
\end{verbatim}

The work on \angelix~\citep{angelix} was the first to demonstrate automated repair of security vulnerabilities via program repair. This application of program repair has huge promise since currently even after vulnerabilities are detected, and marked as CVEs, they remain un-fixed, increasing the exposure of software systems. Subsequent to the work of \angelix, a number of works have developed symbolic execution based methods for security vulnerability repair, such as \senx~\citep{senx}, \extractfix~\citep{Gao21} and \cpr~\citep{cpr}. The reader is referred to these works for the latest results on security vulnerability repair technology, which remains a promising and exciting direction of research.

\subsection{Discussion}
Constraint-based repair techniques first encode the requirement to satisfy the correct specification (e.g., passing given test cases) as a set of constraints.
Then, they synthesize a patch by solving the constraints.
The repairability of constraint-based repair is not restricted by pre-defined transformation operators.
Besides, representing a collection of patch candidates as a symbolic constraint enables constraint-based repair to explore huge candidate patch space.
On the other hand, constraint-based repair suffers from scalability problems.
The reason is two-fold: 1) heavy symbolic execution is usually used to explore program paths and collect constraints (e.g., \angelix and \extractfix), which cannot scale to large programs, and 2) solving constraints using Satisfiability Modulo Theories (SMT) solver could be time- and resource-consuming, limiting constraint-based repair to scale to large programs.

%% file: chapters/4_learning.tex
\section{Learning-based Repair}\label{sec:learning}

Another line of repair is to learn repair strategies from human patches.
The workflow of those techniques is as follows: (1) mine human patches that fix software bugs from software repositories (e.g., open-source projects), (2) learn a code repair model from the mined human patches, and then (3) apply the learned model to the buggy programs to produce patches.
Compared to the search-based and semantic-based APR techniques shown in the previous chapters, the repairability of learning-based APR techniques does not rely on predefined transformation operators/rules.
Instead, they automatically learn repair strategies from available human patches.

\subsection{Sequence-to-Sequence Translation}

First, APR can be treated as a Neural Machine Translation (NMT) problem, which translates buggy programs to a corresponding fixed version.
NMT is widely used in natural language processing (NLP), which translates text from one language (e.g., English) to another (e.g., Chinese).
At its core, NMT is a sequence-to-sequence (seq2seq) neural network model that predicts the output sequences for a given input sequence.
In recent years, by treating APR as a translation problem, seq2seq models have been applied to fix software bugs.


\subsubsection{Sequence-to-Sequence Model}
\begin{figure}[!t]
    \centering
    \includegraphics[width=0.9\textwidth]{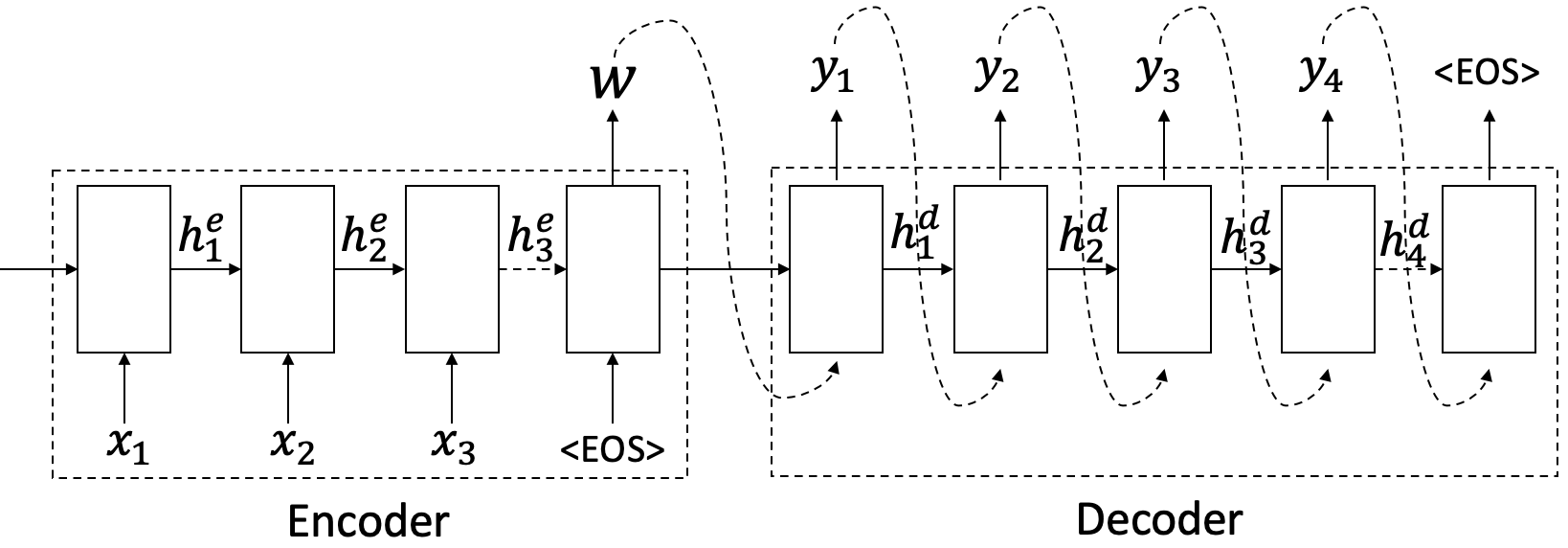}
    \caption{The architecture of neural machine translation model}
    \label{fig:S2SModel}
\end{figure}

A sequence-to-sequence model usually uses a recurrent neural network (RNN) to read an input sequence and generate an output sequence.
Figure~\ref{fig:S2SModel} shows the architecture of a seq2seq model.
Let us consider the input token sequence denoted by \{$x_1, x_2, \dots, x_n$\} , where $n$ is the length of input sequence, and the output token sequence denoted by \{$y_1, y_2, \dots, y_m$\}, where $m$ is the length of output sequence.
At the end of the input and output sequence, a special {<}\texttt{EOS}{>} token is used, representing the \textit{end of sequence}.
A seq2seq model consists of an \texttt{encoder} and a \texttt{decoder}, where encoder encodes input tokens \{$x_1, x_2, \dots, x_n$\} as an intermediate representation $w$ in each training phase.
The outputs of the encoder are then fed into the decoder to predict proper output tokens.

As shown in Figure~\ref{fig:S2SModel}, both encoder and decoder consist of $N$ recurrent units (e.g.,LSTM, GRU).
In each training phase, the encoder reads the input sequence and summarizes the information as ``states'', which is then used to guide the decoder to make accurate predictions.
The state $h^e_t$ of the $t$-th unit in the \textbf{e}ncoder is computed as follows:
\begin{align}
h^e_t=f_e(W^{hx}x_t, W^{hh}h^e_{t-1}),\ t=1, \dots, n
\end{align}
where $W^{hx}$ is a weight matrix computing how input $x_t$ affects state $h^e_t$, and $W^{hh}$ is a weight matrix computing how the previous state affects the current state (i.e., recurrence).
The decoder is also an RNN whose first state is initialized as the final state of the encoder. 
In other words, the states of the encoder’s final unit are the input to the first unit of the decoder network.
Using the states produced by encoder, the decoder starts generating the first output token $y_1$ and state $h^d_1$. 
Formally, in each training phase, the $t$-th state of decoder $h^d_t$ is computed as follows:
\begin{align}
h^d_t = f_d(W^{hh}h^d_{t-1}, W^{hy}y_{t-1}),\ t=1, \dots, m
\end{align}
where $W^{hy}$ is a weight matrix computing how previous output affects the current hidden state, and $W^{hh}$ is a weight matrix that is related to recurrence.
Decoder predicts $y_t$ according to previous output $y_{t-1}$ and hidden state $h^d_t$, the intermediate representation $w$ produced by encoder.
\begin{align}
P(y_t | y_{t-1}, y_{t-2}, y_{t-3}, \dots, y_1, w) = g(h^d_t, y_{t-1}, w)\label{equ:s2s_output}
\end{align}
In high level, the value of $y_t$ should be predicted according to input sequence \{$x_1, x_2, \dots, x_n$\} and the previous output sequence \{$y_1, y_2, \dots, y_{t-1}$\}.
Function $g$ calculates the value of $y_t$ according to the hidden state and previous outputs.
The above formalization shows the process of one training phase.
In multiple training phases, all weights are learned with back-propagation mechanism and supervised learning.

\subsubsection{Program Representation as Token Sequence}
To use seq2seq models for program repair, a program is required to be represented as a sequence.
Typical programs consist of different kinds of tokens, including keywords, variables, literals, special characters (e.g., dots), functions, types, etc.
The most straightforward approach is to extract such tokens from program to form token sequences.

DeepFix~\citep{DeepFix} extracts token sequences as follows.
Some tokens (including types, keywords, special characters, and library functions), which share the same vocabulary across different programs, are retained when representing a program.  
The other types of tokens may have different vocabularies across different programs, for instance, different programs may use different variable names, define functions using different styles, and use specific literals (e.g., ``New York'').
Therefore, such tokens cannot be retained when learning a general model from multiple programs, so they are modeled as follows.

A fixed-size pool of names is first defined and then an encoding map is constructed for each program.
The distinct identifiers (e.g., variable or function name) in the program are mapped to a unique name in the defined pool. 
As for literals, except for some special values (e.g., 0, MAX\_INT, MIN\_INT), the actual values of literals may not be important for learning task. 
Therefore, except for the literals with special values, the literals are mapped to special tokens based on their type, i.e., map all integer literals to \texttt{NUM} and all string literals to \texttt{STR}.
Moreover, the end of a token sequence is denoted using a special token {<}\texttt{EOS}{>}.

With the above encoding strategy, the original buggy program is treated as a sequence of tokens $X$. 
Similarly, the fixed version of the program is represented as another token sequence $Y$.
Given $X$ and $Y$, a seq2seq network can be then used to train a translation model that translates a buggy program to its fixed version.

However, a typical program usually has hundreds of tokens or even more, therefore, predicting long target sequences accurately is very challenging.
To address this problem, when encoding the program, one solution is to also encode line numbers in the program representation.
Specifically, a statement $s$ at a line $l$ in a program is encoded as a pair ${(T_l, T_s)}$ where ${T_l}$ and ${T_s}$ are tokenizations of $l$ and $s$, respectively.
A program \texttt{prog} with $k$ lines is represented as $\{(T_{l1}, T_{s1}), (T_{l2}, T_{s2}), \dots, (T_{lk}, T_{sk}), {<}\mathtt{EOS}{>}\}$ where $(T_{li}, T_{si})$ is the encoded line number and program at line $li$.
Instead of encoding the entire fixed program as output token sequence, the output sequence $Y$ could be represented as a pair ${(T_{li}, T_{si})}$ or a set of pairs $\{(T_{li}, T_{si}), \dots, (T_{lj}, T_{sj})\}$, meaning that the bug is fixed by changing statement $si$ at line $li$ or changing multiple lines between $\{li, \dots, lj\}$.
Compared to the sequence representing the entire fixed program, this output sequence is much smaller and easier to predict.
Given the input sequence $X$ and output sequence $Y$, a repair model can be then trained using seq2seq network.

Since seq2seq models are known to struggle with long sequences, learning with the entire program as an input token sequence is also challenging.
To solve this problem, one simple idea is to (1) just consider the code surrounding the buggy statement~\citep{SequenceR} (i.e., context) and (2) abstract away the program details from the code.
Constructing the abstracted buggy context needs to balance: (1) retain as much information as possible to keep enough context for enabling the model to predict a likely correct fix, and (2) reduce the context into an as much more concise sequence of tokens as possible.

To construct the abstracted buggy context, given a buggy program, the first step is to identify the buggy location.
Fault localization techniques, which are commonly used by existing APR tools, could identify the suspicious buggy lines $b^l$ and the buggy method $b^m$.
Given bug locations $L = \{l_1, l_2, \dots\}$ identified by fault localization technique, where $l_i$ is represented as a tuple $\{b^c_i, b^m_i, b^l_i\}$, representing the buggy class or file ($b^c_i$), the buggy method ($b^m_i$), and the buggy line ($b^l_i$), respectively. 
For each $l_i \in L$, the context around $l_i$ is abstracted and encoded as follow:
\begin{itemize}
    \item \textbf{Buggy Line} To indicate which line is buggy, before the first token and after the last token in the buggy line $b^l_i$, a <\texttt{BUG\_START}> and <\texttt{BUG\_END}> is inserted, respectively.

    \item \textbf{Buggy Method} Since buggy method $b^m_i$ including the buggy line has important information about where $b^m_i$ is and how $b^m_i$ interacts with the rest code in this method, the $b^m_i$ is kept.

    \item \textbf{Buggy Class} Buggy class $b^c_i$ that includes the buggy method also contains important information about the bug. To extract relevant context from $b^c_i$ and keep the context concise, (1) all instance variables are kept, (2) except for buggy method $b^m_i$, the body of the non-buggy methods are stripped out, while their signatures are kept. This is because the buggy code could use instance variables and call the non-buggy methods.
\end{itemize}
With this design, many program details that are likely not relevant to the bug could be abstracted away.
Hence, the scalability of NMT-based repair techniques can be significantly increased.


\subsubsection{Programming Language Model}
Although existing approaches have shown good performance on some datasets, they are still prone to producing incorrect or even uncompilable patches.
One of the reasons that cause uncompilable patches is that NMT-based models have limited knowledge about the strict syntax of programming languages and how developers write code.
To enable NMT models to generate source code that are similar to developers' code, the large number of available source code can be used to pre-train a general programming language model~\citep{cure}.
In the domain of NLP, the pre-training process learns a model over a large number of sequences of natural language text written by human.
Because a pre-trained language model is trained on a huge dataset, it presents the distributions regarding human-like sentence structures.
On top of the pre-training model, one can fine-tune it for a specific task, such as language translation.
The pre-trained model can significantly improve the performance of fine-tuned models for specific tasks.

Inspired by the success of language model in the NLP domain, and considering the existence of a large number of open-source programs, pre-training can also be used to train a \texttt{programming language (PL) model} with the goal of learning general programming language syntax and how developers write code.
Formally, given a sequence of code tokens from open-source programs $X=\{x_1, x_2, \dots, x_n\}$, training a PL model is unsupervised with the goal of maximizing the average likelihood:
\begin{align}
    L_{PL} = \frac{1}{n}\sum^n_{i=1}\mathtt{log} P(x_i|x_1, \dots, x_{i-1}; \theta)
\end{align}
In the above formula, $\theta$ represents the weights of the PL model.
Given a sequence of $\{x_1, \dots, x_{i-1}\}$, the trained model is used to calculate the probability $P(x_i|x_1, \dots, x_{i-1}; \theta)$ that token $x_i$ is the next token.
Generally, PL model training is to find a probability distribution, where real code sequence tokens $\{x_1, \dots, x_{i}\}$ obtained from open-source programs, get a higher $L_{PL}$ probability than other sequences.
Fine-tuning the PL model with the repair dataset generates repair models, that can produce patched programs that are more likely compilable and similar to developer code.

\subsubsection{Copy Mechanism}
The program encoding mechanisms described above have a common issue: 
\textit{only tokens that appear in the training set are available for constructing patches~\citep{SequenceR}}.
For instance, assume a correct patch is to insert an if condition \texttt{if(city==``Singapore'')}, literal ``Singapore'' is not a common used word in programs.
When encoding programs to token sequence, such words (e.g., "Singapore") will be mapped to \texttt{STR}.
In other words, such words (e.g., "Singapore") are not included in the training vocabulary,  however, they are necessary for generating the correct patch.
\textit{Copy mechanism} is one of the successful approaches that are proposed to overcome the vocabulary problem.
The main intuition of copy mechanism is that \textit{the rare words that are not available in training vocabulary can be directly copied from the input sequence to the output sequence}.
This simple idea can be very useful especially when the buggy code and patch contain certain identifier names or literals (e.g., ``Singapore'') - meaning these tokens can be copied from the buggy code to the patch.
To release this idea, a link between the input and output sequence is built by implementing a copy mechanism.
Formally, the copy mechanism contributes to Equation~\ref{equ:s2s_output} when producing token candidates by introducing a probability parameter $p^{gen}$.
Parameter $p^{gen}$ represents the probability that a token is generated by decoder using initial vocabulary, hence $1 - p^{gen}$ is the probability that the token is generated by copying a token from input sequence.
$p^{gen}$ is defined as a learnable function that can be learned during training time:
\begin{align}
p^{gen}_t &= g_c(h^d_t, y_{t-1}) \label{equ:pgen}
\end{align}
Given the hidden state $h^d_t$ and previous token $y_{t-1}$, $p^{gen}_t$ represents the probability of generating $y_t$ from its initial vocabulary.
Therefore, according to $p^{gen}_t$, output token $y_t$ is generated by either: (1) using tokens from the training vocabulary or (2) copying tokens from the input token sequence. Formally, 
\begin{align}
P(y_t) &= p^{gen} * g(h^d_t, y_{t-1}, w) + (1 - p^{gen}) * \sum_{i:x_i=y_t}a_i^t \label{equ:output_with_copy}
\end{align}
where $a_i^t$ represents a learnable attention weight measuring whether input token $x_i$ is same as output token $y_t$.

\subsection{Code Transformation Learning}
Instead of treating program repair as a translation problem, another line of learning-based repair techniques treats APR as a code transformation learning problem.
The transformation learning takes as inputs a set of concrete transformations that fix bugs \{$b_1 \mapsto p_1,b_2 \mapsto p_2, \dots, b_n \mapsto p_n$\}, where $b_i$ is the buggy code (pre-transformation) and $p_i$ is the patched code (after-transformation), and learns a generalized transformation rule $TR$, where $TR$ is expected to be generalized to other buggy code, i.e., transform the buggy code in the wild to their corresponding patched code.


\subsubsection{Learn Transformation Rules via Program Synthesis}
Given an input domain $\mathbb{I}$ and an output domain $\mathbb{O}$, \emph{program synthesis} takes a set $\{i_0 \mapsto o_0, . . . , i_n \mapsto o_n\}$ of input-output pairs and synthesizes a program $P : \mathbb{I} \rightarrow \mathbb{O}$ such that $P(i_k) = o_k$ for $k \in 0...n$.
For the purpose of synthesizing program \textit{transformation rule} $TR$, $\mathbb{I}$ and $\mathbb{O}$ are fixed as $\mathbb{AST}$.
Given a set of history patches \{$b_1 \mapsto p_1,b_2 \mapsto p_2, \dots, b_n \mapsto p_n$\}, the aim is then to synthesize $TR$ such that $TR(b_i) = p_i$ for $i \in 0...n$.
The synthesized $TR$ then serves to transform an input AST into an output AST.
In general, the goal is to synthesize a transformation rule that is the generalization of the history patches, and expect the synthesized rule can be used to fix other bugs.

\begin{figure}
\begin{Verbatim}[commandchars=\\\{\}]
rule         := (guard, transformer)
guard        := pred | Conjunction(pred, guard)
pred         := IsKind(node, kind)
                | Attribute(node, attr) = value          
                | Not(pred)
transformer  := select | construct
construct    := Tree(kind, attrs, childrenlist)
childrenlist := EmptyChildren | select | construct
                | Cons(construct, childrenlist)
                | Cons(select, childrenlist)
select       := Match(guard, node)
node         := ...
\end{Verbatim}
\caption{Domain-specific language for transformation rules}
\vspace{-8pt}
\label{fig:dsl}
\end{figure}

The transformation rules can be expressed using a domain-specific language defined in Figure~\ref{fig:dsl}.
In the DSL, transformation rule $TR$ can be represented as ($\mathtt{guard}, \mathtt{trans}$)~\citep{refazer} which is defined as follows:
\begin{itemize}
    \item \texttt{guard}: $\mathbb{AST}\rightarrow \mathtt{Boolean}$: \texttt{guard} is a set of conjunctive predicates over AST nodes.
    Specifically, given an AST node, a $\mathtt{guard}$ checks whether the node's type, code, and other attributes satisfy its predicate, and it then returns a corresponding Boolean value;
    \item \texttt{trans}: $\mathbb{AST}\rightarrow\mathbb{AST}$: \texttt{trans} transforms an input AST to output AST. The \texttt{trans} builds the output AST using the following two operations: (1) $\mathtt{select}$: using an existing node from input AST (similar to the copy mechanism shown above), and (2) $\mathtt{construct}$: creating a new AST node from scratch.
\end{itemize}
Typically, $\mathtt{guard}$ determines AST that should be transformed, while $\mathtt{trans}$ specifies how to transform an AST.
Thus, for an AST $t$, if $\mathtt{guard}(t)$ is $\mathtt{true}$, $TR(t) = \mathtt{trans}(t)$, otherwise, $TR(t)$ = $\bot$.
In general, given a set of patches \{$b_1 \mapsto p_1,b_2 \mapsto p_2, \dots, b_n \mapsto p_n$\}, synthesizing $TR$ is actually inferring an abstracted transformation accordingly to the concrete examples, such that \texttt{TR}($b_i$) = $p_i$ for all $i \in 0 \dots n$.

Consider the following two patches adapted from APIFix~\cite{gao2021apifix} $p_1$: \texttt{handler.}\texttt{Handle(request)} $\mapsto$ \texttt{handler} \texttt{.Handle(request}, \texttt{token)} and $p_2$: \texttt{TestSubject.Handle} \texttt{(request)} $\mapsto$ \texttt{TestSubject.Handle} \texttt{(request}, \texttt{token)}.
Given these two transformation examples, a rule $(\mathtt{guard}, \mathtt{trans})$ will be synthesized.
The \texttt{guard} of the transformation rule is represented by a set of conjunctive predicates
\begin{align*}
\mathtt{IsKind}(node, &``\mathtt{InvokeExpr}") \\
    \wedge\mathtt{IsKind}&(node.children[0], ``\mathtt{MemberAccess}")\\
        &\wedge \mathtt{IsKind}(node.children[0].children[0], ``\mathtt{Identifier}") \\
        &\wedge \mathtt{IsKind}(node.children[0].children[1], ``\mathtt{DotToken}") \\
        &\wedge \mathtt{IsKind}(node.children[0].children[2], ``\mathtt{Identifier}") \\
        &\ \ \ \ \ \ \ \ \wedge (\mathtt{Attribute}(node.children[0].children[0], ``\mathtt{Text}")=``Handle")\\
    \wedge\mathtt{IsKind}&(node.children[1], ``\mathtt{ArgumentList}")\\
        &\wedge \mathtt{IsKind}(node.children[1].children[0], ``\mathtt{LeftParenthesesToken}") \\
        &\wedge \mathtt{IsKind}(node.children[1].children[1], ``\mathtt{Idenfifier}") \\
        &\ \ \ \ \ \ \ \ \wedge \mathtt{Attribute}(node.children[1].children[1], ``\mathtt{Text}")=``request")\\
        &\wedge \mathtt{IsKind}(node.children[1].children[2], ``\mathtt{RightParenthesesToken}")
\end{align*}
which means that this rule can be applied to \textit{node} only if its kind is ``\texttt{InvokeExpr}'', its first child's (\texttt{node.children[0])} kind is $``\mathtt{MemberAccess}"$, the kind of \texttt{node.children[0].children[0]} is an $\mathtt{Identifier}$, the kind of \texttt{node.children[0].children[2]} is also an $\mathtt{Identifier}$ with attribute value as ``Handle", \texttt{node.children[1]}'s' kind is $``\mathtt{ArgumentList}"$ etc.
Note that, different from \texttt{node.children[0].children[2]}, the attribute value of \texttt{node.children[0].children[0]} is not included in the constraint.
This is because the corresponding attribute values on these two given patches are different (\texttt{handler} vs \texttt{TestSubject}), so their concrete values are abstracted.
Such abstraction ensures this synthesized \texttt{guard} produces \texttt{true} on both given patches.

The \texttt{trans} of the transformation rule is:
\begin{align*}
    \mathtt{Tree}&(``\mathtt{InvokeExpr}", [\ ], [\\
                 &\ \ \ \ select_1, \\
                 &\ \ \ \ \mathtt{Tree}(``\mathtt{ArgumentList}", [],\\ 
                 &\ \ \ \ \ \ \ \ \ \ \ \ \ \ \ [\mathit{select_2}, \mathit{construct_1}])\\
                 &]\\
    )
\end{align*}
The $\mathtt{trans}$ describes how the input AST is transformed into the new AST.
Specifically, it transforms the input AST by creating a new tree in the form of \texttt{Tree(kind, attrs, childrenlist)}.
First, the output AST's type is \texttt{InvokeExpr} with two children. 
The first child ($\mathit{select_1}$) is copied/selected from input AST (\texttt{handler.Handle} of $p_1$ and \texttt{TestSubject.Handle} of $p_2$) according to $\mathit{select_1}$.
The sub-rule $\mathit{select_1}$ extracts a node from input that satisfies:
\begin{align*}
\mathtt{IsKind}&(node, ``\mathtt{MemberAccess}") \\
&\wedge \mathtt{IsKind}(node.children[0], ``\mathtt{Identifier}")\\
&\wedge\mathtt{IsKind}(node.children[1], ``\mathtt{Identifier}")\\
&\wedge~\mathtt{Attribute}(node.children[1], ``\mathtt{Text}")=``\mathit{Handle}")
\end{align*}
and the second child, an argument list, is a newly created tree of type $``\mathtt{ArgumentList}"$.
This node is constructed with two subnodes: $\mathit{select_2}$ and $\mathit{construct_1}$, where $\mathit{select}_2$ copies a node from input AST satisfying:
\begin{align*}
\mathtt{IsKind}(node, ``\mathtt{Identifier}")\wedge
(\mathtt{Attribute}(\mathit{node}, ``\mathtt{Text}")=``\mathit{request}")
\end{align*}
and $\mathit{construct_1}$ constructs a constant AST node from scratch: 
$\mathtt{Tree}(``\mathtt{Identifier}", [``\mathtt{token}"], [])$, which is an $``\mathtt{Identifier}"$ with the name as $``\mathtt{token}"$.
Given a set of input-output examples (history patches in our setting), there are a lot of program synthesis algorithms that can synthesize such a transformation rule ($\mathtt{guard}$, $\mathtt{trans}$).
For the detailed synthesis algorithms, readers are referred to read the tutorial for program synthesis~\footnote{https://people.csail.mit.edu/asolar/SynthesisCourse}.

\subsubsection{Clustering Transformation Rules}
The above synthesis process could produce a lot of transformation rules, which correspond to different fix strategies for different bugs.
To find the most appropriate transformation rule to apply for a given bug, searching from a potentially large space of candidate fixes is expensive.
To mitigate this problem, the transformation rules could be organized into a hierarchical structure~\citep{getafix}.

As we mentioned above, synthesizing transformation rules is actually a process of generalizing concrete transformations (the history patches).
Given a set of transformations, inferring a proper level of generalization is challenging.
If a transformation rule is \textit{under-generalized}, it can cause false negatives, i.e., it cannot transform some programs that should be patched.
For instance, we say transformation rule \texttt{IDENTIFIER.Handle}(\textit{request}) $\mapsto$ \texttt{IDENTIFIER.Handle}(\textit{request, token}) (the rule should be represented via DSL, for simplicity, we show it using code skeleton where \texttt{IDENTIFIER} could match any identifier) is under-generalized since it can be applied only if the argument is \textit{request}, so it fails to patch buggy code \textit{handler}.\texttt{Handle}(\textit{query}).
In contrast, an \textit{over-generalized} transformation rule produces false positives: it may generate patches for some locations that should not be patched.
For instance, we say transformation rule \texttt{IDENTIFIER.IDENTIFIER2}(\texttt{IDENTIFIER3}) $\mapsto$ \texttt{IDENTIFIER.IDENTIFIER2}(\texttt{IDENTIFIER3}, \textit{token}) is over-generalized, since this rule can be applied to any method invocation with one argument.
In general, it is hard to infer whether to (1) generalize the variable, e.g. $\mathit{request}$, (2) generalize the expressions, or (3) even the whole statement.
Instead of inferring a certain level of generalization, the hierarchical structure represents different levels of generalization ranging from general to specific transformation rules.
When fixing a certain bug, the most appropriate transformation rule to apply can be properly selected from the hierarchical structure.

\begin{figure}[!t]
        \centering
        \includegraphics[width=0.8\textwidth]{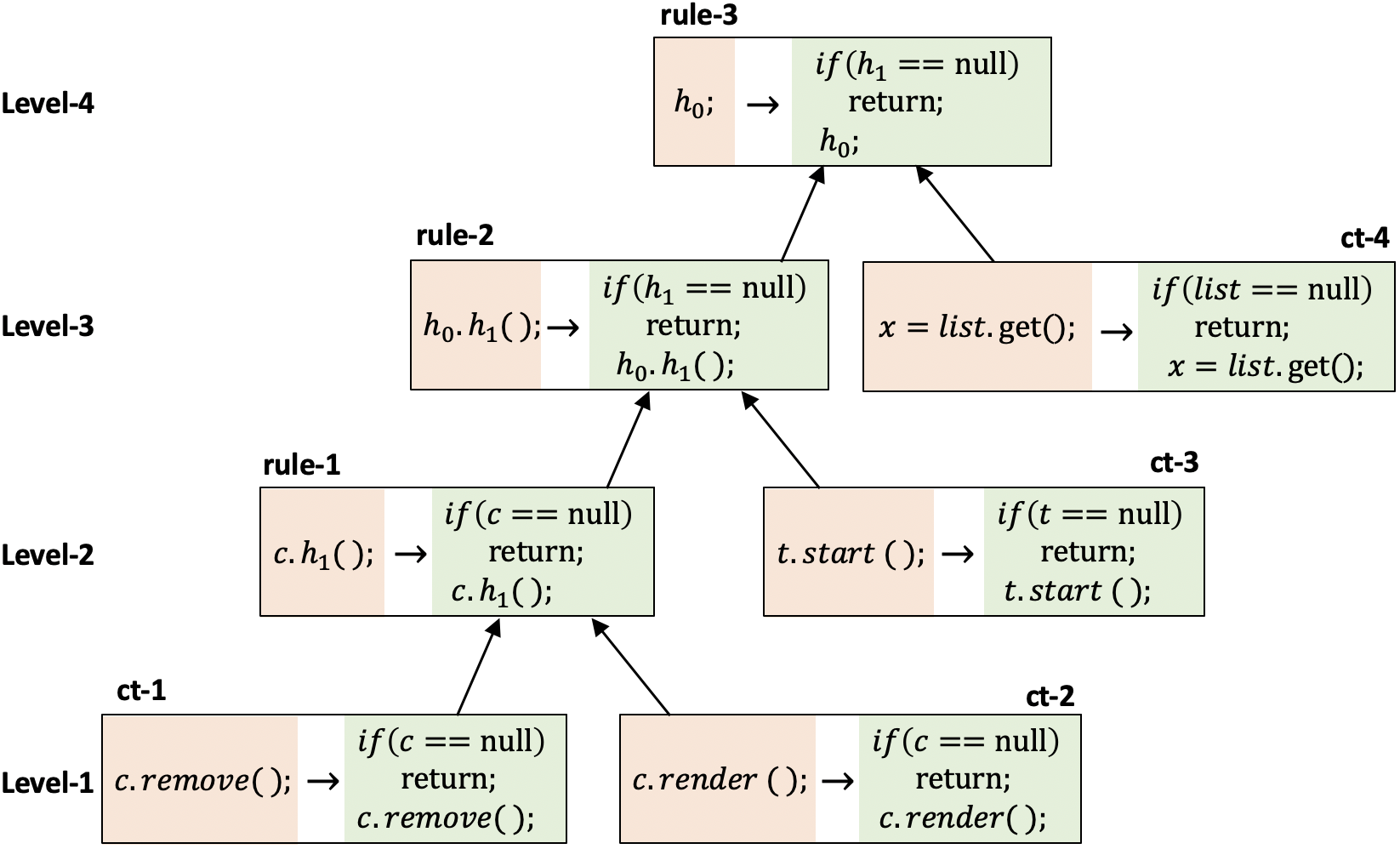}
        \caption{Dendrogram showing concrete transformations merged into more abstract transformation rules. The leaf nodes are concrete transformations and non-terminal nodes are the generalized transformation rules.}
        \label{fig:hierarchy}
\end{figure}

The basic idea to generate the hierarchical structure is to start from concrete transformations and iteratively generalize two similar transformations using the synthesizer until all transformations are generalized to one single transformation rule.
In each iteration, a pair of transformations are picked up to generalize, which will produce the least general generalization of this pair of transformations.
The series of generalization steps finally produces a set of rules in the form of a binary tree, where the parent node is the generalization of two child nodes, and leaf nodes are given concrete transformations.
Figure~\ref{fig:hierarchy} presents an example, which demonstrated a tree of concrete and abstracted transformation rules. 
Each transformation (either concrete transformation or abstracted transformation rule) is denoted with pair of red and green boxes.
Among them, the root of the tree \texttt{rule-3} is the most generalized transformation, while the other rules (\texttt{rule-1}, \texttt{rule-2}) represent different levels of abstractions that are generated by generalizing different concrete transformations.
For instance, the \texttt{rule-1} shows an generalization that abstract the method names (\textit{remove} of \texttt{ct-1} and \textit{render} of \texttt{ct-2}) to $h_1$.
The \texttt{rule-2} further generalizes variable $c$ and $t$ of \texttt{rule-1} and \texttt{ct-3}, respectively.
Keeping intermediate rules, instead of producing the most generalized rule by generalizing all concrete transformations, enables us to apply the most suitable rule to transform a given code.

Given a set of concrete transformations, a novel clustering algorithm has been proposed by GetaFix~\cite{getafix} to generate a tree of transformation rules.
This algorithm maintains a working set $S$ of transformation rules, which is initialized with all the given concrete transformations. 
Then, the following steps are repeated until the size of $S$ is reduced to one:
\begin{itemize}
    \item [(1)] Choose two transformation rules $e_1$, $e_2$ from $S$.
    \item [(2)] Generalize $e_1$, $e_2$ using synthesizer and yields $e_3$ = \texttt{Synthesize}($e_1$, $e_2$).
    \item [(3)] Add $e_3$ to $S$ to $S$, and remove $e_1$ and $e_2$ from $S$.
    \item [(4)] Set $e_3$ as the parent of $e_1$ and $e_2$ in the tree and continue the loop.
\end{itemize}
How to pick the pair $e_1$, $e_2$ from the working set to generalize is crucial for the clustering algorithm.
To minimize the loss of concrete information at each step, the pairs that are similar to each other are always given high priority to be selected.
For instance, for the example shown in Figure~\ref{fig:hierarchy}, the transformations that are applied to \textit{c.remove()} and \textit{c.render()} are first picked up because they are similar and share one same identifier ($c$).

\begin{figure}[!t]
    \centering
\begin{subfigure}{0.45\textwidth}
\begin{lstlisting}
void onDestroyView() {
    (*@\textbf{c.clearListeners()}@*);
    c = null;
}
\end{lstlisting}
$\texttt{\textit{rule-1}}: \mathtt{c.h_0()}$\\
$\mathtt{where\ h_0} \mapsto \mathtt{clearListeners}$
\end{subfigure}
\begin{subfigure}{0.45\textwidth}
\begin{lstlisting}
void onDestroyView() {
    if (c == null) 
        return;
    (*@\textbf{c.clearListeners()}@*);
    c = null;
}
\end{lstlisting}
\end{subfigure}
\caption{Example source code and its corresponding patched code produced by learned transformation rule}
\label{fig:buggy_source_code_example}
\end{figure}

Given a tree of learned rules and a buggy source code $C$ need to be transformed, the last step is to find a proper rule to apply and produce a fixed version of $C$.
In case the over-generalized learned rule produces incorrect results, the most specific rules that can be applied to $C$ are preferred.
For instance, the left part of Figure~\ref{fig:buggy_source_code_example} shows an example source code that needs to be patched by the learned transformation rules in Figure~\ref{fig:hierarchy}.
All the \texttt{rule-3}, \texttt{rule-2} and \texttt{rule-1} are applicable on $C$.
Among them, the most specific transformation rule is \texttt{rule-1}:
\[
\mathtt{c.h_0;\ \mapsto\ if (c\ {=}{=}\ \mathtt{null})\ return;\ c.h_0}();
\]
This transformation rule can be matched to the buggy code by just substituting $h_0$ with \texttt{clearListeners}.
While \texttt{rule-2} and \texttt{rule-3} require substituting more code elements.
By applying this rule, we would get the patched code as shown in the right part of Figure~\ref{fig:buggy_source_code_example}.

\subsection{Discussion}
The learning based repair techniques do not rely on pre-defined transformation operators, enabling them to generate abundant kinds of patches by learning from history patches.
In case of generating uncompilable or incorrect patches, the auto-generated patches by learning-based APR can also be validated using compilers and available test cases just like traditional APR techniques.
However, the current learning-based APR also has a main limitation.
They learn repair strategies across different projects, therefore, they can only learn the common programming language features shared by different projects.
This property causes the learning-based APR can only learn syntactical features, since the semantic features of different projects are usually different.
Besides, the learning-based APR can only fix common bugs shared by different projects, e.g., null-pointer dereference, divide-by-zero, etc.

%% file: chapters/5_overffint_in_synthesis.tex
\section{Overfitting in Program Repair}\label{sec:overfitting}
Although automated program repair techniques have shown their ability in fixing software bugs, their overall fix rate is still pretty low.
According to a recent study, modern APR systems can only fix 11–19\% of the defects in real-world software.
This low fix rate is mainly caused by the fact that specifications driving APR are usually incomplete.
APR aims to fix buggy programs with the goal of making it satisfy given specifications, usually in the form of test cases.
If a patch makes the buggy program satisfy the given incomplete specifications but in an incorrect way, such a patch is called \textit{overfitted} patch.
The overfitted patches can either partially fix the bug or introduce new regression errors.

\subsection{Incomplete Specification}
\label{sec:specification}
We first review the types of specifications driving APR and discuss their impact on patch quality.

\subsubsection*{Test Suites as Specification}
In well-studied test-driven APR techniques, test-suite is treated as a correctness specification.
The test suite is usually composed of a set of passing tests $pT$ and a set of failing tests $fT$.
Then, the repair process aims to fix the buggy program to make it pass both $pT$ and $fT$.
The main advantage of using test suites as specifications is that test cases are widely available.
However, test cases can only specify part of intended program behaviors, hence, they are incomplete specifications. 
Even if a patched program passes all the given tests, it does mean that the patch completely fixes the bug since the patched program may still fail on the inputs outside $pT$ and $fT$.
For example, here is a buggy code snippet (adapted from \cite{TrustAPR}).
This buggy program is intended to copy $\mathtt{n}$ characters from $\mathtt{src}$ to $\mathtt{dest}$, and then returns the number of copied characters.
\begin{lstlisting}[linewidth=\columnwidth,language=C]
int lenStrncpy(char[] src, char[] dest, int n){
    if(src == NULL || dest == NULL)
        return 0;
    int index = -1;
    while (++index < n)
        dest[index] = src[index]; // buffer overflow
    return index;
}
\end{lstlisting}
At line 6, a buffer overflow bug could occur if $\mathtt{n}$ is greater than the size of $\mathtt{src}$ or $\mathtt{dest}$.
\begin{table}[!h]
\vspace{-6pt}
\centering
\small
\begin{tabular}{lllrrr}
\hline
\textbf{Type} & $\mathtt{\mathbf{src}}$ & $\mathtt{\mathbf{dest}}$ & $\mathtt{\mathbf{n}}$ & \textbf{Output} & \textbf{Expected Output}  \\\hline\hline
Passing & \texttt{SOF}      & \texttt{COM}       & 3            & 3      & 3                \\
Passing & \texttt{DHT}      & \texttt{APP0}      & 3            & 3      & 3                \\
Failing & \texttt{APP0}     & \texttt{DQT}       & 4            & *crash & 3                \\
\hline
\end{tabular}
\label{tab:tests}
\vspace{-6pt}
\end{table}
If the above these three tests are provided to APR (the third test can trigger this buffer overflow) as specification, APR tool produces a patch ($\mathtt{{++}index} {<} \mathtt{n}\mapsto$ $\mathtt{{++}index} {<} \mathtt{n}\ \&\&\ \mathtt{index} {<} 3$) that can pass these three tests.
Obviously, the patched program is still buggy.
For instance, the buffer overflow can be triggered when \textit{src}\texttt{ = APPO}, \textit{src}\texttt{ = COM1}, and \textit{n}\texttt{ = 5}.

\subsubsection*{Constraints as Specification}
Besides tests, another line of APR research takes constraints as correctness specifications.
Different from test cases, constraints could represent a range of inputs or even whole input space. 
Taking constraints as specifications, APR aims to patch buggy programs to satisfy the given constraints.
For instance, the constraint on the output of a \texttt{sort} function is ($\mathit{output[0] \leq output[1] \leq output[2] \leq ...}$).
For a buggy implementation of \texttt{sort}, the repair goal is then to ensure the patched program satisfies the above constraint for any input.

On the other hand, in practice, constraints to fix a bug are not always available.
To solve this problem, constraint inferring techniques, e.g., \angelix~\citep{angelix} and \semfix~\citep{semfix}, extract constraints from tests.
Specifically, they first formulate the requirement to pass all given tests as constraints. Second, they synthesize patches with the objective of satisfying the inferred constraints.
Theoretically, those approaches are still driven by the given take cases, hence they also suffer from the overfitting issue.
Besides, some APR techniques, e.g., ExtractFix~\citep{Gao21}, take coarse-grained constraints as input.
Examples of coarse-grained constraints include vulnerability-free constraints, crash-free constraints, assertions, etc.
Such coarse-grained constraints specify some general program properties like enforcing a buffer cannot be overflowed, enforcing a pointer cannot be used after \texttt{free}, etc.
Unfortunately, guarantees from such constraints do not guarantee the patched program is functionally correct, i.e., they just ensure crash-free or vulnerability-free.


\subsubsection*{Static Analyser as Specification}
Apart from test suites and constraints, repair systems can also take static analyzers as specifications.
Static analysis is a source code analysis tool that automatically examines codes and finds bugs before a program is run.
Recently, designing effective and efficient static analyzers has gained a lot of attention.
Expressive and high-quality static analyzers have been designed and implemented to effectively find real bugs.
A lot of companies including Ebay, Microsoft, and Facebook are using static analysis tools in engineering practice.
Typically static analyzers, e.g., Infer and Findbugs, can detect many kinds of potential bugs, such as null pointer dereference, concurrency issues, heap property violations, etc.
Given a bug detect by a static analyzer, then, APR aims to fix the program to pass the check of this static analyzer.
Here is a memory leak (adapted from \cite{footpatch}) detected by Facebook's Infer tool.
\begin{lstlisting}[linewidth=\columnwidth,language=C]
swHashMap *hmap = sw_malloc(sizeof(swHashMap));(*@\label{line:malloc}@*)
if (!hmap) {
    swWarn("malloc[1] failed.");
    return NULL;
}
swHashMap_node *root = sw_malloc(sizeof(swHashMap_node));
if (!root) {
    swWarn("malloc[2] failed.");
    return NULL; (*@{\color{red}// returns, hmap is not freed}\label{line:return}@*)
}
\end{lstlisting}
If mallocing \textit{hamp} succeeds and mallocing \textit{root} fails, a memory leak on \texttt{hamp} happens because the developer forgot to call \textit{free} before returning \texttt{NULL} at line~\ref{line:return}.
This bug can be fixed by adding \textit{free(hmap)} before line \ref{line:return}.
The repair goal is to ensure this bug cannot be detected again by the same static checker.
Driven by this static analyzer, the APR systems search for a patch to satisfy the semantic effect defined by this static analyzer, i.e., if a memory block is allocated in the precondition, it should be freed in the postcondition.
However, similar to the coarse-grained constraints, static analyzers can only ensure general program properties, but they cannot ensure functionality correctness.
For instance, deleting the \texttt{malloc} expression at line~\ref{line:malloc} can also fix the memory leak, which is obviously an incorrect patch.
The produced patches \textit{overfit} to the static analyzers, i.e., the patched program passes the checks of static checkers but it is not functionally correct.

\subsection{Alleviate Overfitting via Heuristic Ranking}

Due to the incomplete specification, the test-driven search-based APR tools are prone to generate overfitted patches.
To increase the possibility of finding correct patches from huge search space, many repair approaches proposed heuristic strategies to give higher priorities to the candidate patches that are likely to be correct.

\subsubsection{Predefined Patch Patterns}
The most straightforward strategy to alleviate the overfitting problem is to use the search space that is likely of being correct.
This approach is useful, especially when fixing a specific class of bugs/errors, such as memory leaks, use-after-free bugs, concurrency bugs, data race, integer overflows, etc.
For certain types of bugs, we could design a specific candidate patch space.
For instance, inserting a \texttt{free} statement is likely fixing memory leaks.
Deleting a \texttt{free} statement or moving the location of a \texttt{free} statement is likely able to fix Use-After-Free bugs.
Adding a pair of locks has a high chance to fix a data race.
By just considering the specifically designed set of patches for each type of bug, APR systems have a great chance to generate correct patches.
For a certain type of bug, we could define a set of templates to generate candidate patches that are likely to be correct.
The templates can be defined either manually or by referring to the patches of the same type of bugs.
For instance, from the open-source repositories, we could search for the patch patterns that fix an integer overflow and then apply the same pattern to generate patch candidates for fixing other integer overflows.

\begin{table}[!t]
\centering
\begin{tabular}{m{0.31\textwidth}|m{0.65\textwidth}}\hline
Anti-patterns                    & Example \\\hline\hline
\textbf{Anti-delete CFG exit node:} disallows removal of return statements, exit calls, and assertions                 &  \begin{lstlisting}[linewidth=0.65\textwidth,language=C,caption={The patch removes the erroneous exit call.\vspace{-10pt}},numbers=none,xleftmargin=0em, framexleftmargin=0em]
static void BadPPM(char* file) {
    fprintf(stderr, "%s:Not a PPM file\n", file);
-     exit(-2);
  }
\end{lstlisting} \\\hline
\textbf{Anti-delete Control Statement:} disallows removal of control statements, e.g., if-statements, switch-statements   &  
\begin{lstlisting}[linewidth=0.65\textwidth,language=C,caption={The patch removes the whole if-then-else statement that checks for the return value of a function call.\vspace{-10pt}},numbers=none,xleftmargin=0em, framexleftmargin=0em]
call_result = call_user_function_ex(...);
-  if (call_result == SUCCESS && retval != NULL && ...) {
-     if (SUCCESS == statbuf_from_array(...))
-         ret = 0;
-  } else if (call_result == FAILURE) {
-           php_error_docref(...);
-         }
\end{lstlisting} \\\hline
\textbf{Anti-delete Single-statement CFG:} disallows deletion of the statement within a CFG node that has only one statement &  
\begin{lstlisting}[linewidth=0.65\textwidth,language=C,caption={The patch removes the statement that assigns the return value of 1 which indicates a failure.\vspace{-10pt}},numbers=none,xleftmargin=0em, framexleftmargin=0em]
fail:{
- ret = 1; 
}
\end{lstlisting} \\\hline
\textbf{Anti-append Early Exit:} disallows insertion of return/goto statement in the middle of a basic block       &  
\begin{lstlisting}[linewidth=0.65\textwidth,language=C,caption={The patch adds a conditional return statement before a function call that throws an error.\vspace{-10pt}},numbers=none,xleftmargin=0em, framexleftmargin=0em]
+  if ((type != 0))
+     return;
zend_error((1<<3L),"Uninitialized string offset:",...);
\end{lstlisting} \\\hline
\end{tabular}
\caption{Sample anti-patterns (adapted from \cite{anti-pattern}) with examples that illustrate the usage of each anti-pattern}
\label{tab:anti-patterns}
\end{table}

Instead of providing templates to generate the search space which may unduly restrict the repair space, Anti-Pattern~\citep{anti-pattern} proposes a set of ``anti-patterns'', i.e., a set of forbidden transformations that should not be used by search-based APR tools.
The main intuition of anti-patterns is that the produced patches by APR often modify programs by deleting functionality.
Although, such patches are sufficient to pass the given test suite but unacceptable to developers in general.
For instance, for fixing a memory leak, APR may directly delete the allocation of the leaked memory, which resolves the memory leak but is obviously not acceptable.
To solve this problem, anti-patterns summarize the disallowed modifications that are not acceptable.
That is, even if such a modification produces a fixed program that passes all given test-suite, it will not be treated as a correct patch.
Table~\ref{tab:anti-patterns} lists four examples of pre-defined anti-patterns.
For instance, the \textit{Anti-delete CFG exit node} anti-pattern does not allow removal of return statements, erroneous exit calls, or assertion, since such patches likely repair bugs by hiding the errors instead of really fixing them.
For each anti-pattern, Table~\ref{tab:anti-patterns} also presents an example patch, where ``-'' denotes deleted statements by the patch, while ``+'' marks added statements.
The unchanged statements are donated as code without any leading symbol.
For instance, the first anti-pattern does not allow the patch that removes the erroneous exit call \texttt{exit(-2)}, which is very likely unacceptable.

\subsubsection{Ranking Based on Syntactic and Semantic Distance}
The correct patch may not be ranked at the top even though they are included in the search space.
To efficiently find the correct patch among all the plausible patches, the candidate patches could be ranked based on their syntactic and semantic distances to the original program.

\paragraph{Syntax-Based Ranking}

The syntax-based ranking is built on top of the following assumption: \textit{small patches are less possible to change the correct behavior of the program than more complex patches. Thus, the patches that modify minimal original program behaviors are prioritized.}~\citep{directfix}
With this hypothesis in mind, existing approaches designed strategies to prioritize likely correct candidate patches according to syntactic distance.

A simple approach to find the ``minimal'' patch is to enumerate all the search space and select the simplest patch.
Typically, given the candidate patch $\mathtt{N_{patch}}$ (donated as AST node) and original buggy code $\mathtt{N_{buggy}}$, the syntactic distance $\mathtt{syntactic\_distance}(\mathtt{N_{patch}}, \mathtt{N_{buggy}})$ is usually measured by considering various features.
For instance, here are some syntactic features used by existing APR tools~\citep{S3}:
\begin{itemize}
    \item \textbf{AST Edit Distance} Edit distance is a way of quantifying how dissimilar two structures are to one another by counting the minimum number of operations required to transform one into the other.
    Here, the similarity of the candidate patch with the original code can be measured by edit distance.
    Specifically, given the AST of a candidate patch $\mathtt{N_{patch}}$ and AST of original buggy code $\mathtt{N_{buggy}}$, we calculate operations that can transform $\mathtt{N_{patch}}$ to $\mathtt{N_{buggy}}$ in the form of actions on AST nodes including insertion, deletion, update, or move. The syntactic distance is measured by the number of actions needed to transform $\mathtt{N_{patch}}$ to $\mathtt{N_{buggy}}$.
    \item \textbf{Cosine similarity} The AST is ``abstract'' since it cannot reflect all the details of the real syntax. Hence, the Edit Distance over AST node cannot capture all the syntactical details, such as the type information (Integer, Boolean, String and etc). Vectors of type occurrence counts (e.g., [{``Integer'', 1}, {``Boolean'', 2}, {``String'', 5}...], meaning integer appears once, boolean appears two times and etc.) can be used to represent the syntax of the candidate patch and original buggy code. The similarity of two pieces of codes can then be calculated using the cosine similarity of their type occurrence count vectors.
\end{itemize}
We prefer the candidate patches that have shorter syntactic distances to the original code with the goal of maximally preserving the correct behavior of the original program.

Apart from measuring the distance between the candidate patches and the original code, APR tools also consider the syntactic similarity between the contexts surrounding the patched code~\citep{context_aware_apr}.
Many repair techniques, e.g., GenProg~\citep{GenProg}, generate patch search space by using the code fragment extracted from the same application or across multiple applications.
More specifically, a bug is fixed by replacing the buggy code $\mathtt{N_{buggy}}$ in the buggy program with code $\mathtt{N_{patch}}$ extracted from somewhere else.
Note that the bug can be fixed by inserting $\mathtt{N_{patch}}$ or deleting $\mathtt{N_{buggy}}$, for simplify, we just consider replacements.
As shown above, the candidate patches can be ranked according to the $\mathtt{syntactic\_distance}(\mathtt{N_{patch}}, \mathtt{N_{buggy}})$.
Besides, they can also be ranked according to the context of the $\mathtt{N_{patch}}$ and $\mathtt{N_{buggy}}$, donated as $\mathcal{C}_\mathtt{patch}$ and $\mathcal{C}_\mathtt{buggy}$, respectively.
The main intuition is that $\mathcal{C}_\mathtt{patch}$ should be syntactically similar with $\mathcal{C}_\mathtt{buggy}$.
Thus, $\mathtt{syntactic\_distance}(\mathcal{C}_\mathtt{patch}, \mathcal{C}_\mathtt{buggy})$ is measured using the following features:

\begin{itemize}
    \item \textbf{Context similarity} The contexts $\mathcal{C}$ of a node \texttt{N} is defined as a set of pairs (\texttt{node\_type}, \texttt{count}), which are the number of occurrences of different types of node (e.g., Expression, Statement and etc) of \texttt{N}'s ancestor and siblings. The distance of context between $\mathcal{C}_\mathtt{patch}$ and $\mathcal{C}_\mathtt{patch}$ is defined as 
    \[
    1 - \frac{\sum_{(\mathtt{type},\mathtt{count}) \in \mathcal{C}_\mathtt{buggy}}{\mathtt{min}(\mathtt{count}, \mathcal{C}_\mathtt{patch}[\mathtt{type}])}}{\sum_{(\mathtt{type},\mathtt{count}) \in \mathcal{C}_\mathtt{buggy}}{\mathtt{count}}}
    \]
    Basically, the above formula measures the proportion of $\mathcal{C}_\mathtt{buggy}$ that are included in $\mathcal{C}_\mathtt{patch}$.
    
    \item \textbf{Dependency context similarity} Given an AST node \texttt{N}, the dependency contexts $\mathcal{C}^d$ capture the information about the dependency nodes that affect \texttt{N} and dependent nodes that are affected by \texttt{N}.
    Specifically, dependency nodes are extracted via intra-procedure backward analysis based on the def-use relationship of variables used from \texttt{N}, while dependent nodes are extracted via a forward analysis.
    $\mathcal{C}^d$ is defined as a set of (\texttt{node\_type}, \texttt{count}) pairs, which are the number of occurrences of different types of all the dependency and dependent nodes.
    The distance in terms of dependency context is similar to the measurement metrics above.
    The formal definition of dependency context similarity between $\mathcal{C}^d_\mathtt{buggy}$ and $\mathcal{C}^d_\mathtt{patch}$ as defined as follows:
    \[
    1 - \frac{\sum_{(\mathtt{type},\mathtt{count}) \in \mathcal{C}^d_\mathtt{buggy}}{\mathtt{min}(\mathtt{count}, \mathcal{C}^d_\mathtt{patch}[\mathtt{type}])}}{\sum_{(\mathtt{type},\mathtt{count}) \in \mathcal{C}^d_\mathtt{buggy}}{\mathtt{count}}}
    \]
\end{itemize}
We prefer the candidate patches whose contexts have shorter syntactic distances to the context of the original code.
The $\mathtt{syntactic\_distance}(\mathcal{C}_\mathtt{patch}, \mathcal{C}_\mathtt{buggy})$ can be combined with \[
\mathtt{syntactic\_distance}(\mathtt{N_{patch}}, \mathtt{N_{buggy}})\] to rank the candidate patches in a more proper way.

Enumerating all the search space and selecting the simplest repair is not efficient.
In practice, even finding a single repair usually takes a lot of time.
To find simple repairs more efficiently, DirectFix~\citep{directfix} proposes a semantics-based repair approach that integrates the two phases of program repair (1) fault localization and (2) repair search into a single step.
Then, component-based program synthesis and partial MaxSAT constraint solving are used to produce the simplest patches directly.

\begin{figure}[!h]
    \centering
\begin{subfigure}{\textwidth}
\begin{lstlisting}
int foo(int x, int y){
    if (x > y) // Fault: the condition should be x >= y
        y = y + 1;
    else
        y = y - 1;
    return y + 2;
}
void test_foo() { assert (foo(0,0 == 3); }
\end{lstlisting}
\caption{A buggy function and its test}
\label{list:directfix_bug}
\end{subfigure}

\begin{subfigure}{\textwidth}
\[
\phi_\mathit{buggy} \equiv (\mathtt{if} (x_1 > y_1)\ \mathtt{then}\ (y_2 = y_1+1)\ \mathtt{else}\ (y_2 =y_1-1)) \wedge (\mathit{result} = y_2 + 2)
\]
\caption{The trace formula $\phi_\mathit{buggy}$ for foo; variables $x_i$ and $y_i$ correspond to the program variables $x$ and $y$, respectively, and $\mathit{result}$ to the return value of the program.}
\label{list:formula}
\end{subfigure}
\caption{A trace formula is constructed from a buggy program and its tests, which is adapted from ~\cite{directfix}}
\end{figure}

More specifically, DirectFix first translates a buggy program into a formula.
Figure~\ref{list:directfix_bug} shows a buggy function and its test, this function fails on \texttt{test\_foo}. 
Figure~\ref{list:formula} presents the produced formula for this bug and test.
The given test can be translated into the following constraint:
\[ \mathcal{O} \equiv x_1 = 0 \wedge y_1 = 0 \wedge \mathit{result} = 3 \]
Since the conjunction $\mathcal{O}$ and $\phi_\mathit{buggy}$ is unsatisfiable, the test fails.
Then, the repair goal is to minimally change $\phi_\mathit{buggy}$ to produce a modified formula $\phi_\mathit{repair}$ such as $\phi_\mathit{repair} \wedge \mathcal{O}$ is satisfiable.
The correct patch should change $\phi_\mathit{buggy}$'s condition $x_1 > y_1$ to $x_1 >= y_1$ .

To generate the correct patch, DirectFix reduces the problem of generating the simplest patch as constraint solving problem.
Given a trace formula $\phi_\mathit{buggy}$, the repair condition (constraint) $\phi_\mathit{rc}$ is constructed as follows:
\begin{align*}
\phi_\mathit{rc} \equiv &(\mathtt{if}\ v_1\ \mathtt{then}(y_2 = v_2)\ \mathtt{else}\ (y_2 = v_3)) \wedge (\mathit{result}=v_4)\\
&\wedge \mathtt{cmpnt}(v_1 = x_1 > y1)\wedge \mathtt{cmpnt}(v_2 = y_1 + 1)\\
&\wedge \mathtt{cmpnt}(v_3 = y_1 - 1)\wedge \mathtt{cmpnt}(v_4 = y_2 + 2)    
\end{align*}

The formula $\phi_\mathit{rc}$ shown above is semantically equivalent to $\phi_\mathit{buggy}$. 
Differently, fresh variables $v_i$ is introduced to substitute the r-value expressions of $\phi_\mathit{buggy}$. 
Meanwhile, each $v_i$ is kept the equality relationship with its represented rvalue expression (e.g., $v_i$ = $x_1$ > $y_1$) in a \texttt{cmpnt} function.
The \texttt{cmpnt} function componentizes its parameter expression into a component-based synthesis (see Chapter~\ref{cha:semantic_repair}) problem.
To obtain the smallest patch, a partial maximum satisfiability (pMaxSMT) solver is used.
The formula of pMaxSMT is split into two categories: \textit{hard clauses} that must be satisfied and \textit{soft clauses} that is not necessarily to be satisfied.
The hard clauses constrain the semantics of the component and the specification of the program (oracle), and soft clauses express the program expressions that can be changed or kept.
Satisfying a soft clause means that the corresponding expression keeps as is, while not satisfying a soft clause means that the expression needs to be changed.
After splitting $\phi_\mathit{rc} \wedge \mathcal{O}$, the hard and soft clauses are then fed into pMaxSMT solver.
To solve the provided constraint, the solver minimally removes or changes some soft clauses (if necessary), such that the modified $\phi_\mathit{rc} \wedge \mathcal{O}$ is satisfiable, and returns a model corresponding to a fix.
In general, DirectFix does not have an explicit ranking algorithm.
Instead, it encodes the test cases and syntactic information into a set of constraints and directly produces the simplest patches by solving a partial MaxSAT program.

\paragraph{Semantics-Guided Ranking}
Apart from patch syntax, the patch semantics are also used to rank candidate patches~\citep{S3}.
The intuition is that we prefer the patch that minimally changes original program behaviors and maximally preserves the correct behavior of the original program.
Given the candidate patch $\mathtt{N_{patch}}$ and original buggy code $\mathtt{N_{buggy}}$, the semantic distance $\mathtt{semantic\_distance}(\mathtt{N_{patch}}, \mathtt{N_{buggy}})$ is measured using the following features.
\begin{itemize}
    \item \textbf{Semantic Anti-Patterns} Just like the anti-pattern shown in Table~\ref{tab:anti-patterns}, semantic anti-patterns aim to avoid generating patches that are likely to be semantically incorrect. 
    For instance, the anti-duplicate pattern prevents meaningless patches like $a < a$, $0 == 0$, etc. Expressions containing semantic anti-patterns are thus likely to be incorrect. The priority of these patches is decreased.
    
    \item \textbf{Model Counting} Model counting is a technique that counts how many models satisfy a given constraint. Model counting is used to calculate the ``disagreement'' level between $\mathtt{N_{patch}}$ and $\mathtt{N_{buggy}}$.
    That is, we say that $\mathtt{N_{patch}}$ and $\mathtt{N_{buggy}}$ disagree with each other if the value produced by $\mathtt{N_{patch}}$ is different from the value produced by $\mathtt{N_{buggy}}$. 
    The disagreement level between $\mathtt{N_{patch}}$ and $\mathtt{N_{buggy}}$ is measured as the number of inputs that produce different truth values on $\mathtt{N_{patch}}$ and $\mathtt{N_{buggy}}$, which is regarded as their semantic distance.
    For instance, assuming that the original expression $\mathtt{N_{buggy}}$ is $a < 10$, a candidate patch $\mathtt{N_{patch1}} = "a <= 13"$, and another candidate patch $\mathtt{N_{patch2}}$ is $a <= 15$.
    The semantic distance between $\mathtt{N_{patch1}}$, $\mathtt{N_{patch2}}$ and $\mathtt{N_{buggy}}$ is 4 and 6 (representing the number of integers on which the original expression and patch produce different results), respectively. In this case, we would prefer $\mathtt{N_{patch1}}$ since it changes fewer program behaviors than $\mathtt{N_{patch2}}$.
    
    \item \textbf{Value Distance} The value distance measures the difference between the produced values by $\mathtt{N_{patch}}$ and $\mathtt{N_{buggy}}$ on certain tests. For example, assuming that the original expression $\mathtt{N_{buggy}}$ is $a - 10$, and a candidate patch $\mathtt{N_{patch1}}$ is $a - 13$, and $a - 15$ as another candidate patch $\mathtt{N_{patch2}}$. Given a test that assign the value of $a$ as 20, the produced values by $\mathtt{N_{buggy}}$, $\mathtt{N_{patch1}}$ and $\mathtt{N_{patch2}}$ are 10, 7 and 5, respectively. So, under this test, the value distance between $\mathtt{N_{buggy}}$ with $\mathtt{N_{patch1}}$ and $\mathtt{N_{patch2}}$ are 3 and 5, respectively. 
    To generate the value distance, there are three steps: (1) generating a set of inputs $I$ that produce different results on the candidate patches and original code; (2) collecting the values of original code and each candidate patch for each $i_n \in I$; (3) calculating the value distance between $\mathtt{N_{buggy}}$ with each candidate patch.
    Finally, the patch with the shortest distance is selected.
\end{itemize}
Similar to syntactic distance, the candidate patches that have a shorter semantic distance with the original code are preferred.
Semantic and syntactic features are sometimes used together to rank the candidate patches~\citep{S3}.

\subsubsection{Learning to Rank}\label{sec:learn-ranking}
Apart from ranking algorithms based on syntactic and semantic analysis, the candidate patches can also be ranked by learning how history patches fixed similar bugs.
This is based on an assumption that history patches often share some common patterns, and they are similar in nature.
Hence, past patches and their common fix patterns can potentially provide some useful information about what kinds of candidate patches are likely to be correct.
This assumption is similar to that of learning-based program repair techniques.

The history patches can help assess the quality and fitness of the candidate patches~\citep{history_driven_apr}.
To do so, we first mine a dataset of developer-produced patches that fixed real bugs from version control systems, e.g., GitHub.
The dataset is then used to mine common patch patterns by generalizing and combining similar patches.
Specifically, the original patches are generalized by abstracting the project-specific details, such as variable names, method names and etc.
The generalized patches are then grouped according to their code structures.
For instance, assume we have two patches \texttt{if (array==null) return;} and \texttt{if (address==null) return;} that fixed null pointer dereferences, by abstracting their identifier names, they are generalized to \texttt{if (VAR==null) return;} and \texttt{if (VAR==null) return;}, respectively.
Since the generalized patches share the same code structure, these two patches are grouped together and their patch pattern would be 
inserting \texttt{if (VAR==null) return;} before a pointer access.
Later, candidate patches are ranked using the extracted common patterns.
The candidate patches that appear frequently in the mined pattern are prioritized.
For instance, when fixing a null pointer dereferences, a candidate patch that matches the \texttt{if (VAR==null) return;} pattern would be given higher priority. 

Apart from learning common patterns, a probabilistic model can be trained to rank the candidate patches according to their possibility of being correct~\citep{prophet}.
For a buggy program $\mathtt{prog}$ and a set of candidate patches, a trained probabilistic model assigns a probability $P(p | \mathtt{prog},\theta)$ to each candidate patch $p$, indicating how likely $p$ can correctly fix $\mathtt{prog}$. 
model parameter $\theta$ is trained in the training process, which is an offline phase.
Once the probabilistic model is trained offline, it can be used online to rank candidate patches based on their probability of being correct.

The design of the probabilistic model is based on the main hypothesis:\textit{ correct code, even across different applications, shares common correctness properties}.
To learn an effective model, it is necessary to keep the important and common correctness code properties, while abstracting away the unimportant shallow details (e.g., variable names) that are associated with certain applications. 
It is crucial to identify which features to use when training probabilistic models.
Two types of features are considered by existing techniques: \textit{modification features}, and \textit{program value features}.
\begin{itemize}
    \item \textbf{Modification features}: Modification features express modifications and interactions between the modified code and the surrounding statements. Two types of modification features are considered. The first type expresses the modification types of a patch, such as ``inserting control flow'', ``adding guard condition'', ``replacing condition'', ``inserting statement'', ``replacing statement'', etc. The second type expresses the types of statements near the patched statement and the modification kind. Specifically, the second type of the modification features is formalized as a set of pairs (\texttt{StatementKind}, \texttt{ModificationKind}), where the \texttt{StatementKind} represents the type of the buggy statement, or the statement before/after the buggy statement.
    \item \textbf{Value features}: Value features express the similarities and differences about how variables/constants are used in the original program and in the patch. 
    To avoid application-specific information introducing noise to the feature space, it is necessary to abstract away the syntactic details, such as names of variables and values of constants. 
    For instance, assume a candidate patch changes an assignment $a=b+c \mapsto b = a+c$, the value feature of $a$ in the original code and patch would be \texttt{<=, L>} and \texttt{<=, R>}, respectively. Meaning that variable $a$ is the left value of an assignment in the original code, and variable $a$ is the right value of the assignment in the patch. Similarly, the respective value feature of $b$ in the original code and patch would be \texttt{<=, R>} and \texttt{<=, L>}. Based on the abstraction, value features would model how the variables/constants are used differently or identically before and after applying the patch.
\end{itemize}
The learned probabilistic model is then applied to the candidate patches to predict the probability of each patch of being correct.

\subsection{Alleviate Overfitting via Test Generation}

Since the test cases driving APR are usually incomplete, the other straightforward idea to alleviate the overfitting problem is automatically generating more tests.
Overfitted patches pass the given tests, but they still fail on some tests outside the given tests.
Therefore, the goal of automatically generating more tests is to find such tests that fail the overfitted patches.
There are two main challenges in efficiently generating useful test cases 1): how to generate inputs that can drive the program execution to the patch location \textbf{and} fail the overfitted patch 2): how to define the oracle (i.e., the expected outputs) of the newly generated inputs.
There are quite a lot of existing test generation techniques, e.g., symbolic execution, grey-box fuzzing, and evolutionary algorithm.
These techniques are designed for the program testing purpose, i.e., maximize code coverage and maximize bug finding.
Because of the fact that they have no knowledge about the patch semantics, they are not efficient in generating useful tests for ruling out overfitted patches.
Therefore, people designed customized test generation approaches for alleviating the patch overfitting issue.

\subsubsection{Encode patch semantics into program}

As we mentioned above, existing test generation techniques are not efficient in ruling out overfitted patches since they are guided by coverage information and they have no knowledge about the patch semantics.
For the purpose of test patches, one idea is to encode the patches into the original program by inserting a dummy statement as the coverage goals~\citep{xin2017identifying}.

Let us look at an example program in Figure~\ref{fig:difftgen} that converts a string (e.g., ``Yes'', ``True'') into a boolean value.
\begin{figure}[!h]
    \centering
    \begin{lstlisting}[linewidth=\columnwidth,language=C]
public static boolean toBoolean(String str) { 
    if (str==``true'') return true;
    if (str==null) return false;
    switch (str.length()) {
        case 2: { ... } 
        case 3: {
            char ch = str.charAt(0); 
            if (ch==`y')
                return (str.charAt(1)==`e'||str.charAt(1)==`E')
                    && (str.charAt(2)==`s'||str.charAt(2)==`S');
            if (ch==`Y') (*@{\color{red}// Changed to ``if (str!=null)'' (Overfitting Patch)}@*)
                return (str.charAt(1)==`e'||str.charAt(1)==`E')
                    && (str.charAt(2)==`s'||str.charAt(2)==`S');
            (*@{\color{green}//Inserted "return false;" (Correct Patch)}@*)
        }
        case 4: {
            char ch = str.charAt(0); 
            if (ch==`t')
                return (str.charAt(1)==`r'||str.charAt(1)==`R')
                    && (str.charAt(2)==`u'||str.charAt(2)==`U') 
                    && (str.charAt(3)==`e'||str.charAt(3)==`E');
            if (ch==`T') { ... } 
        }
        return false; 
    }
}
\end{lstlisting}
    \caption{An \texttt{IndexOutOfBoundsException} from Lang\_51 in the Defects4J Bug Dataset}
    \label{fig:difftgen}
\end{figure}
This program has an \texttt{IndexOutOfBoundsException} bug at line 21 if it is executed with ``tru'' as input.
With ``tru'' as input, this program is expected to return false as the output. 
This bug happens because the developer forgot to break the switch statement after executing case 3.

Automatic repair technique NoPol~\citep{nopol} produces a patch at line 11 by modifying the if-condition from $\mathit{ch}==$`$\mathtt{Y}$' to $\mathit{str}!=\mathtt{null}$.
The patched program now works correctly on the input ``tru'' since it returns false at Line 12.
Hence, the patched program passes the given test suite.
However, this bus has not been correctly fixed by this patch.
To efficiently generate a test that can still fail the patched program, we could encode the patch semantics into the program for the purpose of guiding test generation.
Specifically, we can insert the following dummy statement before line 11.
\begin{lstlisting}[linewidth=\columnwidth,language=C]
if (((ch==`Y')&&!(str!=null))||(!(ch==`Y')&&(str!=null)))
    int delta_syn_3nz5e_0 = -1;  // A dummy statement
\end{lstlisting}
Then, test inputs can be generated with the guidance of the newly inserted control flow.
If a test input satisfies the inserted coverage goal, i.e., explores the inserted control flow, it will lead to a differential execution between the original program and the patched program.
Such input could find the divergent behaviors of two programs and has the potential to expose an overfitting behavior of patched programs.
For this example, a new test case with the input string ``@es'' could be generated.
This new test triggers a new failure on the patched program, i.e., the patched program returns true but the expected output is false. 
So, we will figure out that this patch is an overfitted patch.

More formally, this approach takes a buggy program $\mathtt{prog_{buggy}}$ and patched program $\mathtt{prog_{patch}}$ as inputs. 
It first calculates syntactic differences $\Delta$ between the $\mathtt{prog_{buggy}}$ and $\mathtt{prog_{patch}}$.
Each $\delta\in\Delta$ is a tuple <$\mathtt{stmt_{buggy}}$, $\mathtt{stmt_{patch}}$>, where $\mathtt{stmt_{buggy}}$ and $\mathtt{stmt_{patch}}$ are buggy statements and the corresponding fixed statements, respectively.
Based on $\mathtt{prog_{buggy}}$, $\mathtt{prog_{patch}}$ and $\Delta$, a program called test program $\mathtt{prog_{test}}$ is generated.
To obtain $\mathtt{prog_{test}}$, a dummy statement is inserted into $\mathtt{prog_{patch}}$.
For each modification $\delta\in\Delta$, the dummy statement checks whether $\mathtt{stmt_{buggy}}$ and $\mathtt{stmt_{patch}}$ produce the same output.
If they produce different outputs, the dummy statement will lead the program execution to a new control flow.
A test input that covers the new control flow is likely to expose the overfitting behaviors of plausible patches.

\subsubsection{Guide test generation via patch partition analysis}\label{sec:testGenPatchPartition}
\begin{figure}[!t]
        \centering
        \includegraphics[width=0.45\textwidth]{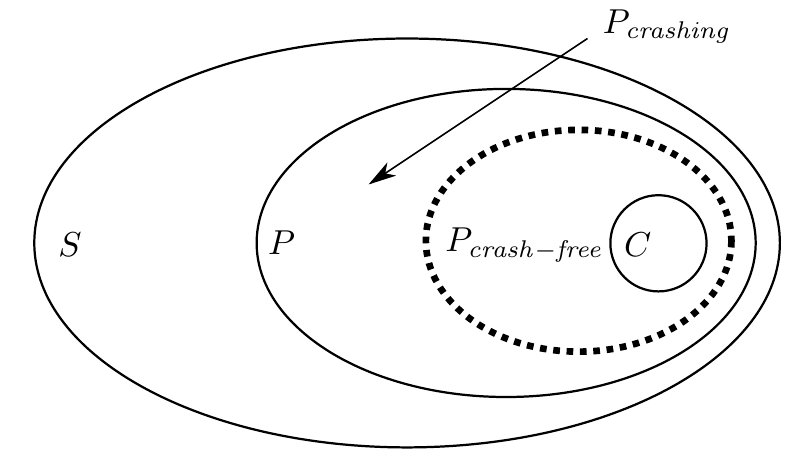}
        \caption{Structure of program repair search space, where $S$ is a space of candidate patches, $P$ is a set of plausible patches, $P_\mathit{crashfree}$ is a set of crash-free patches, $C$ is a set of correct patches, $P_\mathit{crashing} = P \setminus P_\mathit{crashfree}$ is a set of crashing patches.\label{fig:plausible-crashing}. Figure is adapted from ~\cite{fix2fit}}
\end{figure}

Similar to the idea of exposing semantic differences between different program versions, another idea generates useful tests by analyzing patch equivalence~\citep{fix2fit}.
At a high level, the patch candidates can be divided into different partitions according to their semantic behaviors, i.e., semantically equivalent patches are put into the same partition.
Figure~\ref{fig:plausible-crashing} presents the initial program repair search space.
The full search space firstly can be divided into two partitions: plausible partition (set $P$, the patches that pass given tests) and incorrect partition(set $S-P$).
Just as we mentioned in Section~\ref{sec:specification}, test suite is an incomplete specification.
Therefore, only part of plausible patches may be correct (set $C$), while the remaining patches merely overfit the tests (set $P-C$).
In rare cases, all the plausible patches are correct.
Moreover, when fixing program crashes, the tests outside of the given test suite may still crash the patched programs that are fixed by overfitted patches.
To solve this problem, one idea is to use test generation to filter out the plausible patches that still crash the programs~\citep{fix2fit}.
The main idea is to separate the set of plausible patches into two partitions: $P_\mathit{crashfree}$ which represents the crash-free plausible patches, and $P_\mathit{crashing}$ which represents the plausible patches still causing crashes.
Crash-free plausible partition is composed of patches that not only pass given tests, but also not causing crashes on the inputs outside of the given test suite.
Later, the patch should be chosen from the crash-free plausible patches $P_\mathit{crashfree}$.

\begin{figure}[!h]
    \centering
\begin{multicols}{2}
\begin{itemize}
    \item[$p1$:] $\mathit{ch}==$`$\mathtt{Y}$' $\mapsto\mathit{ch}\ !=$`$\mathtt{W}$'
    \item[$p2$:] $\mathit{ch}==$`$\mathtt{Y}$' $\mapsto\mathit{ch}\ !=$`$\mathtt{X}$'
    \item[$p3$:] $\mathit{ch}==$`$\mathtt{Y}$' $\mapsto\mathit{ch}\ $`$!=\mathtt{Y}$'
    \item[$p4$:] $\mathit{ch}==$`$\mathtt{Y}$' $\mapsto\mathit{ch}\ $`$!=\mathtt{Z}$'
\end{itemize}
\end{multicols}
    \caption{Four plausible but incorrect patches for fixing the bug in Figure~\ref{fig:difftgen}}
    \label{fig:patches}
\end{figure}

Let us revisit the above example. Suppose we have four plausible patches shown in Figure~\ref{fig:patches}.
For the given failing input ``tru'', all of these four patches make the program pass.
All these patches are classified into the same patch partition since these four patch statements are evaluated as true under input ``tru'' (value-based test-equivalent~\citep{TOSEM18Sergey}).
To filter out overfitted patches, the test generation goal is then to break this patch partition, i.e., find out a test input that makes these statements produce different values.
For instance, test input ``Wru'' can break this partition because the patch statement in $p1$ is evaluated as \texttt{false}, while other patches are still evaluated as \texttt{true}.
Besides, with ``Wru'' as input, the patched program using $p1$ still throws ``IndexOutOfBoundException'', hence $p1$ will be put into the crashing partition $P_\mathit{crashing}$.

Relying on an evolutionary algorithm, traditional greybox fuzzing techniques, e.g., AFL~\citep{afl}, give high priority to the tests that improve code coverage for further mutation.
By mutating the tests that improve code coverage, greybox fuzzing has a higher chance of further improving code coverage.
Following this idea, we also give the test inputs that break patch partitions with higher priority
The main intuition is \textit{by mutating the test inputs that break patch partitions, we hope to generate more tests that further break the patch partitions}.
For instance, by mutating ``Wru'', it is very likely to generate new inputs ``Xru'', ``Yru'', and ``Zru'' which can further distinguish the patch partition and discard more overfitting patches.

Formally, given a buggy program $\mathit{buggyprog}$ and a set of patch candidates $S$, the patch validation and test generation are fused into a single process.
The test generation produces new tests with the objective of differentiating patches from $S$ and distinguishing patch partitions \{$P_1, P_2,\dots$\}, and patch validation maintains the patch partitions with the reference to all the available tests (including the given test cases and the newly generated tests).
Hence, a test suite is generated with the goal of (1) covering the functionalities of the original program, (2) covering the functionality of the program that is modified by candidate patches, and (3) covering functionality that differs between different candidate patches, i.e., breaking patch partitions \{$P_1, P_2,\dots$\}.
Since the generated test suite has a higher chance to find divergences of candidate patches, it has a higher chance to differentiate $P_\mathit{crashfree}$ and $P_\mathit{crashing}$.

\subsubsection{Oracle of newly generated tests}
To determine whether a patched program is semantically correct, it is necessary to know the expected outputs (test oracle) of the newly generated tests.
However, obtaining test oracles is a notoriously difficult problem in program testing.
One straightforward approach is to ask developers to provide test oracles, but it is a time-consuming task.
In the context of program repair, fortunately, it is not necessary to query developers for the expected outputs of all the test inputs.
Instead, only if \texttt{buggyprog} and \texttt{patchedprog} produce different outputs on a test input, the developer is queried to determine which output is correct~\citep{xin2017identifying, Bohme2020}.
If the output of \texttt{patchedprog} is incorrect, the corresponding patch will be discarded as an overfitted patch.
Similarly, if multiple plausible patches drive the program to generate different outputs, the developer's feedback can also help to rule out at least one of the plausible patches, since at most one of the different outputs is correct on a deterministic program.

Although developer feedback can help discard overfitted patches, it also adds a heavy burden to developers.
Existing approaches turn to use some general oracles (like crash-free~\citep{fix2fit}, memory safety~\citep{yang2017better}, etc).
Specifically, if a patched program crashes on some test inputs, the corresponding patch must be overfitted.
For the patched program by $p1$ shown in Figure~\ref{fig:patches}, test input ``Wru'' can still trigger the \texttt{IndexOutOfBoundsException}. 
Therefore, patch $p1$ can be determined as an overfitted patch without developers' participation.
The absence of program crashes cannot guarantee the correctness of the patched program.
To deal with this problem, Fix2Fit~\cite{fix2fit} proposes to use sanitizers to enhance patch checking.
Sanitizers can detect various vulnerabilities, such as buffer overflow/underflow, and integer overflow at run-time.
Generally, software vulnerabilities are converted into normal crashes by sanitizers, for instance, with the help of AddressSanitizer, triggering a buffer overflow would crash the program execution.
By using sanitizers, the patches that introduce vulnerabilities can be ruled out.
Compared to just relying on crashes, more overfitted patches can be filtered out.


\subsection{Alleviate Overfitting via Semantic analysis}
Test case generation helps discard overfitted patches that introduce crashes or cause memory errors, but it does not provide formal guarantees.
In other words, test case generation cannot guarantee that there is no unknown test that can still crash the after-filtering patches.
To solve this problem, approaches based on semantic reasoning are applied to completely fix program crashes/vulnerabilities.

\begin{figure}[!t]
        \centering
        \includegraphics[width=0.6\textwidth]{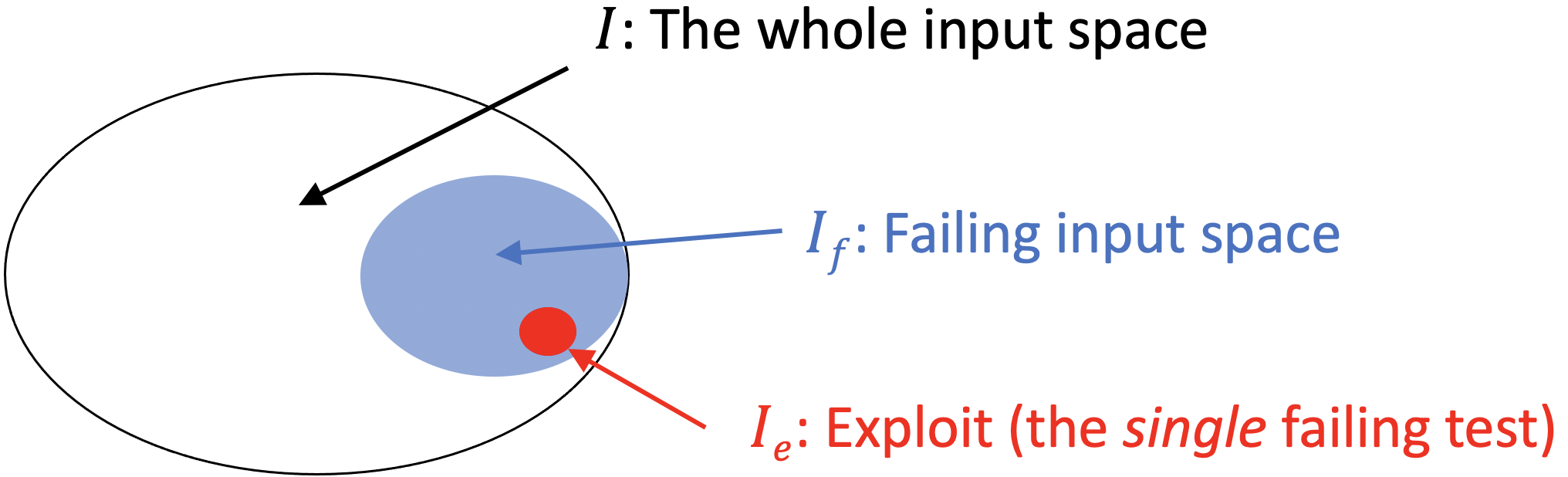}
        \caption{ The input space of a vulnerable program, where $I_e$ represents one given exploit and $I_f$ is the set of inputs that can trigger the same vulnerability. \label{fig:input_space}}
\end{figure}

\subsubsection{Guide patch generation via extracted constraints}\label{sec:extractfix}
\paragraph{High-level Idea} When a vulnerability is detected, people usually attach a test case that can trigger this vulnerability along with the bug/vulnerability report.
Such a test case that can trigger vulnerabilities is also called \textit{exploit}.
As demonstrated in Figure~\ref{fig:input_space}, except for the given exploit ($I_e$), there could be many other inputs (called failing inputs $I_f$) that can trigger the same vulnerability.
Driven by exploit $I_e$, APR system may just fix the vulnerability on $I_e$, but not be able to completely fix the bug on all failing inputs.
To solve this problem, we can consider generalizing $I_e$ and inferring a representation of all the failing input space $I_f$~\citep{Gao21}.
The generalized representation represents not only the given exploit, but also all the failing inputs.
If APR is driven by the generalized representation, it can then fix all the failing test cases and hence completely fix this vulnerability.

Figure~\ref{fig:coreutils} shows an example bug in \texttt{Coreutils}, which is adapted from~\cite{Gao21}.
In this example, an undefined behavior bug of \verb+memcpy+~\footnote{https://debbugs.gnu.org/cgi/bugreport.cgi?bug=26545} happens at line 7.
This undefined behavior will crash the program on some platforms.
This bug will be triggered when the memory spaces of the source and target overlap with each other.
For instance, if the value of $\mathtt{size}$ is 13, the \emph{for}-loop at line 4 will terminate in the second iteration when $\mathit{size}/2{=}6$ (integer division) and $i{=}6$.
Hence, the memory space of \verb+memcpy+'s destination is $[r, r+7)$, while the memory space of \verb+memcpy+'s source is $[r+6, r+13)$.
There is one byte overlap between the memory space of the destination and source.
Taking $\mathtt{size}{=}13$ as input, UndefinedBehaviorSanitizer (UBSAN) will crash the program.

\begin{figure}[!t]
\begin{lstlisting}[linewidth=\columnwidth,language=C]
void fillp (char *r, size_t size){
  ...
  r[2] = bits \& 255;
  for (i = 3; (*@{\fbox{i < size / 2}}@*); i *= 2)
    memcpy(r + i, r, i);
  if (i < size)
    (*@{\fbox{memcpy(r + i, r, size - i)}}@*);
}
\end{lstlisting}
\caption{An undefined behavior bug from Coreutils\label{fig:coreutils} adapted from~\cite{Gao21}}
\end{figure}

Starting with the single crashing input ($\mathtt{size}{=}13$), we first infer a generalized representation of all the failing inputs in the form of \textit{constraint}.
According to the specification of memory, the source and destination region should not overlap.
In other words, for memcpy invocation $\texttt{memcpy}(\texttt{p}, \texttt{q}, \texttt{s})$, the following constraint $\texttt{p}{+}\texttt{s} \leq \texttt{q} \lor \texttt{q}{+}\texttt{s} \leq \texttt{p}$ must be satisfied.
In this example, the corresponding constraint is
\begin{align*}
    (\texttt{r}{+}\texttt{i}{+}\texttt{size}{-}\texttt{i}{\leq}\texttt{r}
     \lor
     \texttt{r}{+}\texttt{size}{-}\texttt{i}{\leq}\texttt{r}{+}\texttt{i})
    \equiv
    (\texttt{size} \leq 0 \lor \texttt{size}{\leq}2{\texttt{*}}\texttt{i})
\end{align*}
This constraint is called \textit{Crash-free Constraint} (\textit{CFC}).
Any input, e.g., \texttt{size}=13, that makes program execution violate CFC at line 7, is a failing input.
Accordingly, the negation of CFC represents all the failing test cases.
Then, the repair goal is to correct the program to ensure this constraint is satisfied on any test input.

Given CFC, a patch should be then introduced at the fix location to ensure CFC is always true at line 7.
Since the fix location (line 4 in this example) is usually different from the crash location.
$\mathit{CFC}$ is first backward propagated from the crash location to the fix location along with all the feasible paths.

In the above example, $\mathit{CFC}$ is propagated backward from line 7 to line 4 along with one path (the truth branch of the if-statement at line 6).
After propagation, the constraint \textit{CFC'} at line 4 is $i{<}\mathit{size} \mapsto (\mathit{size} \leq 0 \lor \mathit{size}{\leq}2{*}i)$.
This means that with CFC' as the precondition at line 4, executing the program along with $i{<}\mathit{size}$ path will ensure CFC is always satisfied at line 7. 
According to the constraint at the fix location, a patch $f$ can be synthesized to replace the condition of \emph{for}-loop.
In order to completely fix this bug, it is necessary to ensure CFC' is always satisfied at line 4 after applying $f$.
In this example, $f$=``${i <= size/2}$'' is produced by synthesizer. Finally, this buggy program can be patched by:
\begin{verbatim}
- for (i = 3; i < size / 2; i *= 2)
+ for (i = 3; i <= size / 2; i *= 2)
\end{verbatim}
Fortunately, this patch is semantically equivalent to the patch provided by developers.
On the other hand, just driven by the failing case, it is very likely that APR generates overfitting patches.
For instance, Fix2Fit~\citep{fix2fit} generates the following patch which fixes the bug when $\mathtt{size}{=}13$.
However, the patched program can still crash on other inputs, such as $\mathtt{size}{=}7$.
\begin{verbatim}
+ for (i = 3; i < size / 2 || i == 6; i *= 2)
\end{verbatim}

\paragraph{Formal Treatment}
Let us now present the formal patch generation process based on the extracted constraint.
Overall, an exploit is first generalized into a \emph{crash-free constraint} (\emph{CFC}), which represents an abstracted constraint that is violated by the witnessed vulnerability.
A crash is broadly defined to be any program termination caused by violating some properties.
A crash could be in the form of (1) violation of a user-defined assertion (i.e., $\mathtt{assert}(C)$), (2) violation of assertion enforced by operating system, such null pointer dereference, or (3) violation of check inserted by \emph{sanitizers} for enforcing safety properties, e.g., AddressSanitizer (ASAN)~\footnote{https://clang.llvm.org/docs/AddressSanitizer.html} for memory safety checks and UndefinedBehaviorSanitizer (UBSAN)~\footnote{https://clang.llvm.org/docs/UndefinedBehaviorSanitizer.html} for integer overflow protection.
\emph{The condition that should be satisfied to prevent the crash} is represented as CFC at the crashing location.
Once a generated patch ensures $\mathit{CFC}$ is always satisfied at the crash location, then the crash will not be triggered by any program input.
The basic workflow to generate patches is as follows.
\begin{table}[!thb]
\centering
\caption{Basic crash classes, crash expressions/statements, and the corresponding \emph{Crash-Free Constraint} $\mathit{CFC}$-template adapted from~\cite{Gao21}.
Seven types of crashse: explicit developer assertion violation, sanitizer-induced crash such as buffer overflows/underflows, integer overflows, API constraint violations.}
\begin{tabular}{|l|c|l|l|}
\hline
Class & Template ID & Expression & $\mathit{CFC}$ Template \\
\hline
\hline
developer & $T_1$ & $\mathtt{assert}(C)$ & $C$ \\
\hline
\multirow{6}{*}{sanitizer} & \multirow{2}{*}{$T_2$} & \multirow{2}{*}{$\texttt{*p}$} & $\texttt{p}{+}\texttt{sizeof}(\texttt{*p}) \leq \mathit{base}(\texttt{p}){+}\mathit{size}(\texttt{p})$ \\
                           &                        &                                & $\texttt{p} \geq \mathit{base}(\texttt{p})$ \\
\cdashline{2-4}
& $T_3$  & $\texttt{a}~\mathit{op}~\texttt{b}$ &
    $\texttt{MIN} \leq \mathtt{a}~\mathit{op}~\mathtt{b} \leq \texttt{MAX}$ (over $\mathbb{Z}$) \\
\cdashline{2-4}
& $T_4$  & $\texttt{memcpy}(\texttt{p}, \texttt{q}, \texttt{s})$ &
    $\texttt{p}{+}\texttt{s} \leq \texttt{q} \lor
     \texttt{q}{+}\texttt{s} \leq \texttt{p}$ \\
\cdashline{2-4}
& $T_5$     & $\texttt{*p}$ (for $\texttt{p}{=}0$) &
    $\texttt{p} \neq 0$ \\
\cdashline{2-4}
& $T_6$ & $\mathtt{a}~/~\mathtt{b}$ (for $\mathtt{b}{=}0$) &
    $\mathtt{b} \neq 0$ \\
\hline
\end{tabular}
\label{tab:template}
\end{table}
\begin{enumerate}
\item \textbf{Constraint Extraction}. \label{step:cfc}
    Given a program and an exploit that triggers a crash, we extract the ``crash-free constraint'' ($\mathit{CFC}$) according to a predefined template.
    Table~\ref{tab:template} defines seven types of templates for some common bugs/vulnerabilities, which formulate the underlying cause of each defect. For instance, the template for a buffer overflow is $\mathtt{buffer\_access}-\mathtt{buffer\_base}<\mathtt{buffer\_size}$.
\item \textbf{Fix Localization}.\label{step:fix}
    Given the crash location and the extracted $\mathit{CFC}$, \emph{fix localization} technique is then applied to decide one (or more) fix location(s).
    The fix locations can be determined using a dependency-based approach.
    The dependency-based fix localization determines statements that have a \emph{control} or \emph{data}-dependency with the statement at the crash location.
    The determined statements could affect the truth of $\mathit{CFC}$.
\item \textbf{Constraint Propagation}.\label{step:wpc}
    Given CFC at the crash location, we then propagate it to a given fix location and calculate $\mathit{CFC}'$ by solving the following Hoare triple:
    \begin{align}\label{eq:cfc}
        \{\mathit{CFC'}\}P\{\mathit{CFC}\}
        \tag{\textsc{CFC-Propagation}}
    \end{align}
    The program between crash and fix location is donated as $P$.
    CFC' is the weakest (least restrictive) precondition that guarantees postcondition CFC~\citep{chandra2009snugglebug}.
\item \textbf{Patch Synthesis}.\label{step:patch}
    Once CFC' is calculated, patch candidates are then synthesized at the fix location.
    Patch synthesis is to construct expression $f$ to substitute the expression $\rho$ at the fix location, with the goal of making the following Hoare triple holds:
    \begin{align}\label{eq:repair}
        \{\mathit{true}\}~[\highlight{\rho \mapsto f}]~\{CFC'\}~P~\{\mathit{CFC}\}
        \tag{\textsc{CFC-Repair}}
    \end{align}
    The synthesized patch $\rho \mapsto f$ fixes the buggy program by substituting expression $\rho$ with the synthesized expression $f$.
    The precondition before the patch is set as $\mathit{true}$ to ensure the patch fix the bug regardless of the context.
    The generated patch ensures that $\mathit{CFC}'$ is satisfied at the fix location.
    Therefore, in the patched program, $\mathit{CFC}$ will also be always satisfied at the crash location.
\end{enumerate}
In general, we rely on the extracted constraints to make programs satisfy safety properties including fixing buffer overflow, integer overflow, null pointer dereference, and so on.
Once the patched program satisfies this constraint, it is guaranteed that such crashes/vulnerabilities are completely fixed.

\subsubsection{Patch exploration via Concolic repair}\label{sec:cpr}
To tackle the overfitting problem, the concept of \textit{Concolic Program Repair} (CPR)~\citep{cpr} introduces a repair process that (1) performs the co-exploration of input space and patch space, (2) handles an abstract patch space, and (3) incorporates additional user-provided specifications.
\begin{figure}[!t]
\centering
\includegraphics[width=0.5\textwidth]{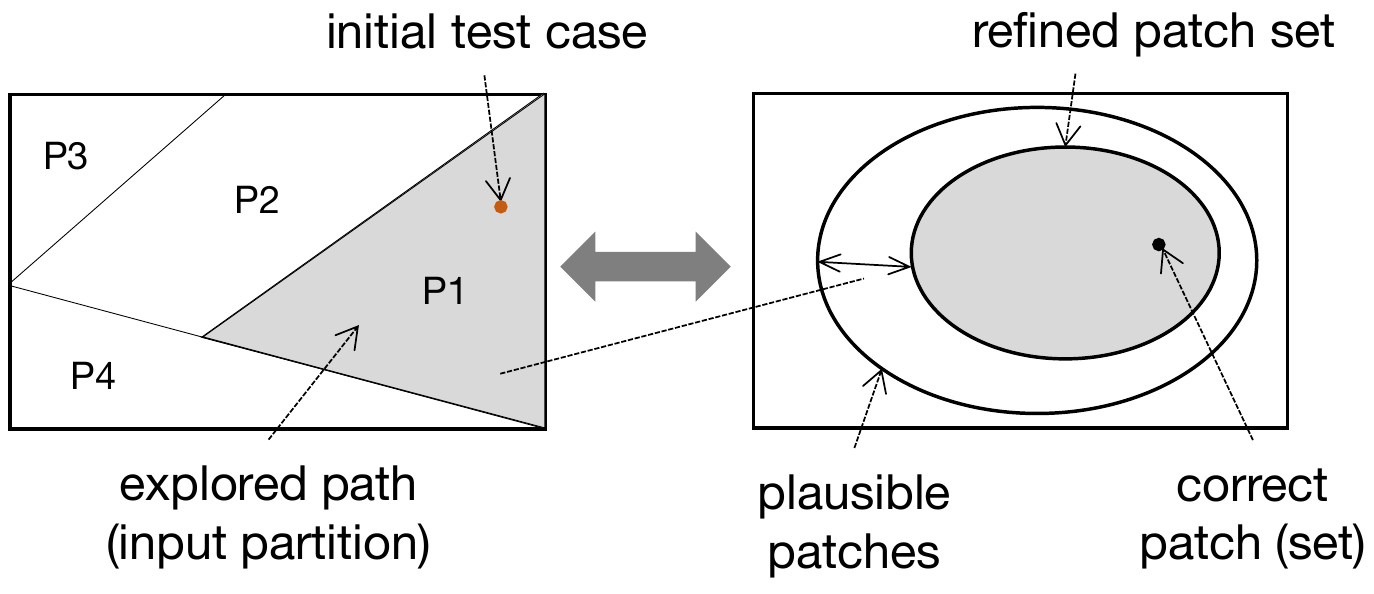}
\caption{Illustrates the concept of co-exploration in \textit{Concolic Program Repair}~\citep{cpr}. The left part shows the input space, the right part shows the patch space. By incrementally covering the input space, the patch space gets refined gradually. \label{fig:cpr_coexploration}}
\end{figure}
Figure~\ref{fig:cpr_coexploration} illustrates the concept of co-exploration. The left side shows the input space, while the right side shows the patch space. Starting from an initial failing test case, CPR generates a set of plausible patches. Therefore, it limits the patch space to patches that can pass the initial failing test case. CPR further refines the patch space by exploring not only the single test input but the complete input partition exercised by the test. In addition to the existing tests, CPR generates new test inputs via generational search~\citep{sage} and uses the user-provided constraint to reason about the patch within the corresponding input partitions. Over time, this process will incrementally cover more of the input space and further refine the patch space, leading to a gradual improvement of the generated patches.
A key insight of this work is that even simple user-provided constraints can help refine patches regarding newly generated inputs.

\paragraph{Overview: Concolic Repair Algorithm}

\renewcommand{\algorithmicrequire}{\textbf{Input:}}
\renewcommand{\algorithmicensure}{\textbf{Output:}}
\begin{algorithm}[h!]
\caption{Concolic Repair Algorithm~\citep{cpr}}
\label{alg:cpr_algorithm}
\begin{algorithmic}[1]
\REQUIRE set of initial test cases $I$, buggy locations $L = (patchLoc, bugLoc)$, budget $b$, \\ specification $\sigma$, language components $C$
\ENSURE set of ranked patches $P$
\STATE $P \gets$ \textsc{Synthesize}(C, I, L)
\WHILE{$P \neq \emptyset$ and \textsc{CheckBudget}(b)}
    \STATE $t$, $\rho \gets$ \textsc{PickNewInput}($P$)
    \IF{no test $t$ available}
        \RETURN $P$
    \ENDIF
    \STATE $\phi_t$, $hit_{patch}$, $hit_{bug} \gets$ \textsc{ConcolicExec}($t$, $\rho$, $L$)
    \IF{$hit_{patch}$}
        \RETURN $P \gets$ \textsc{Reduce}($P$, $\phi_t$, $\sigma$, $hit_{bug}$)
    \ENDIF
\ENDWHILE
\RETURN $P$
\end{algorithmic}
\end{algorithm}
	
Algorithm~\ref{alg:cpr_algorithm} shows the general repair process applied by CPR. It starts with synthesizing a pool of plausible patches based on the provided test cases and the given fix location. The remaining process focuses on eliminating and refining patches based on additional observations with new inputs. These inputs are generated via generational search~\citep{sage}, i.e., by flipping the terms in the already collected path constraints. Via concolic execution, CPR determines the path constraint for a new input and whether the execution touches the relevant patch location. Only if this is the case the path constraint is used to refine the patch pool. In addition to reducing the patch space, CPR uses dynamic execution information to rank the remaining patches. Patches are ranked higher (1) if they are exercised with the generated input (i.e., there is supporting evidence for their correctness in terms of generated test cases) and (2) if the input in combination with the patch exercises the original bug observation location. The patches are deprioritized if the original control flow is changed significantly.

The construction of a rich search space and its efficient exploration are general technical challenges of APR techniques. CPR is tackling these challenges by maintaining an \textit{abstract} patch space and applying efficient \textit{infeasability checks} between the input and patch space to avoid unnecessary computations.

\paragraph{Abstract Patch Space}
Concolic program repair uses an abstract patch representation to (1) generate and maintain a smaller set of patch candidates and (2) enable the refinement of patches. An abstract patch $\rho$ in this patch space is defined as the 3-tuple $(\theta_\rho, T_\rho, \psi_\rho)$ with the set of program variables $X_P$, the corresponding subset of input variables $X \subseteq X_P$, and the set of template parameters $A$.
$\theta_\rho(X_P, A)$ is the repaired (boolean or integer) expression,
$T_\rho(A)$ denotes the conjunction of constraints $\tau_\rho(a_i)$ on the parameters $a_i \in A$ included in $\theta_\rho$: $T_\rho(A) = \bigwedge\limits_{a_i \in A} \tau_\rho(a_i)$,
and $\psi_\rho(X, A)$ is the \textit{patch formula} induced by inserting the expression $\theta_\rho$ into the buggy program.
For example, assume the patch $\rho$ is the expression inside a conditional statement. Assuming that \texttt{x} is an available program variable, concrete patches could be \texttt{x > 0}, \texttt{x > 1}, \texttt{x > 2}, etc. In the abstract patch space of CPR such a patch would be represented with a template like $x > a, a \in [0, 10]$, and hence, would subsume a set of concrete patches. With regard to the previously mentioned 3-tuple, such abstract patch would be described with:
$\theta_\rho := x > a$, 
$T_\rho = tau_\rho(a) := (a \ge 0) \land (a \leq 10)$,
$\psi_\rho := x > a$.
Abstract patches can be refined by modifying the parameter constraint $T_\rho$, which is part of one of the infeasibility checks discussed in the next paragraph.

\paragraph{Infeasability Checks}
The co-exploration concept of concolic program repair includes infeasibility checks in both directions: (1) from input space to patch space and (2) from the patch space to the input space. Such infeasibility checks help to further efficiently limit both spaces.

The check from the input space to the patch space is called \textit{\textbf{patch} reduction}. It means that CPR tries to reduce the patch space based on the collected information from the input space. The formula to check is as follows:
$\forall a_i \in A \; \forall x_j \in X: \phi(X) \land \psi_\rho(X,A) \land T_\rho(A) \implies \sigma(X)$. It describes that if there is an input $x_i$ and parameter $a_i$ that is feasible under the obtained path constraint $\phi$ and the patch formula $\psi_\rho$ with its parameter constraint $T_\rho$, then it needs to satisfy the user-provided specification $\sigma$. The formula can be filled with the corresponding values and negated to check for counterexamples. Each counterexample leads to the refinement of the parameter constraint $T_\rho$, i.e., to a reduction of the patch space.

The check from the patch space to the input space is called \textit{\textbf{path} reduction}. Before exploring a partition in the input space, CPR checks whether there is at least one patch left in the patch space that can exercise the corresponding path. If no such patch is left, the input partition does not need to be explored.

\subsection{Human in the Loop}\label{sec:human-in-the-loop}
We presented several approaches to alleviate the patch overfitting issue.
Unfortunately, none of them could perfectly solve this problem.
This is because complete program specification is usually not available and inferring developers' intent is error-prone and fundamentally difficult.
So, the ultimate solution is to involve human in the loop: \textit{asking developers to validate the auto-generated patches before applying them in the program}.
In this setting, instead of fully replacing developers in fixing bugs, APR systems play the role of helping developers by generating patch suggestions.

To validate the correctness of patches, developers must understand the bug to be fixed and fully understand patch's semantics.
Even for expert programmers, understanding the root cause of bug and its corresponding patch is not easy and straightforward.
To help developers understand the bug and the patches, people proposed workflows to enable developers to review patches just with a few clicks~\citep{APR20}.
Technically, the patch semantics are first translated into a set of \textit{questions}, e.g., ``what is the expected value of variable \textit{index} at line 46?'', ``whether statement \texttt{return NULL;} should be executed or not for a test?''.
To review patches, developers just need to select the answers to those questions.
Such developers' feedback can then help filter overfitted patches and find the correct patch among the plausible patches.

Let us take a look at the example shown in Listing~\ref{lst:example1}, which is adapted from \cite{APR20}.
In this example, incorrect bound checking causes a buffer overflow vulnerability~\footnote{https://bugs.chromium.org/p/oss-fuzz/issues/detail?id=1345}.
This buffer overflow write occurs at line 10 when the value of \textit{remaining\_space} is equal to \textit{width} at line 3.
This is because the program will illegally rewrite the memory data after \textit{frame\_end}. 
To correctly fix this vulnerability, one solution is to modify the assignment from $\mathtt{frame\_end-frame}-3$ at line~\ref{line:assignment} to $\mathtt{frame\_end-frame}-4$.

\begin{figure}[!t]
{
\begin{lstlisting}[basicstyle=\footnotesize,caption={Buffer overflow vulnerability in \textit{FFmpeg} adapted from \cite{APR20}},upquote=true,label=lst:example1,escapechar=!]
    int remaining!\_!space = frame!\_!end-frame - 3; !\color{green}//correct patch: "frame\_end-frame-3" -> "frame\_end-frame-4"\label{line:assignment}!
    if (remaining!\_!space < width) 
        return AVERROR!\_!INVALIDDATA;
    frame[0] = frame[1] = frame[width] =
            frame[width+1] = bytestream!\_!get!\_!byte(gb);
    frame += 2;
    frame[0] = frame[1] = frame[width] =  !{\color{red}// buffer overflow location}!
    frame[width +1] = bytestream!\_!get!\_!byte(gb);
    frame += 2;
\end{lstlisting}}
\vspace*{-15pt}
\end{figure}

Given a failing test triggering this buffer overflow, current repair systems could generate several plausible patches.
For example, here are four patches that can fix the failing test by replacing the assignment at line~\ref{line:assignment}.
\begin{itemize}
    \item[$p1$:] $\mathtt{frame\_end-frame}-3$
    $\mapsto\mathtt{frame\_end{-}frame{-}4}$
    \item[$p2$:] $\mathtt{frame\_end-frame}-3$ $\mapsto\mathtt{frame\_end{-}frame{-}5}$
    \item[$p3$:] $\mathtt{frame\_end-frame}-3$ $\mapsto\mathtt{frame\_end{-}frame{-}6}$
\end{itemize}
Among these patches, only one of them is correct and the other two are overfitted patches.
To filter out the overfitted patches, the following question will be constructed: ``What is the correct value of $\mathtt{remaining_space}$ at line~\ref{line:assignment} to pass the failing test?''. 
Then, a set of candidate answers to the above question will be generated by this tool and visually provided to developers.
Asking and answering questions in terms of variable values are designed based on the assumption that \textit{developers are usually better at reasoning about program outputs}.
This is because program correctness is usually defined in terms of outputs.
For developers, answering those questions is an analogy with the debugging process.
Therefore, developers should be able to answer those questions without exhaustively analyzing the program and patch semantics.
The candidate answers could be generated with references to plausible patches.
For the example question above, the candidate answers could be 8, 7, 6, corresponding to the above three plausible patches $p1$, $p1$, and $p2$ when ($\mathtt{frame\_end-frame}=12$).
Suppose developer indicates that the expected value is 8, then, another question will be created: ``How to produce the expected value (8) at line~\ref{line:assignment}?''. 
The candidate patches are the answers to this question.
In the above example, patch $\mathit{frame\_end{-}frame{-}4}$ is the only answer as it can produce the correct value 8.
Later, developer determines whether this candidate patch is correct or not.
Instead of directly answering developers to review candidate patches one by one, the proposed process enables developers to review via answering questions.
Relying on the interactive question-answering process, the overfitted patches, such as $\mathit{frame\_end{-}frame{-}5}$, can be ruled out.
Involving human in the patch generation and review process actually provides a way to alleviate the over-fitting problem.

%% file: chapters/7_tools.tex
\section{Program Repair Technology}\label{sec:technology}
In this chapter, we detail the program repair tools available, as well as certain known industrial deployment of program repair technology.

\subsection{Program Repair Tools}

Automated program repair has gained increasing attention from both academia and industry researchers.
Many automated repair tools have been developed or even deployed in the past few years.
Table \ref{tab:repair_tools} lists some representative automated program repair tools.
Existing repair tools are mainly designed for fixing the bugs or vulnerabilities in C/C++ and Java programs.
The widely used benchmarks for evaluating C/C++ repair tools are ManyBugs~\citep{manybugs}, IntroClass~\citep{manybugs}, Codeflaws and ExtractFix benchmark~\citep{Gao21}.
Existing repair tools have shown good performance on those benchmarks.
For instance, according to the study shown in~\citep{TOSEM18Sergey}, among 105 bugs in ManyBugs benchmark, GenProg, Prophet, and Angelix (only evaluated on partial benchmark) automatically generate patches for 27, 42, and 28 bugs respectively.
Among the generated patches, 3, 15, and 10 of them are equivalent to human patches.
Besides general bugs, existing tools also showed great performance in fixing vulnerabilities.
For instance, ExtractFix~\citep{Gao21} could automatically fix 16 vulnerabilities out of 30 subjects in the ExtractFix benchmark.
In contrast, the widely used benchmark for evaluating Java repair is Defects4J~\citep{defects4j}, which includes more than 400 real-world defects.
On the Defect4J benchmark, the best Java repair tools can correctly fix more than 50 bugs.
Besides the C/C++ and Java repair tools, few tools (e.g., Clara, Refactory) are designed for fixing the bugs in introductory Python programs, which are mainly used for programming education.
For instance, Refactory is designed to generate real-time patches for student programs with the goal of helping students to learn programming.

As shown in Table~\ref{tab:repair_tools}, all the repair tools are also classified according to their underlying technique.
The listed tools are search-based, semantic-based, or learning-based repair.
When the automated program repair was proposed, researchers mainly focused on search-based (e.g., GenProg) and semantic-based (e.g., SemFix) tools.
In recent years, we have seen a trend of tools and techniques that apply deep learning to the field of program repair.
Learning-based repair tools have gained a lot of attention and achieved great performance in generating correct patches.

Most repair tools are from academia, while some of them (e.g., SapFix~\citep{sapfix}, GetaFix~\citep{getafix} and Fixie~\citep{fixie}) are from industry and have been deployed to fix  real-world bugs.
Note that, Table~\ref{tab:repair_tools} shows an incomplete set of repair tools, a complete list of repair tools can be found at \url{http://program-repair.org/tools.html}.

{\small
\begin{table}[h]
\caption{Representative automated program repair tools}
\label{tab:repair_tools}
\begin{tabular}{p{0.1\linewidth}|p{0.1\linewidth}|p{0.1\linewidth}|p{0.6\linewidth}}\hline
Tool      & Language & Technique  & Short description                                                               \\\hline\hline
AllRepair & C/C++  & Search       & mutation-based repair tool for C programs equipped with assertions in the code  \\\hline
GenProg   & C/C++  & Search       & automated program repair tool based on genetic programming                      \\\hline
Fix2Fit   & C/C++  & Search       & Combining patch exploration with greybox fuzzing to alleviate overfitting       \\\hline
Prophet   & C/C++  & Search \& Learning & automated program repair that learns from correct patches                 \\\hline
RSRepair  & C/C++  & Search       & a modification of GenProg that uses random search                               \\\hline
SPR       & C/C++  & Semantic     & automated program repair tool with condition synthesis                          \\\hline
SemFix    & C/C++  & Semantic     & automated program repair tool based on symbolic analysis                        \\\hline
Angelix   & C/C++  & Semantic     & automated program repair tool based on symbolic analysis                        \\\hline
CPR       & C/C++  & Semantic     & program repair based on concolic execution                                      \\\hline
DeepFix   & C/C++  & Learning     & tool for fixing common programming errors based on deep learning                \\\hline
MemFix    & C/C++  & Static       & static analysis-based repair tool for memory deallocation errors in C programs  \\\hline
ACS       & Java   & Search \& Learning  & repair tool with accurate condition synthesis                            \\\hline
jGenProg  & Java   & Search       & the Java version of GenProg for bugs in Java programs                           \\\hline
NPEFix    & Java   & Search       & generates patches for Null Pointer Exceptions with meta-programming             \\\hline
SapFix    & Java   & Search       & the first deployment at Facebook of automated end-to-end fault fixing           \\\hline
SimFix    & Java   & Search       & fixing Java bugs by leveraging existing patches and similar code                \\\hline
TBar      & Java   & Search       & template-based automated program repair                                         \\\hline
Nopol     & Java   & Semantic     & automated program repair tool for conditional expressions                       \\\hline
Coconut   & Java   & Learning     & tool for fixing bug by learning from history patches                            \\\hline
Genesis   & Java   & Learning     & infer code transforms from history patches for automatic patch generation       \\\hline
GetaFix   & Java   & Learning     & automated program repair that learns from history patches                       \\\hline
SequenceR & Java   & Learning     & program repair based on sequence-to-sequence learning                           \\\hline
Clara     & Python & Search       & repair tool for introductory programming assignments                            \\\hline
Refactory & Python & Search \& inference & tool for generating real-time program repairs of buggy student programs  \\\hline
Fixie     &        & Learning     & a repair tool developed and used at Bloomberg                                     \\\hline
\end{tabular}
\end{table}}

\subsection{Industrial Deployment}
\label{sec:industry}

We now discuss known industrial deployment of program repair.

\begin{figure}[h]
    \centering
    \includegraphics[width=0.9\textwidth]{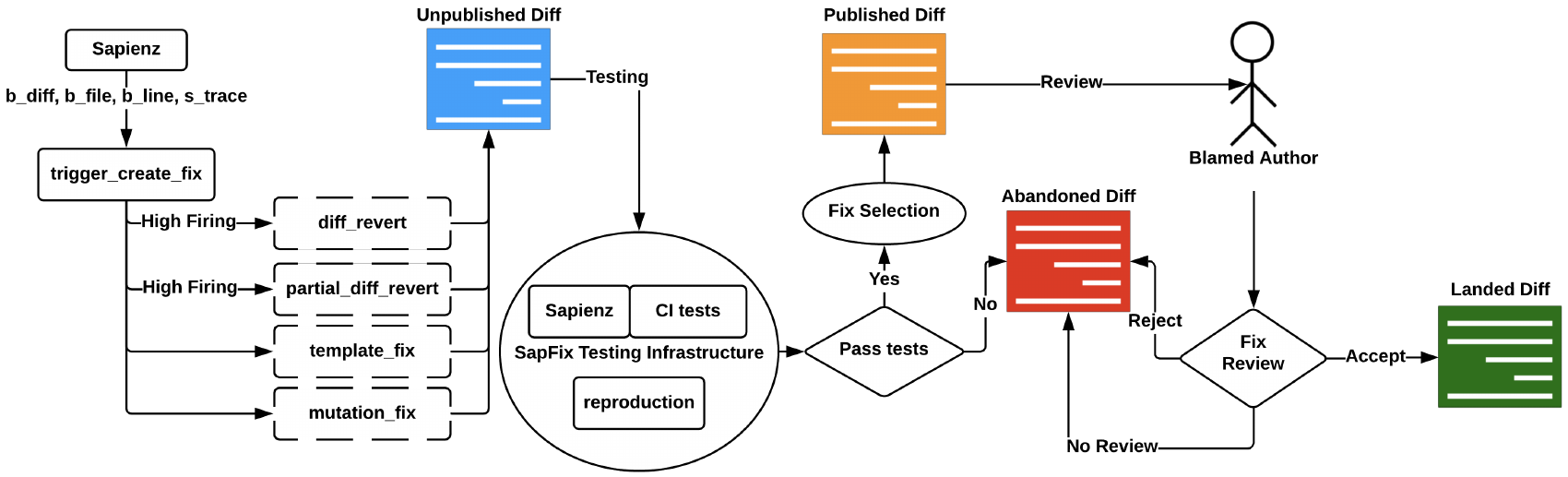}
    \caption{Facebook's \sapfix Workflow adopted from~\cite{sapfix}.}
    \label{fig:sapfix}
\end{figure}

The ﬁrst end-to-end deployment repair in an industrial context was presented by SapFix~\cite{sapfix} at Facebook and is called \sapfix. Its workflow (see Figure~\ref{fig:sapfix}) starts with testing the code changes added to the continuous integration with \sapienz~\citep{sapienz}, a search-based testing tool deployed at Facebook. It specifically searches for null-dereference faults. \sapfix generates repairs by using template-based and mutation-based approaches. If no patch can be found, it will revert the code change. All generated patches are cross-checked using the static analyzer \infer~\citep{infer}. After heuristically selecting one patch candidate, it will notify the software developer to review the patch. SapFix is a repair tool that is integrated as part of continuous integration, where SapFix monitors test failures, reproduces them, and automatically looks for patches. 

\begin{figure}[h]
    \centering
    \includegraphics[width=0.8\textwidth]{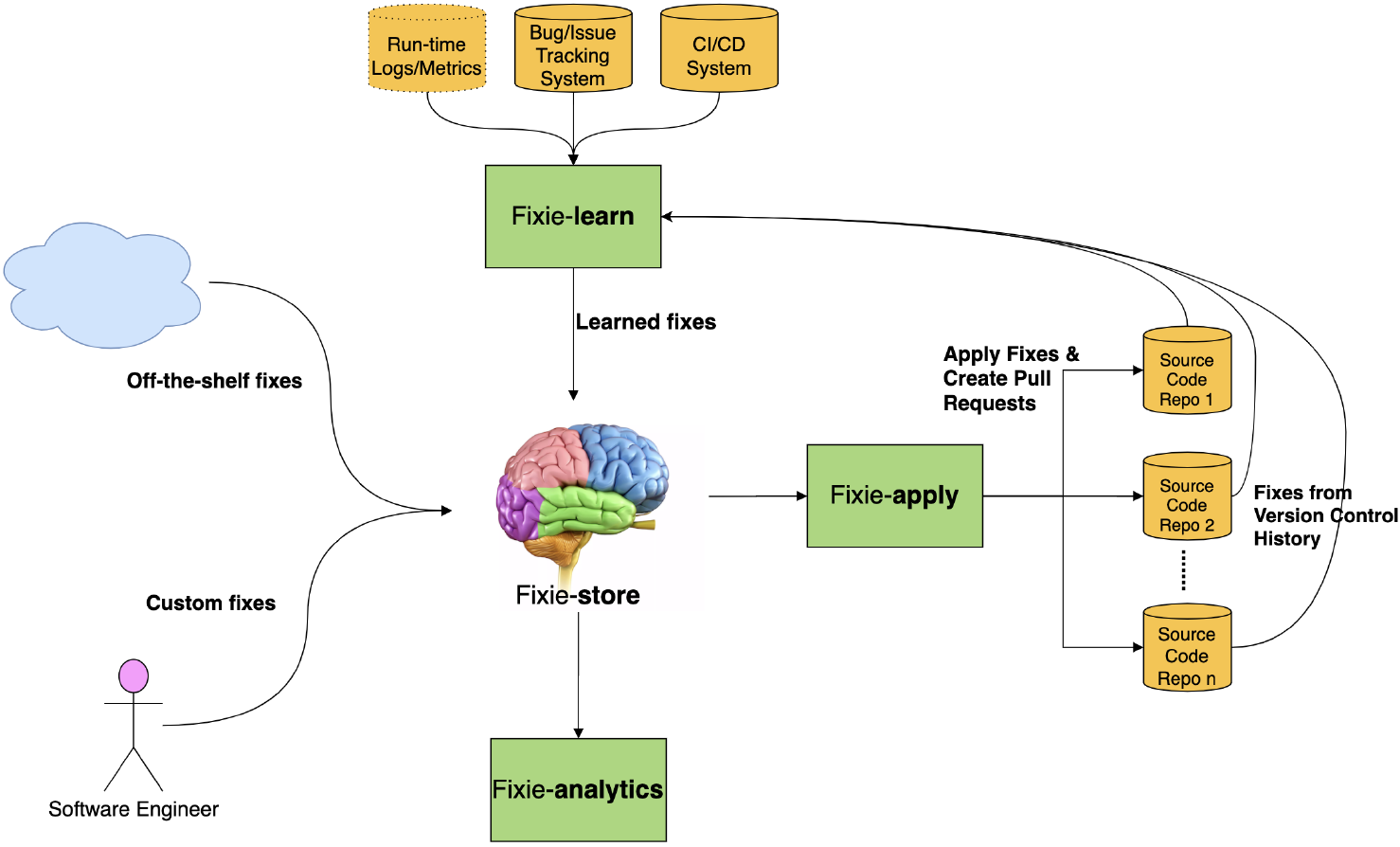}
    \vspace{-3pt}
    \caption{Bloomberg's \fixie Architecture adopted from~\cite{fixie}.}
    \label{fig:fixie}
    \vspace{-10pt}
\end{figure}

After Facebook, other companies like Bloomberg~\citep{fixie} have started looking into concrete ways to integrate APR into their development processes. Their tool \fixie combines three different fix types (see Figure~\ref{fig:fixie}). Firstly, they use off-the-shelf linter tools like clang-tidy\footnote{\url{https://releases.llvm.org/10.0.0/tools/clang/tools/extra/docs/clang-tidy}}. Such tools are broadly accepted and deployed in practice and can provide fixes with high confidence for, e.g., style violations and interface misuses.
Secondly, they allow software developers to propose their custom fixes that can then be proposed for other scenarios as well.
Thirdly, they learn fixes and fix patterns from the version-control history linked with bug/issue trackers and CI/CD pipelines.
\fixie collects and stores patches from these three sources and proposes fixes by creating pull requests.
Fixie~\cite{fixie} reports that after some initial backlash by developers and improvement iterations, they received positive feedback from the users. In particular, after putting the user in control of reviewing and applying auto-generated patches. Software developers see \fixie as a development "assistant", which frees up their time to focus on tasks like cleaning up the code, for which there is usually not much time left.

There exist interest in the industry (e.g. Fujitsu\cite{Fujitsu}) on combining program repair with static analysis systems. Companies which have shown interest in this line of work, that we are aware of, include Oracle Labs. 

\subsection{Achieved Results So Far}\label{sec:results}

The research in automated program repair has already created a vast portfolio of techniques and methods.
Community-wide datasets like the Defects4J \citep{defects4j}, CodeFlaws~\citep{codeflaws}, and ManyBugs~\citep{manybugs} benchmark have been established to foster fair comparisons and to identify common limitations.
Table~\ref{tab:repair_tools} shows a subset of the created tools, which have been developed for different programming languages and different application scenarios.
These tools have shown the ability to fix many bugs.
Specifically, out of the ~400 bugs from Defects4J benchmark, when the perfect fix location is given, around 20\% of bugs can be correctly fixed by at least one APR tool. Suppose the perfect fix location is unknown, relying on the realistic fault localization output, i.e., spectrum-based fault localization, existing tools correctly fix around 10\% of bugs~\citep{tbar}. Moreover, with the exception of some tools like Angelix \citep{angelix} which can produce multi-line fixes, many existing automated repair tools are limited to fixing simple single-line patches, such as inserting a null-check for fixing null pointer dereference.

This means that while automated program repair is a promising and emerging technology - significant engineering and technical challenges remain in automatically synthesizing large-scale patches. At the same time, the adoption of automated program repair in the industry has started (Section~\ref{sec:industry}) with first deployments, and companies like Facebook/Meta and Bloomberg shared their identified benefits and lessons learned with the researchers. In this regard, we consider the recent development of language model based code generation \citep{codex,codewhisperer}from industry (which nevertheless tends to generate buggy code) with some satisfaction. We feel that there remain significant prospects in terms of leveraging automated repair to improve automatically generated code snippets - which may provide a way to leverage the research advances in program repair for enhancing programmer productivity.

Apart from enhancing programmer productivity - there remain other possible areas where automated program repair has shown promising results. One such area is automated security vulnerability repair. Early on, tools like Angelix~\citep{angelix} have shown promise in this regard, by repairing complex vulnerabilities like the well-known Heartbleed vulnerability~\footnote{\url{http://heartbleed.com}}. More recently, the works on semantic reasoning~\citep{Gao21, cpr, senx} and inductive inference based repair \citep{vulnfix} have shown further promise in this domain. Since security vulnerabilities are typically detected via fuzz testing \citep{fuzzing21}, a security vulnerability repair approach which is seamlessly integrated with fuzz campaign, would be ideal. We posit that this is a difficult but achievable goal, and would urge the research community to work towards this goal.

Last but not the least, program repair approaches are already being used for programming education via a number of point technologies \citep{Clara,Sarfgen,Refactory,verifix}. This shows the power of automated program repair for providing feedback to students struggling to learn to program. In the future, there remains the opportunity of moving from such point technologies to developing widely-used intelligent tutoring systems for programming education that rely on program repair. Due to the wide demand for computer science undergraduate degrees which typically involves learning programming in the first year, this would indeed be an impactful research direction to work towards.

%% file: chapters/6_application.tex
\section{Applications}\label{sec:applications}

The automated program repair techniques discussed in the previous chapters have many applications and use cases in the context of software development. First and foremost, they can be applied to repair functional errors, which would be the most general application of automated program repair techniques. Existing test suites help pinpoint the buggy locations and guide the repair process. \cite{Monperrus2018LivingReview} lists and categorizes the existing research literature in program repair, which spans various repair scenarios like static errors, concurrency errors, build errors, etc. A rather prominent topic is the repair of security vulnerabilities: with 28,695 security vulnerabilities reported in 2021 \citep{vulnerabilities2021} and an average time to fix of 200 days for as critical classified vulnerabilities \citep{timetofix2021}, program repair has enormous potential for the security domain. Another exciting application aspect is the integration into development workflows by accommodating repair features in the Integrated Development Environments (IDE) and Continuous Integration (CI) workflows. With the rising number of students in computer science related studies, the need for automated tutoring aids is rising \citep{senx}. Automated program repair offers solutions to produce high-level feedback that can scale with the high number of new students.
Furthermore, it is not always necessary to generate a new patch from scratch but reuse existing human-written patches for similar but different applications. Automated patch transplantation addresses this matter by extracting, transforming, and integrating patches between programs. It is not only interesting from a technical perspective but also can increase the trust of the developers \citep{TrustAPR}. 
Below we discuss these application scenarios that go beyond general repair and present specific usage scenarios and integrations.

\subsection{Repairing Security Vulnerabilities}
The sheer amount of reported and only slowly fixed security vulnerabilities motivates the application of automated program repair in the security context. Moreover, targeting security repair has its appeal because the bug search and repair generation does not need to rely on test suites compared to functional errors but can take into account general security properties regarding buffer overﬂows, integer overﬂows, null pointer dereferences, or division by zero errors.

\paragraph{General-purpose repair applied for security}
Besides the special use case and its assumptions, it is possible to apply general-purpose repair tools for repairing security vulnerabilities.
For example, \angelix~\citep{angelix} (also see Chapter~\ref{sec:angelix}), although relying on a test suite and not explicitly tailored for security repair, can be applied to repair the popular Heartbleed vulnerability\footnote{\url{http://heartbleed.com}}.
Similarly, concolic program repair~\citep{cpr} is not a security-tailored approach; it uses the available test suite to synthesize plausible patches and later refines them by test generation and a user-provided constraint (see more details in Chapter~\ref{sec:cpr}). In its evaluation, there are many successfully repaired security vulnerabilities. As long as the provided specifications (e.g., test cases and assertions) are sufficient to infer the expected behavior, the general program transformations included in general-purpose repair approaches might also successfully repair security vulnerabilities.

\paragraph{Security-tailored repair}
One of the challenges in security repair is that there is usually only one failing test, the exploit, that acts as repair input. Based on this exploit, the APR techniques must find a suitable repair specification to generate a correct patch. In general, this can be handled by \textit{generating} (i.e., searching for) additional test cases to build up a test suite that helps refine patches. As previously mentioned, techniques like \cpr~\citep{cpr} refine patches by generating more test inputs while using user-provided constraints/properties as test oracles.

Given that security properties can be defined across programs, static analyzers and fuzzers equipped with sanitizers usually search for violations of such properties without much user intervention. When finding a violation, these properties are available for further analysis. APR can use that information and create tailored bug localization and fixing strategies to make the program adhere to these properties. For example, it can analyze the violated property by applying \textit{symbolic and deductive} reasoning.
\cite{AutoPaG} repair out-of-bound vulnerabilities by limiting access to reading and writing arrays. Suppose \texttt{p} is an array and \texttt{i} is an index that would cause an out-of-bounds error \texttt{p[i]} when is executed. First, it uses data-flow analysis to identify vulnerable program statements, and by changing the read statement to \texttt{p[i mod size]} (where \texttt{size} is the inferred allocated buﬀer size), it avoids any exploit of this vulnerability.
\cite{senx} use human-specified safety properties to steer the patching. Their tool \senx takes a vulnerability-triggering input and executes it concolically to detect violations of the safety properties tailored for buffer overﬂow, bad cast, and integer overﬂow vulnerabilities. With the collected information about the execution trace and the affected variables, \senx generates a predicate defining the concrete expression necessary to prevent the safety violation and synthesize a source-level patch. The patch is enforced at the first location that is applicable. While this avoids the vulnerability, the fix is basically placed close to the crash location to disable the vulnerability, which may not address the underlying problem.
\cite{Gao21} incorporate sanitizers to extract a crash-free constraint at the violation location and use dependency analysis to locate potential fix locations (see more details in Chapter \ref{sec:extractfix}). Due to this fault localization step, they build an analytical understanding that helps define repair strategies that can go beyond disabling vulnerabilities at their crash locations. They propagate the crash-free constraint to the identified fix locations by computing the weakest pre-condition, which is used to synthesize a minimal fix.

Instead of deductive reasoning, APR techniques can also follow an \textit{inductive} reasoning approach that attempts to learn a fix pattern or infer a patch invariant.
For example, \citep{Zimin2021} apply neural transfer learning to repair security vulnerabilities. Learning-based program repair (see Chapter \ref{sec:learning}) attempts to extract general repair transformations from human-written patches. Transfer learning means learning from one domain to solve problems in a different domain. This is interesting in the context of security vulnerabilities because the available data for security repair is relatively small, while the data set for general bug fixes is significantly larger. Therefore, \cite{Zimin2021} attempt to learn a neural model on general bug fixes (source domain) and tune it on security vulnerabilities (target domain). Intuitively, it uses source domain training to understand how general code modifications and patch generation can be accomplished, while training in the target domain fine-tunes the model for the specific set of vulnerabilities. Their empirical evaluation shows that the model trained with transfer learning performs better than a model only trained on the target domain.
\begin{figure}
    \centering
    \includegraphics[width=.8\textwidth]{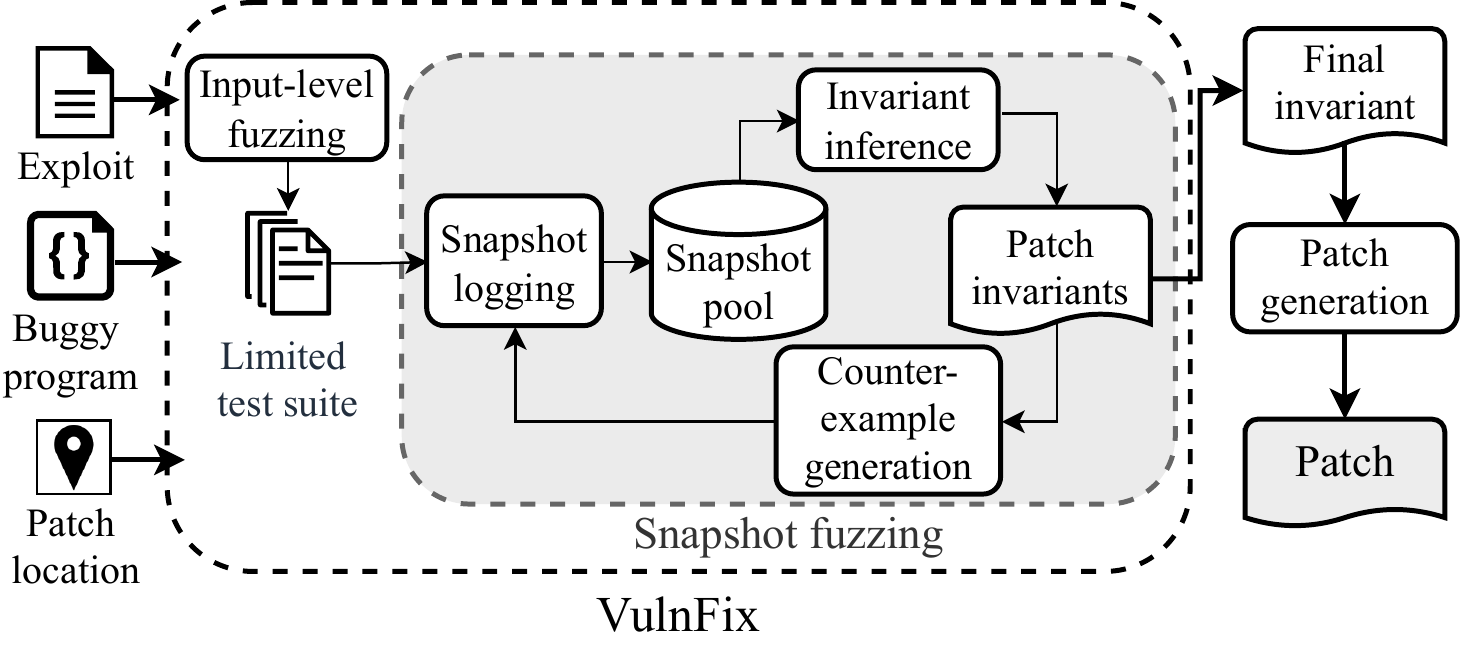}
    \caption{Workflow illustrating inductive inference for security repair~\citep{vulnfix}.}
    \label{fig:vulnfix}
\end{figure}
More recently, \cite{vulnfix} presented \vulnfix to repair security vulnerabilities using inductive inference. Figure~\ref{fig:vulnfix} shows the proposed workflow. It starts with common input-level fuzzing to build a test suite that can reach the crash and patch locations. These initial tests are expected to be diverse; some trigger the vulnerability, and some do not. With these tests, \vulnfix extracts \textit{snapshots}, i.e., the program states at the given patch location. Therefore, the initial test suite leads to a pool of initial snapshots, from which \vulnfix infers initial patch invariants by using dynamic invariant inference systems like Daikon~\citep{daikon} and cvc5~\citep{cvc5}. One of the key elements of \cite{vulnfix}'s work is their subsequent \textit{snapshot fuzzing}, which mutates the states instead of inputs to identify \textit{counterexamples} that prove the current patch invariants as insufficient and which guide their refinement. Snapshot fuzzing may lead to infeasible states. However, the resulting patch invariant is still helpful as it will hold for both feasible and infeasible states. As a general limitation of inductive inference techniques, \vulnfix may disable benign behavior because it over-approximates the necessary patch invariant.

\paragraph{Static analysis driven security repair}
Static analyzers like \infer~\citep{infer} are extensively applied in the industry to detect security vulnerabilities. In particular, \infer uses separation logic to detect violations of pointer safety properties. For example, this covers resource leaks and memory leaks potentially caused by doubly freeing memory and null dereferences. Techniques like \footpatch~\citep{footpatch} and \saver~\citep{saver} use \infer to generate repairs for the identified vulnerabilities.
Depending on the identified bug type, \footpatch uses pre-defined repair templates to generate a repair specification, which defines the error heap configuration $F$ and the fixed heap configuration $F'$. Then \footpatch searches for a patch, here called \textit{repair fragment} $C$, that allows transforming $F$ to $F'$. By filling in the specifics for the considered scenario, this becomes a Hoare triple with a \textit{hole} for $C$. Finally, \footpatch searches over the existing program fragments to identify the desired semantic change. \cite{footpatch} decided to only consider \textit{additive} program transformations based on their observation that the considered bug type (memory and resource leaks) typically are caused by \textit{missing} heap operations like the releasing of resources, freeing of memory, or null checks.
One disadvantage of \footpatch is that it might introduce new double-free or use-after-free vulnerabilities as part of their patches because it only focuses on the original error report.
\memfix~\citep{memfix} instead focuses on adding deallocation statements to repair all double-frees and use-after-frees. As a result, they guarantee not to cause new vulnerabilities with the generated patches but suffer from low scalability.
\saver~\citep{saver} improves on these works and adds support for \textit{conditional} deallocation and also \textit{relocation} of pointer dereferences. They formulate the memory error repair problem as a graph labeling problem and thereby can provide a scalable and safe repair technique.

Based on described trends in security vulnerability repair, it does not sound too futuristic to expect auto-generated security patches during a live programming session to avoid security vulnerabilities even before the corresponding program code is submitted. The following section discusses such integrations into the development workflow in more detail.

\subsection{Integration in Development Workflow}\label{sec:integration}
Integrating APR techniques into the development workflow will be the last impediment to deploying APR in practice. There are two major possibilities, the integration into \textit{development tools} like IDEs, which can give immediate support during the code writing or during unit testing, and the integration into CI workflows, where a \textit{repair bot} can test and repair committed code changes.

\subsubsection{IDE Integration}
\begin{figure}[h]
    \centering
    \includegraphics[width=0.7\textwidth]{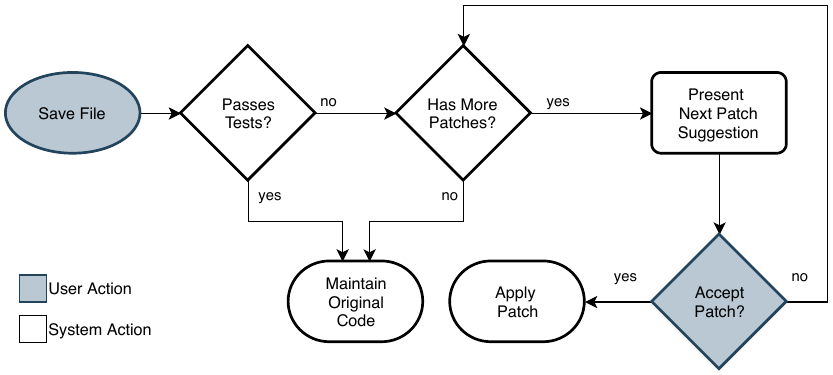}
    \caption{Flowchart proposing an IDE integration \citep{campos2021}.}
    \label{fig:paprika}
\end{figure}

The integration of APR into development tools like IDEs can come in various variants and degrees. For example, the patches can be generated offline but illustrated in the IDE. Alternatively, patches can be generated along with a unit test run and presented as suggestions. However, such interactive and direct integration would require almost \textit{real-time} patch generation and is challenging for current APR techniques.
\cite{campos2021} investigated the second variant; their idea is to provide developers with a repair technique that acts like real-time syntactic code suggestions but for semantic errors. Therefore, they implemented a Visual Studio Code extension. Figure \ref{fig:paprika} illustrates their general flowchart.
  
With saving the changed file, the defined workflow triggers unit testing, and if tests fail, the APR background will attempt to identify faulty locations and propose patches. Since the patch generation must be fast, \cite{campos2021} deploy a mutation-based approach \citep{Debroy2010}. The mutations are limited to replacing arithmetic, relational, logical, increment/decrement, and assignment operators with another operator from the same class. Additionally, they negate decisions in if and while statements. The empirical evaluation in \cite{campos2021} indicates that developers with such an IDE-integrated APR technique can reach a solution faster than developers without such support.
Having an interactive repair approach can help with filtering patches by showing them to the developers and letting them decide to accept or reject them. The interaction can also go beyond that and be more proactive by sending specific queries and questions to the developers. For example, instead of showing the generated patches, \cite{Liang2021} extract attributes related to the patch location and the affected execution trace. Then they formulate questions that can filter multiple patches based on the answers by the developers. Their user study with 30 developers revealed that questions about the affected variable values help most to distinguish and filter patches. Furthermore, their technique helps developers' understanding, can reduce the debugging time on average by up to 28\%, and can increase the success rate of finding a correct patch by up to 62.5\%. Even if no correct patch can be generated, the information about the fault locations and a partially correct patch can be helpful for the developers to study the issue themselves and find a solution. Even deeper integrated, APR can support the interactive debugging of bugs~\citep{Reiss2022}.
In addition to asking the developers about the generated patches, it also can help to query information about the expected behavior as shown in~\cite{Bohme2020} (as discussed in Chapter~\ref{sec:human-in-the-loop}). 
Overall, integrated into the development tools and debugging process, e.g., as an IDE plugin, APR can provide valuable insights to support the developer in understanding and solving the issue.

\subsubsection{Repair Bots}
Instead of direct and integrated support for the developer, APR also can be deployed with existing CI pipelines of development teams. In fact, in a recent developer survey~\citep{TrustAPR}, the integration into these toolchains was a major requirement for developers to accept the APR technology in practice.

\paragraph{Human-Competitive Repair Bots}
To be a real benefit for the developer, who is the user in this scenario, the integrated repair must fulfill two aspects: it needs to be fast, faster than other developers, and the repair must be of high quality so that a human reviewer will accept the patch. These are known challenges in the human-competitive task~\citep{Koza2010} and usually appear when designing task automation, i.e., when automated technologies are applied to solve tasks that humans usually perform. \cite{Monperrus2018} investigated APR on its readiness to tackle the human-competitive task by implementing and deploying a \textit{repair bot} on GitHub. The bot is called \textit{Repairnator} and is specialized in fixing \textit{build} errors. \cite{Monperrus2018} demonstrated that Repairnator can compete with the human developer and successfully generated patches for five reported bugs, which the human developers have merged. Although build failures are just one error type, it showed that successful APR integrations are possible and within reach.

\paragraph{Principles for Program Repair Bots}
\cite{Tonder2019} define general requirements and principles for the engineering of repair bots.
They emphasize that a repair bot needs to generate well-formed syntax for potentially many programming languages, called \textit{syntax generality}. Furthermore, generated patches can be syntactically validated by compiling the modified source code. However, this can become costly if many compiler toolchains are needed to support a multi-language setup. Instead of an analytical approach, constructive syntax validation solutions can be built into mutation operators, e.g., to ensure that general syntax attributes like parentheses are well-balanced so that no syntactically invalid code is produced. After passing the syntax check, the patches need to be validated semantically, e.g., using available test suites and static analysis checks. All patches should be post-processed with auto-formatting tools to adhere to organization-specific coding guidelines and maintain readability. The final "gatekeeper of program changes" will be humans, even for automatically generated patches. Therefore, the developer must be supported for efficiently reviewing the generated patches. Patch explanations and their easy accessibility via patch rankings~\citep{TrustAPR} will be a key aspect for future repair bot developments.

\begin{figure}
    \centering
    \includegraphics[width=0.7\textwidth]{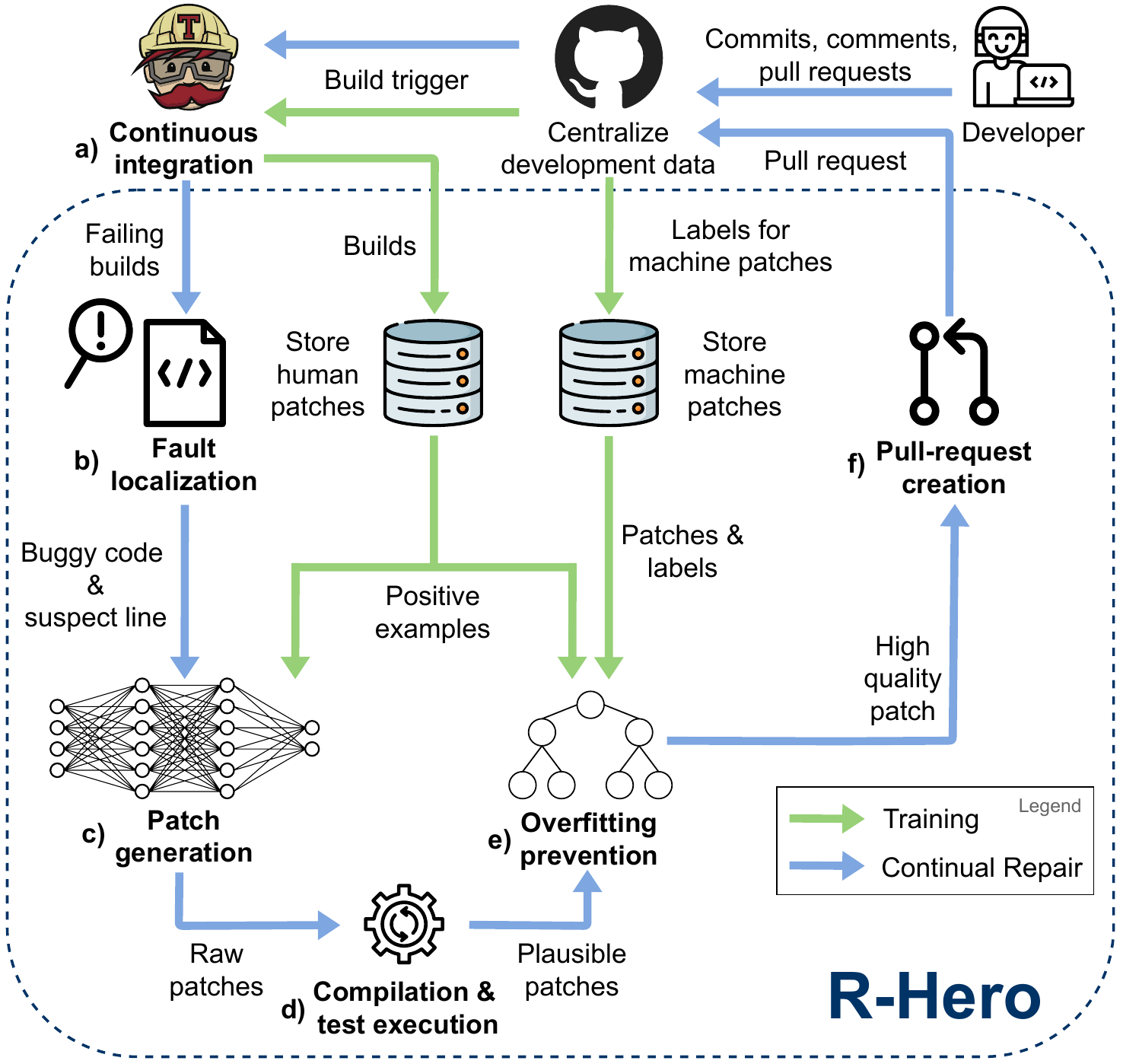}
    \caption{Program repair bot based on continual learning~\citep{Baudry2021}}
    \label{fig:rhero}
\end{figure}

\paragraph{Interactive CI Bots}
Interactive program repair that uses user inputs to guide the repair process applies not only to the IDE scenario but also in the continuous integration context. For example, \cite{Baudry2021} show how a repair bot can be gradually improved by incorporating human patches and feedback on auto-generated patches. The bot is called \textsc{R-Hero} and Figure~\ref{fig:rhero} shows its approach overview.
It maintains two databases: one for human patches collected via GitHub Actions and Travis CI and one for auto-generated patches labeled as correct or incorrect (based on the developers' feedback). Starting from a failing build observed in the continuous integration, \textsc{R-Hero} attempts to localize the error and generates a patch by using a neural network-based repair technique~\citep{SequenceR}. The collected human patches help to train and improve the neural network continually. After testing, the patches are additionally filtered using the overfitting detection system ODS~\citep{StaticPatchClassifcation}. It uses supervised learning leveraging the human patches and the labeled auto-generated patches to distinguish correct patches from overfitting patches. Finally, the generated patch is provided to the developer as a pull request.
So far, continual learning with \textsc{R-Hero} has two major limitations: it is only trained on one-line patches and hence, does not handle complex and multi-line patches. Further, the prototype implementation does not scale and needs further optimizations. Nevertheless, the concept shows what an interactive and integrated program repair approach can look like.

Automated program repair cannot only support developers in their daily work but also can be integrated into other domains, e.g., programming education. The following section discusses these opportunities and the existing advances.

\subsection{Support Programming Education}
With a rising number of CS students, the need for scalable programming education is becoming more and more important. Massive Open Online Courses (MOOCs)~\citep{mooc} will shape the near future of computer science teaching and students will require feedback for their programming assignments. Hiring more and more tutors is not scalable, so that eventually the usual tutor-student interaction will be reduced or completely missing. Providing a personalized feedback in such an impersonal environment is one of the major challenges.

\paragraph{Goal and Ingredients}
APR's goal is to generate patches for incorrect software, so why not run APR on students' solutions for programming tasks? The ingredients for APR are available: a programming task usually comes with a \textit{problem statement}, for which the student submits a \textit{solution} as source code, the lecturer/tutor has a private \textit{reference implementation}, which correctly implements the required functionality, and furthermore, there usually are \textit{test cases} to validate the submitted solution. Additionally, there could be pre and post conditions, specific performance requirements, a specific error model provided by the tutor, and historical student submissions that can be labeled as correct and incorrect solutions. 
While students can be supported by just providing information on failing test cases, it would be way more beneficial to show students how to correct their own code. APR generates patches, i.e., instructions on how to change incorrect software to correct software, and hence, provides a well-fitted basis to support students in their learning endeavour.

\paragraph{Solutions}
\cite{autograder} presented initial ideas on how to construct some auto-grading and auto-feedback mechanisms. They assume to have the reference implementation and some error model. The error model is expressed in their own language \textsc{Eml} and contains a set of rewriting rules for typical student mistakes and their possible corrections. The incorrect student's solution is rewritten with these rules, which leads to multiple patch candidates annotated with the necessary correction costs. By symbolically exploring the patch space and using counter-example guided synthesis (CEGIS)~\citep{sketching}, it identifies the solution with the minimal cost. The messages associated with the error model's rewriting rules and the affected source line numbers are used to generate natural language feedback for the student.

For instance, Figure~\ref{exp:education} shows an example of a student's buggy program and corresponding feedback generated by~\cite{autograder}.
This task requires students to write a program computing the derivative of a polynomial, whose coefficients are represented as a list of integers.
For example, if the input list \texttt{poly} is [2, -3, 1, 4] (denoting $f(x)$ = 4$x^3$ + $x^2$ - 3$x$ + 2), the \texttt{computeDeriv} function should return [-3, 2, 12] (denoting the derivative f'($x$) = 12$x^2$ + 2$x$ - 3).
In this program, student made several mistakes: 1) the return value at line 5 is wrong, which should be \texttt{return[0]}; 2) the range at line 6 should be \texttt{range[1, len(poly)]}, 3) the comparison in line 7 should check not-equation.
To help student fix this bug, the right part of Figure~\ref{exp:education} presents the auto-generated feedback for students.

\begin{figure}[!t]
\begin{subfigure}[c]{0.525\textwidth}
\begin{lstlisting}[basicstyle=\footnotesize,upquote=true,escapechar=!]
def computeDeriv(poly):
    deriv = []
    zero = 0
    if (len(poly) == 1):
        return deriv
    for expo in range (0, len(poly)):
        if (poly[expo] == 0):
            zero += 1
        else:
            deriv.append(poly[expo]*expo)
    return deriv
\end{lstlisting}
\vspace{-6pt}
\end{subfigure}
~
\begin{subfigure}[c]{0.40\textwidth}
\small
\textbf{Generated Feedback:}
\begin{itemize}
    \item In the return statement return deriv in line 5, replace deriv by [0].
    \item In the comparison expression (poly[expo] == 0) in line 7, change (poly[expo] == 0) to False.
    \item In the expression range(0, len(poly)) in line 6, increment 0 by 1.
\end{itemize}
\end{subfigure}
\caption{A student's program and corresponding auto-generated feedback (adapted from~\cite{autograder})}
\label{exp:education}
\vspace*{-15pt}
\end{figure}

Going forward, \cite{Clara} applies a \textit{wisdom of the crowd} approach and builds its repair technique based on collected, correct student solutions. First, it clusters the existing solutions based on their structural and variable mapping. Then, it matches the incorrect student submission to find the best fitting cluster, which is subsequently used to repair the incorrect solution. The repair starts with generating all possible \textit{local} repairs that replace expressions from the incorrect solution with their counterparts in the collected cluster. Each local repair gets assigned a cost value based on the tree edit distance between the abstract syntax trees of the involved expressions. Note that there can be many local repairs, and the goal is to find a \textit{consistent} subset of repairs that eventually lead to a correct patch with minimal total cost. Therefore, \cite{Clara} formulate a 0-1 ILP problem that can be solved with off-the-shelf ILP solvers. After selecting the subset of local repairs, it generates a textual description of the necessary repair steps and presents it to the students.
A user study with students from introductory programming courses showed that feedback could be quickly generated (i.e., within 60 seconds) for most of the assignments and indicated that the generated feedback is indeed helpful for the students.

\paragraph{Related Quality Attributes}
Beyond these existing techniques, \cite{Sarfgen} define various quality attributes crucial for proper student grading and feedback: Ideally, such a system is \textit{fully automated} so that it does not require manual efforts like the provision of patch templates or typical errors. It should provide a \textit{minimal} patch to illustrate the minimal effort needed to repair the student's submission so that the student can focus on the problematic part. The repair operators should allow \textit{simple} and \textit{complex} repair modifications. Furthermore, the system should be \textit{portable} to other programming languages to enable universal support of programming exercises. \cite{Sarfgen} propose a high-level data-driven framework that follows three steps: \textit{search} for closely-related correct solutions in the pool of available submissions that can be the basis for repairing an incorrect solution, \textit{align} the identified programs to extract the necessary repair transformations, and \textit{repair} the incorrect submission by finding a minimal set of transformations. They integrated their system with the Microsoft-DEV204.1x course, an online programming course for C\#, and deployed it in production. The collected feedback suggests its practicality and usefulness.

The overfitting issue in APR (see Chapter \ref{sec:overfitting}) also applies to APR in programming education. When techniques rely on high-quality test suites, they may produce incorrect feedback because large test suites are unavailable or do not have the required quality. While expert programmers may still use partial fixes or repair artifacts~\citep{Liang2021}, this is not expected from students~\citep{Yi17}. Instead, students and novice programmers highly rely on the \textit{correctness} of the auto-generated feedback. \cite{
verifix} is tackling this issue by providing \textit{verified} repairs. They attempt to verify the equivalence between the reference solution and the student's submission, and with each failed attempt, they retrieve a counterexample that guides the repair generation. Therefore, they enforce the semantical equivalence between patch candidate and the reference solution; hence, the finally generated repair can be trusted by the students. Furthermore, the whole repair process operates in a reasonable timescale of less than 30 seconds, which shows potential for a live feedback mechanism.

\paragraph{What is good feedback?}
While the research and tool development continues, e.g., \cite{Lu2021} proposed a lightweight technique that outperforms previous work in generation speed and number of repaired programs, one main issue remains unsolved, which is how to provide \textit{good} student feedback. What makes good student feedback anyway? Certainly, it is not just reporting failing test cases, and certainly, it is also not the provision of a fixed solution. While test cases are helpful, they are not enough to find the correct answer. While the fixed solution can help understand the problem, it robs the students from solving it themselves. The key aspect is to guide the students to the correct answer without revealing the solution upfront. It starts by explaining why something is wrong in their submission, which can be supported by a failing test case and a potential fault location. Then, compiler error messages can be explained, and solution hints can be provided. Step by step, more information can be revealed. Questions instead of direct information can guide the student to identify the problem themselves. Eventually, it is not just the technical aspect of how to quickly generate accurate repairs, but the pedagogical aspects of the students' learning experiences that are crucial for deploying intelligent tutoring techniques.

\paragraph{Effect of Introducing Automated Feedback and Grading}
While the technical contributions had been partially evaluated with user studies as well, the evaluation results usually do not report more than an indication that the generated feedback is helpful. In contrast, \cite{Hagerer2021} performed a survey study and examined the effect of introducing automated grading functionality. Their framework \textsc{Artemis}~\citep{artemis} does not apply automated program repair; however, it provides feedback on failing test cases and involves static code analysis to measure the code quality. The study shows that programming courses that use \textsc{Artemis} receive higher ratings, the students are more satisfied with the tutors, and the students' perceived programming learning experience improved. Possible reasons for these trends are the automated feedback and the reduced correction biases by the automated grading. Moreover, by introducing the automated platform, tutors have more time to focus on their student's needs.

\subsection{Patch Transplantation}
\begin{figure}
    \centering
    \includegraphics[width=0.8\linewidth]{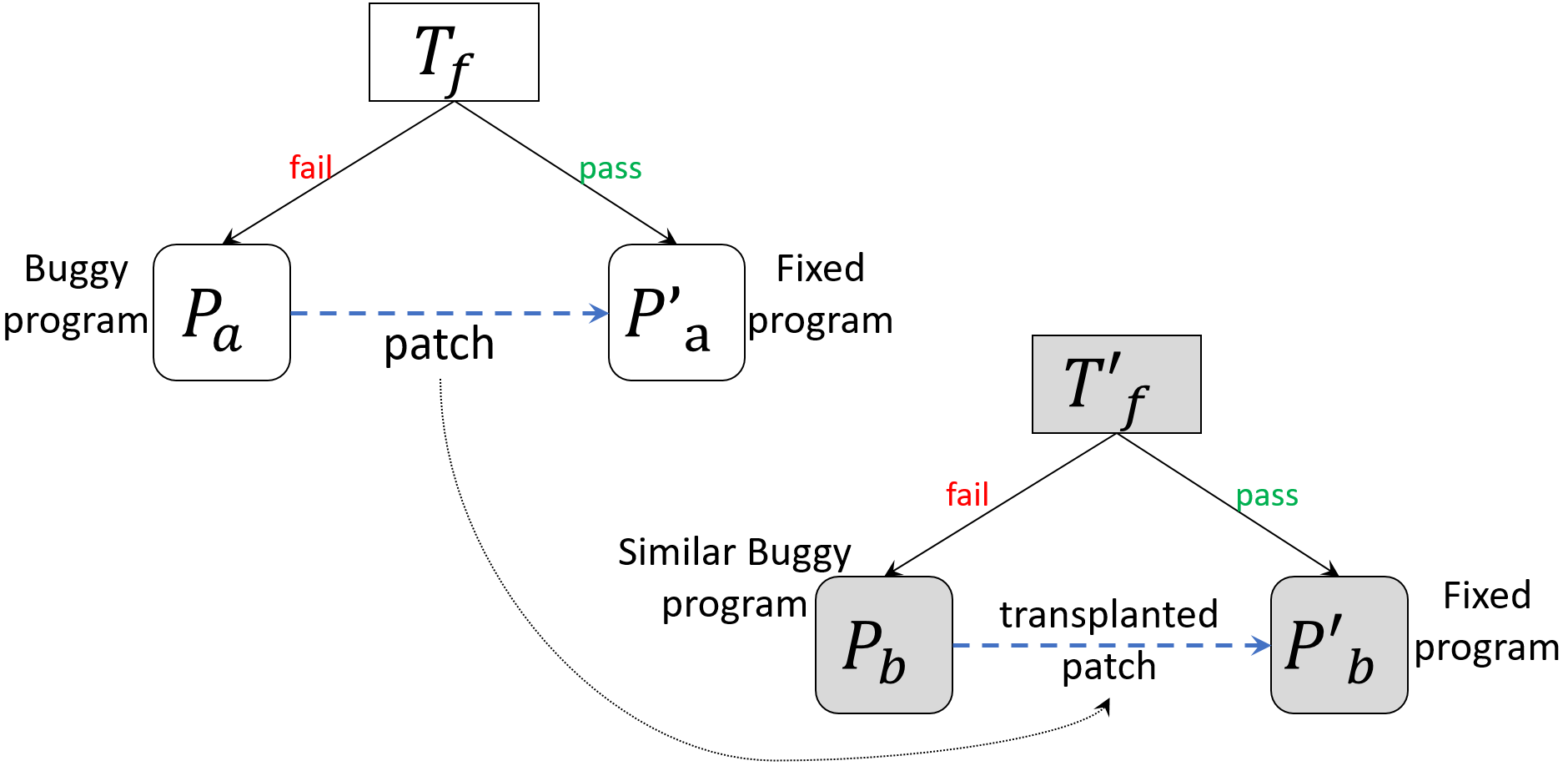}
    \caption{The automated patch transplantation problem (adapted from \cite{patchtransplant})}
    \label{fig:patchtransplant}
\end{figure}

A related but different problem to program repair is \textit{automated patch transplantation}. The key motivation is that if there is already a patch, but for a different, albeit similar, application, one can attempt to \textit{transplant} this patch to the current buggy program instead of generating a patch from scratch.
As an early work in this area, \cite{codephage} developed a code transfer system to transplant missing checks. Their transplantation process assumes two program inputs, one that triggers the error and one which does not. Then they use a database of applications to search for a donor that passes both inputs and includes a check necessary to reject the error-triggering input. Next, by using symbolic execution, they extract the corresponding input constraint and translate it into the namespace of the target application. Subsequently, they identify potential insertion points in the target application, at which the necessary input fields are available as program expressions.
\cite{patchtransplant} systematically explore the general patch transplantation problem and define it formally as follows. As illustrated at Figure~\ref{fig:patchtransplant}, given a so-called donor program with its buggy version $P_a$ and its fixed version $P'_a$, the goal is to modify the buggy host or target program $P_b$ to produce a fixed version $P'_b$. The two programs $P_a$ and $P_b$ are assumed to be ``\textit{similar}''. $P'_b$ is created by transplanting the patch $P_a \rightarrow P'_a$ to $P_b$., i.e., extract, adapt, and integrate the patch.

For instance, Figure \ref{fig:transplantation} presents patch transplantation between LibGD and Libtiff, two image processing libraries.
A bug in Libtiff causes an overflow, since the maximum LZW bits (represented using variable \texttt{datasize} and \texttt{c} in Figure~\ref{fig:transplantation_b} and \ref{fig:transplantation_a}, respectively) allowed in GIF standard is 12, otherwise a bugger overflow will be triggered.
This can be fixed by inserting a check as shown in Figure~\ref{fig:transplantation_b}.
This vulnerability also exists in LibGD and ImageMagick libraries which are also image
processing software similar to Libtiff. 
All three programs are vulnerable to the same exploit because they follow the same standard for GIF image processing. 
Therefore, the patch from Libtiff can be transplanted to LibGD via patch extraction, adaptation, and integration.
The adaptation required for the patch is the namespace mapping from \texttt{datasize} in Libtiff to \texttt{c} in LibGD.

\begin{figure}[!t]
\begin{subfigure}[b]{0.48\textwidth}
\begin{lstlisting}[basicstyle=\footnotesize,upquote=true,escapechar=!]
int readraster ( void ) {
    ...
    datasize = getc(infile);
+   if (datasize>12)
+       return 0 ;
    clear = 1 << datasize;
    eoi = clear + 1;
    ...
}

\end{lstlisting}
\vspace{-6pt}
\caption{Developer patch for Libtiff 4.0.4}
\label{fig:transplantation_b}
\end{subfigure}
~
\begin{subfigure}[b]{0.48\textwidth}
\begin{lstlisting}[basicstyle=\footnotesize,upquote=true,escapechar=!]
static void ReadImage (...) {
    ...
    if (ReadNotOK(fd, \&c, 1)) {
        return;
    }
+   if ( c > 12)
+   return;
    ...
}
\end{lstlisting}
\vspace{-6pt}
\caption{Developer patch for LibGD 2.0.34 RC1}
\label{fig:transplantation_a}
\end{subfigure}
\caption{Patch transplantation example between two image processing libraries}
\label{fig:transplantation}
\end{figure}

\paragraph{Transplantation Patch Classes}
\citet{patchtransplant} define a hierarchy of four patch classes; with increasing levels, it becomes more difficult to transplant the patch successfully.
\textit{Class I: Syntactically Equivalent Transplantation} represents the case where no adaption is needed because the same syntactical patch can be copied from the donor program to the target program. Therefore, class I is the simplest scenario of patch transplantation.
\textit{Class II: Syntactically Equivalent Transplantation with Dependency} means a missing dependency like a function that is not present in the target program. The transplantation process needs to detect and include it in the patch.
\textit{Class III: Semantically Equivalent Transplantation} represents the case where it requires the adoption of the patch to enable the transplantation to the target system because of syntactic differences between both programs. For example, it could be necessary to perform a namespace translation.
\textit{Class IV: Semantically Equivalent Transplantation with Dependency} means the need for adoption and the detection of missing dependencies.

\paragraph{Transplantation Process}
The transplantation process starts with \textit{patch extraction}, which analyzes the source diff between $P_a$ and $P'_a$ along with the execution trace. Next step is to identify the appropriate \textit{insertion point} $P_b$ and adopt the patch accordingly to the context in $P_b$. To identify the best insertion point, one can compare the dynamic execution traces between all programs $P_a$, $P'_a$, and $P_b$, and check which divergence point in $P_b$ matches best the divergence between $P_a$ and $P'_a$. Concolic execution supports this task by extracting the symbolic path constraints, which helps identify the program part that handles the same input partition as the donor program. Furthermore, it collects the symbolic values for the variables relevant to the patch. With this information, the various variables in the three programs can be mapped to each other. It allows us to compare the variable usages and hence, to identify the \textit{candidate function} to transplant the patch between the divergence point and the crash location. After identifying the insertion point, the patch needs to be translated to the namespace of the target program $P_b$. Additionally, one needs to check the context and dependencies of the patching code, e.g., to identify functions that are called in the patch but are not present in the target program. As in general program repair, the program modification must be followed by validation. For example, the generated $P'_b$ can first be syntactically checked, and if successful, a differential fuzzing campaign can search for differences between the $P_b$ and $P'_b$.

\paragraph{Transplantation Use Cases}
Two prominent practical use cases are concerned with patch transplantation: (1) the propagation of patches from a different implementation, e.g., the same protocol or functionality, and (2) the \textit{backporting} of patches from a current version to an older software version.
The second use case is studied by \cite{fixmorph} in the context of security patches in the Linux Kernel. From 2011 to 2019, 8\% of all patches in the Linux Kernel have been backported to an older version, while for 50\% of these patches, it took the developers more than 46 days until the patch made it to the older version. Therefore, there is a need for automated techniques to support the developers in this process. \cite{fixmorph} synthesize the necessary transformation rules by using a domain-specific language (DSL) tailored to define transformation rules in the context of patch backporting.
Other related use cases are the transplantation of \textit{features}~\citep{featuretransplant} and \textit{test cases}~\citep{testtransplant}, which both may require the application of automated program repair on top of them.

%% file: chapters/8_future.tex
\section{Perspectives}\label{sec:perspectives}
After illustrating the state of the art and its achievements in the previous chapters, we end this article by gauging different future-looking perspectives on automated program repair. 
We show the need for repair technology deployments that fit into existing workflows like the continuous integration system. To this end, we present the first results of a field study on developer interest in repair \citep{TrustAPR}. This may provide some perspectives on how the future works on automated program repair can shape up.
Next, we provide a perspective on the recent growth of Language models such as GPT-3 as embodied in the GitHub Copilot. Such language models support the automated generation of code, and it would be worth studying how automated repair can work in tandem with such language models in the future \citep{arxivMay22}.
Finally, we provide perspectives on how future work can explore the synergies between testing and repair to provide a unified quality assurance component in the development workflow.

We conclude the article by revisiting the challenges in program repair including those which have been partially addressed by current research. We provide a forward looking outlook of the field at the end by referring to recent trends in automated coding.

\subsection{Human Study with 100+ Software Developers}
In Chapter~\ref{sec:applications} we illustrated many applications and possible (and already achieved) integrations of automated program repair. However, the eventual success of these applications and integrations in the software development practices depends mainly on the \textit{acceptance} of APR by the software developers. Therefore, to further explore the general developers' interest in automated program repair as their daily development companion and to gauge the possibilities of more intensified field studies, \cite{TrustAPR} conducted a survey with 100+ software developers to identify requirements and trust-related aspects for APR.
The survey focused on three core topics: (1) the general acceptability of APR and how developers envision interacting with APR techniques, (2) the availability of additional specifications or, more general, input artifacts that can support the APR process, as well as the expected impact on the trust of the auto-generated patches given that the additional specifications are taken into account, and (3) the expected evidence and explanations for generated patches.

\begin{figure}
\begin{subfigure}{.49\textwidth}
  \centering
  \includegraphics[width=\linewidth]{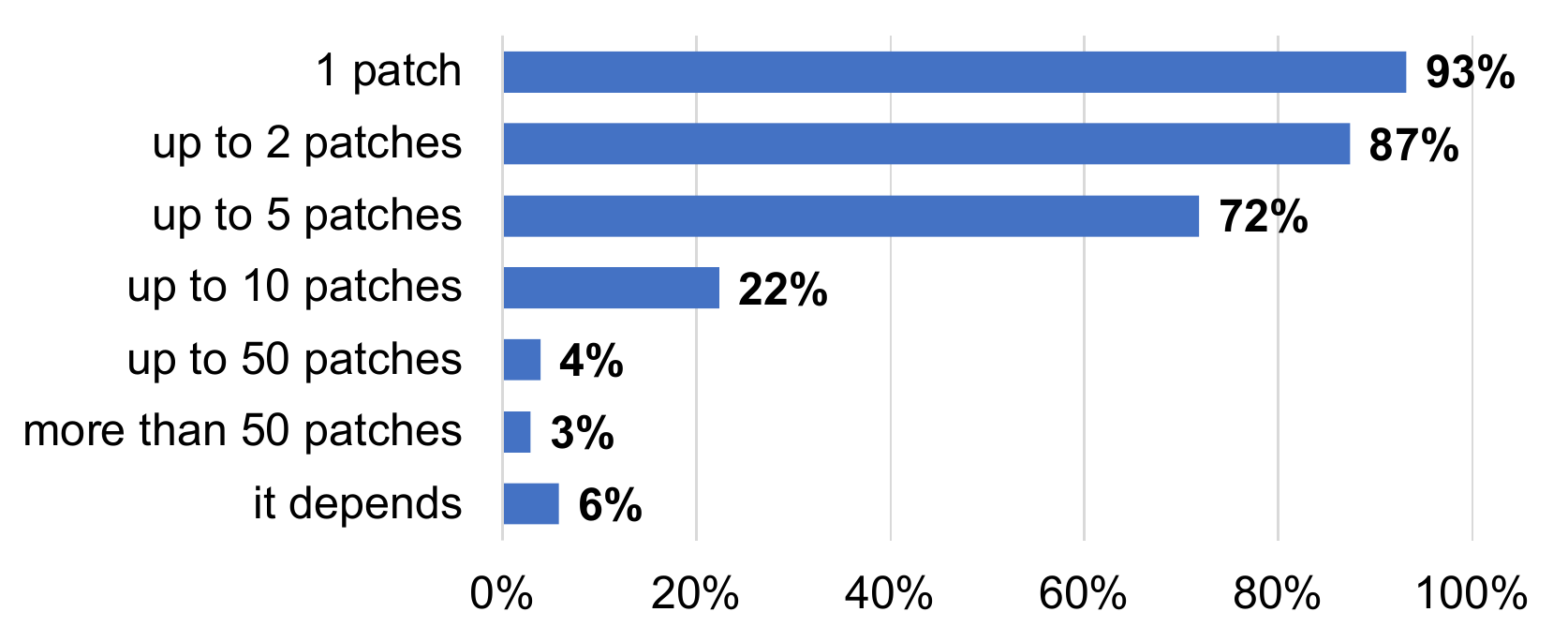}
  \caption{\textit{How many auto-generated patches would you be willing to review before losing trust/interest in the technique?}}
  \label{fig:numberOfPatches}
\end{subfigure}
\hfill
\begin{subfigure}{.49\textwidth}
  \centering
  \includegraphics[width=.9\linewidth]{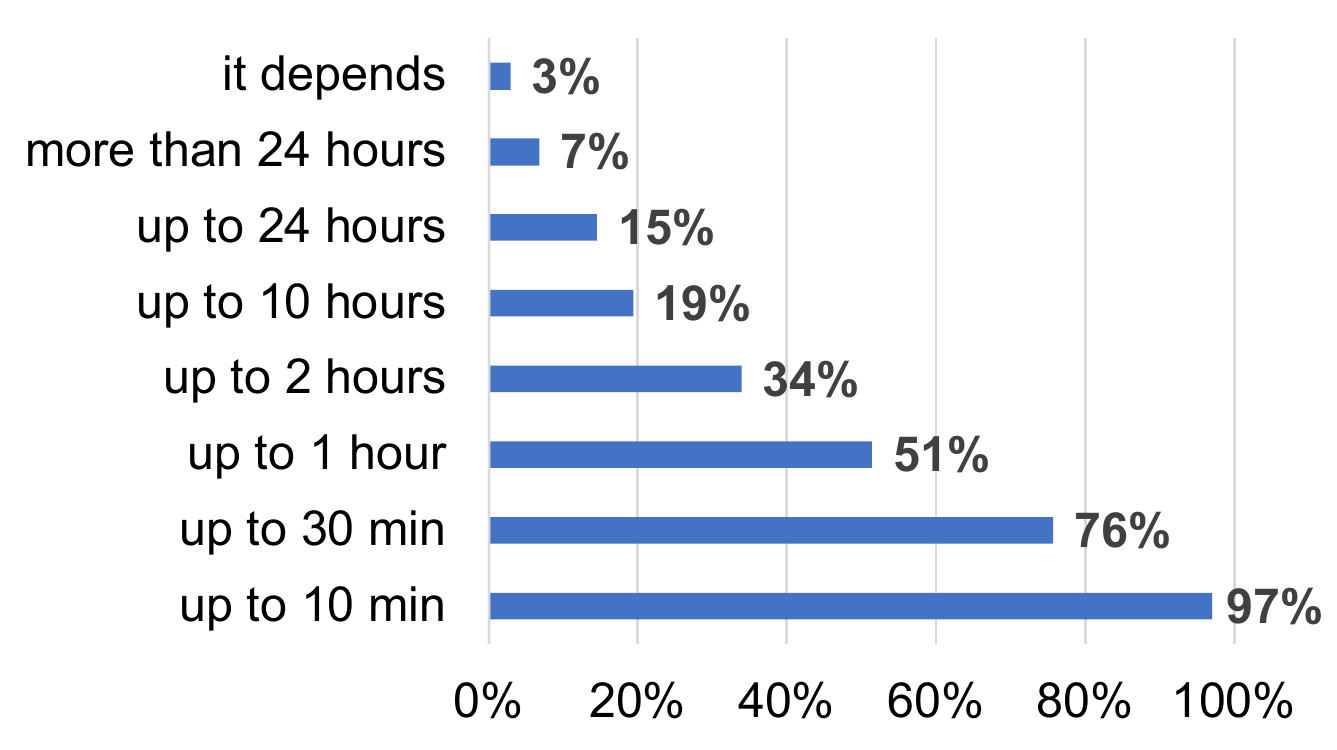}
  \caption{\textit{What is an acceptable timeout for APR?}}
  \label{fig:timeout}
\end{subfigure}
\hfill
\begin{subfigure}{.49\textwidth}
  \centering
  \includegraphics[width=\linewidth]{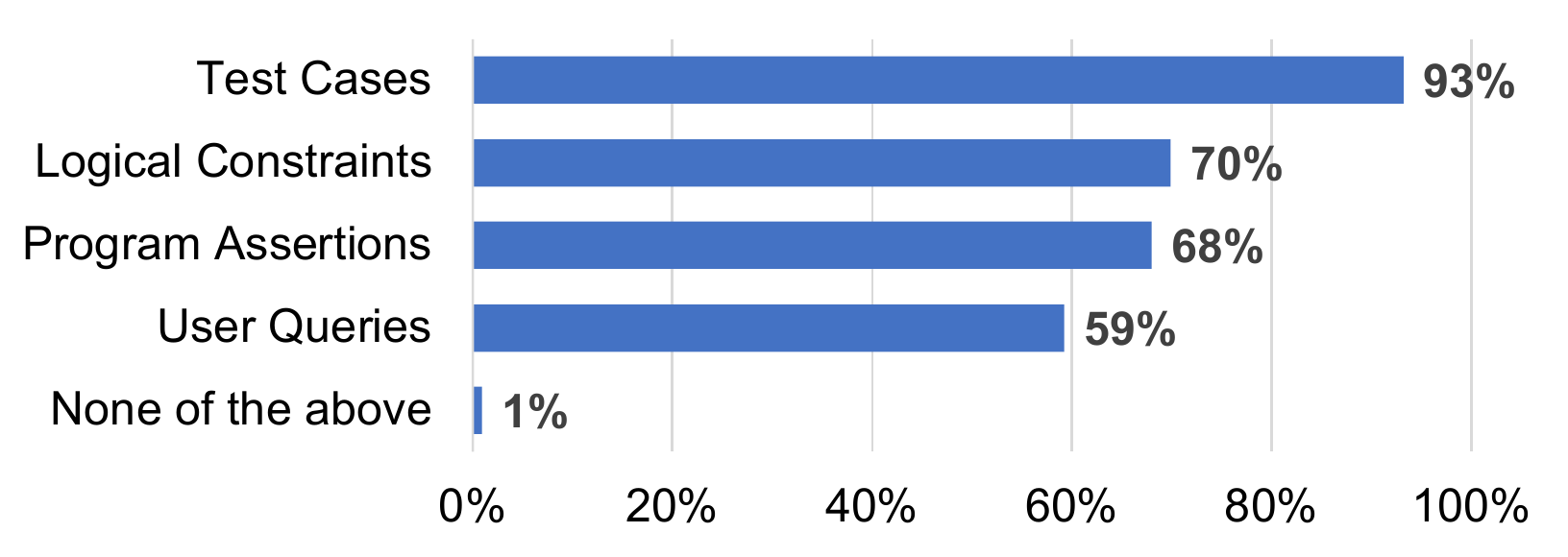}
  \caption{\textit{Which of the following additional artifacts will increase your trust?}}
  \label{fig:trustImpact}
\end{subfigure}
\hfill
\begin{subfigure}{.49\textwidth}
  \centering
  \includegraphics[width=\linewidth]{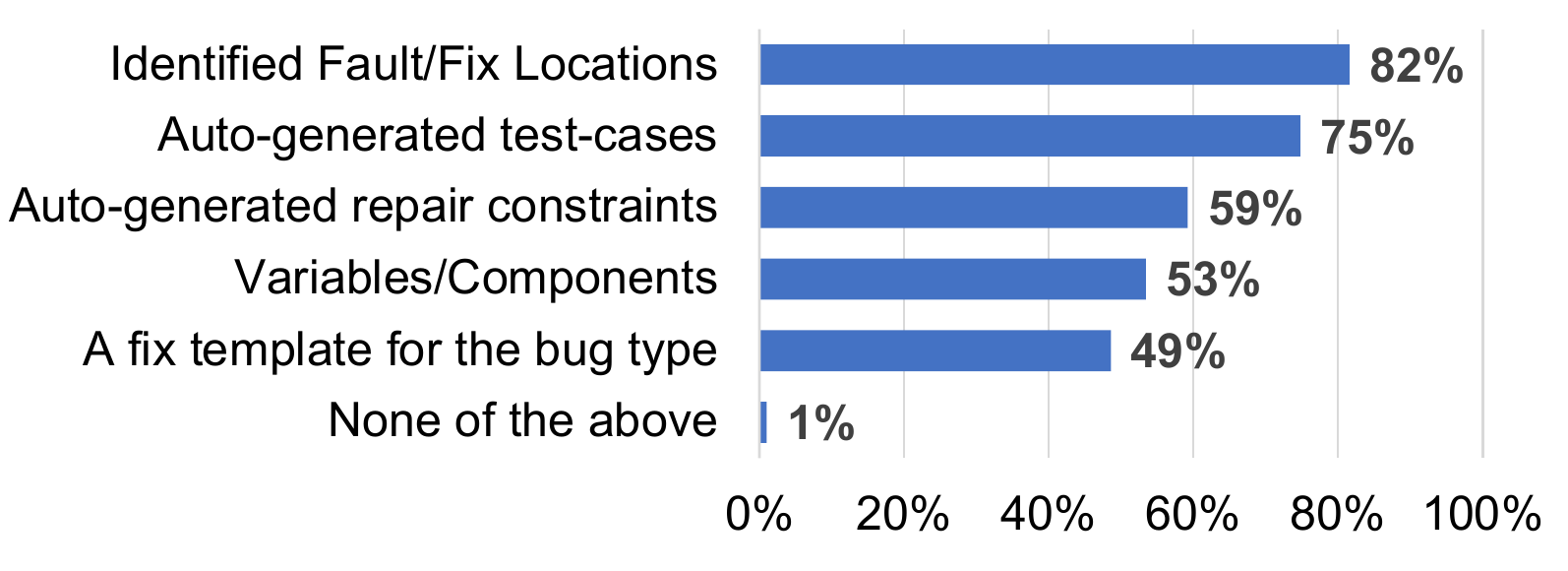}
  \caption{\textit{Which of the following information (i.e., potential side-products of APR) would help you to fix the problem yourself (without using generated patches)?}}
  \label{fig:sideproducts}
\end{subfigure}
\caption{Selected survey responses regarding developers' expectations and trust aspects in APR~\citep{TrustAPR}.}
\label{fig:test}
\end{figure}

\paragraph{Acceptance of APR and the envisioned interaction}
As an encouraging first insight, 72\% of the surveyed developers are willing to review auto-generated patches, which shows that developers are generally open to using APR techniques in their development workflow. However, they would not completely trust patches per default and require a manual review. As evident from the numbers in Figures \ref{fig:numberOfPatches} and \ref{fig:timeout}, APR has only a small margin for maneuver: most developers would only review up to five patches and would expect them within a 1-hour timeout.
Additionally, developers envision a small amount of interaction, e.g., only to provide initial search ingredients. Furthermore, developers expect that APR is fully integrated into existing development workflows like DevOps pipelines. Therefore, the existing works in APR integration (see Chapter~\ref{sec:integration}) are essential steps in achieving the developer's acceptance in practice.

\paragraph{Artifacts and their impact on trust}
The survey responses showed that software developers can provide additional artifacts that can support the APR process: test cases, program assertions, logical constraints, execution logs, and potential fault locations. Moreover, using these user-provided artifacts can positively impact the trustworthiness of the generated patches. Figure~\ref{fig:trustImpact} shows that the developers have the impression that, in particular, test cases bear the potential to establish trust with APR. On the contrary, developers see less potential in repeated user queries as it would be necessary for an intensive human-in-the-loop repair approach (see Section~\ref{sec:human-in-the-loop}).

\paragraph{Patch Evidence and APR side-products}
For software developers, it is essential to see evidence and explanations for the generated patches. Presenting such evidence for the correctness of the patch and explaining the targeted fault and the repair itself can enable the developers to select the best patch candidate efficiently. Additional information, e.g., the code coverage and the ratio of the input space covered by the patch validation, can act as supporting material for these decisions. Even when APR tools cannot generate patches, the APR side-products can help the developers. Figure~\ref{fig:trustImpact} shows that developers find it helpful to receive information about the identified fault locations, the generated test cases, or the inferred repair constraints to patch the bug themselves.

\paragraph{How to get closer to trust?}
The study results show that developers in practice are willing to include APR in their daily work but are not (yet) ready to trust APR completely. The preliminary evaluation by \cite{TrustAPR} shows that the collected requirements by the developers (i.e., the 1-hour timeout and the 5 to 10 patches to review) are tight constraints and are hardly met by the current APR solutions. Therefore, the APR community needs to keep improving APR's general capabilities to generate high-quality patches. Furthermore, we also need to incorporate aspects into APR that enable developers to efficiently review the patches so that APR can effectively be integrated into current development workflows. This way, software development will evolve by using automated repair technology.
To enable the efficient patch review by the developers, an APR technique needs to support three key aspects:
(1) APR needs to give insights into why the patch targets the right issue, e.g., by showing the fault/fix localization results and the inferred repair constraints.
(2) APR needs to show evidence for the correctness of the patch, e.g., additional test cases, test suite coverage information, or input coverage information.
(3) APR needs to provide easy accessibility of the generated patches, e.g., appropriate ranking capabilities and the efficient navigation of patch candidates in the programming environment.
Existing APR side products can already support some of these items, like fault locations and inferred repair constraints. However, the study also urges more research for patch explanations, patch ranking, and the efficient traversal of a possibly abstract patch space.

Beyond the discussed survey, we will require more user studies to identify which workflows can work and be accepted in practice. In particular, the interaction between repair tools and software developers is an area, which will need more focus in the future.

\subsection{Language Model based Code Generation}
Designing AI-based systems to automatically solve programming tasks has gained considerable attention in recent years.
The most notable of these comes in the form of transformer-based large-scale language models, which used to transform natural language text.
Large language models, such as Codex~\citep{codex} and AlphaCode~\citep{alphacode}, have also successfully generated code for many programming tasks in Python, Java, C, etc.

\paragraph{Program Repair for fixing Code Generated by Language Model}
Codex and AlphaCode have shown capability in generating correct solutions for many programming tasks.
However, the success rate of existing language models remains low, especially for complex programming tasks.
One of the reasons is that language models lack awareness of program semantics (e.g., type information, run-time program states, etc.), resulting in incorrect programs.
A large part of bugs made by Codex are syntax errors or misaligned algorithms, i.e., uncompiled programs or programs with incorrect algorithms.
Meanwhile, some bugs require small changes, e.g., changing operators, modifying expressions, or changing statements.
For instance, Figure~\ref{figure:codex} shows an example program produced by Codex for an programming task in LeetCode~\footnote{https://leetcode.com}.
The comments in Figure~\ref{figure:codex} are the program descriptions, which are provided to Codex as prompt, and the code is automatically generated by Codex.
Unfortunately, the produced program has a bug, causing the program to fail on some test cases.
The correct fix is to change statement $i {-}{=} 2$ at line 7 to $i {-}{=} 1$.
Compared to language models, typical repair
tools generate patches by reasoning about the program semantics against the given specification.
Hence, the repair technique has the potential to increase the success rate of language models.
In the above example, Recoder~\citep{recoder}, a recent program repair tool, could automatically fix the bug and make it pass all the test cases.

\begin{figure}[!t]
\centering
\begin{lstlisting}[style = mystyle, escapechar=@]
//A fancy string is a string where no three consecutive characters are equal. Given a string s, delete the minimum possible number of characters from s to make it fancy. Return the final string after the deletion. It can be shown that the answer will always be unique.
public String makeFancyString(String s) {
  StringBuilder sb = new StringBuilder(s);
  for (int i = 2; i < sb.length(); i++) {
    if (sb.charAt(i) == sb.charAt(i-1) && sb.charAt(i) == sb.charAt(i-2)) {
      sb.deleteCharAt(i);
@{\color{red}-\ \ \ \ \ \  i -= 2;}@
@{\color{green}+\ \ \ \ \ \  i -= 1;}@
    }
  }
  return sb.toString();
}
\end{lstlisting}
\vspace{-6pt}
\caption{The program for a LeetCode programming task generated by Codex.}
\label{figure:codex}
\vspace{-6pt}
\end{figure}

\paragraph{Language Model for Program Repair}
Language models could also be used for fixing software bugs.
In March 2022, a new version of Codex edit mode was released.
Instead of just translating program descriptions to programs~\footnote{https://openai.com/blog/gpt-3-edit-insert}, the Codex edit model can change existing code in a complete program.
This new feature makes it practical to use Codex for program repair.
Codex edit mode requires users to provide instructions to guide the code change, such as ``fix the bug at line 2'', or ``fix the index-out-of-bound exception''.
To fix a bug, users need to provide precise and clear instructions.

The repair based on large language models could even produce better performance in fixing software bugs than learning based repair techniques.
Compared to existing learning-based repair, e.g., SequenceR and Recoder, Codex is trained on a much larger dataset than Recoder, which helps Codex to learn more fix patterns (see \cite{arxivMay22} for comparison results).
In fact, large language models learn code edit patterns from huge existing programming artifacts (including code, commits, comments and etc.).
For instance, the existing search-based approaches
like GenProg and TBar, may not be able to fix bugs that require either (1) additional fix patterns, or (2) a large search space for fix ingredients (e.g., specific literal). 
This limitation shows that a pattern-based APR tool is hard to scale. 
Instead of manually adding more patterns to a new APR tool, future APR research on designing fixing operators should shift to a more scalable way (automatically learn fix patterns from huge programming artifacts). 

We thus postulate that language model based repair approaches could play a significant role in future, in achieving the capabilities of search-based, pattern-based and learning-based repair techniques. At the same time, the relationship of the language model based approach with respect to program synthesis is not well-understood today (see \citep{Jain22} for an initial work). Since semantic repair approaches (or constraint based repair approaches) rely on a program synthesis back-end, there exist opportunities in combining semantic repair approaches with language model based repair in the future.

\subsection{Synergies of Testing and Repair}
To conclude our perspectives, we want to allude to the integration of repair into the development workflow by merging testing and repair. In Section~\ref{sec:cpr}, we already discussed the concept of Concolic Program Repair (CPR)~\citep{cpr}, which proposes the co-exploration of input and patch space to prune overfitting patches, and hence, to establish a notion of gradual improvement. The \textit{simultaneous exploration} of test \textit{inputs} and \textit{patches} hold the potential to amplify the overall search. Over time, additional tests are systematically generated, and patch suggestions are systematically refined. Developing the ability to refine the patches gradually, it will become crucial to have an efficient representation of patches and sets of patches. Eventually, such a co-exploration repair system aims at converging to a correct patch. Throughout the search process, we can start \textit{engaging} with the developer. Thus, the developer can guide the search, and the generated artifacts provide insights to increase patch comprehension.

\begin{figure}[h]
    \centering
    \includegraphics[width=0.6\textwidth]{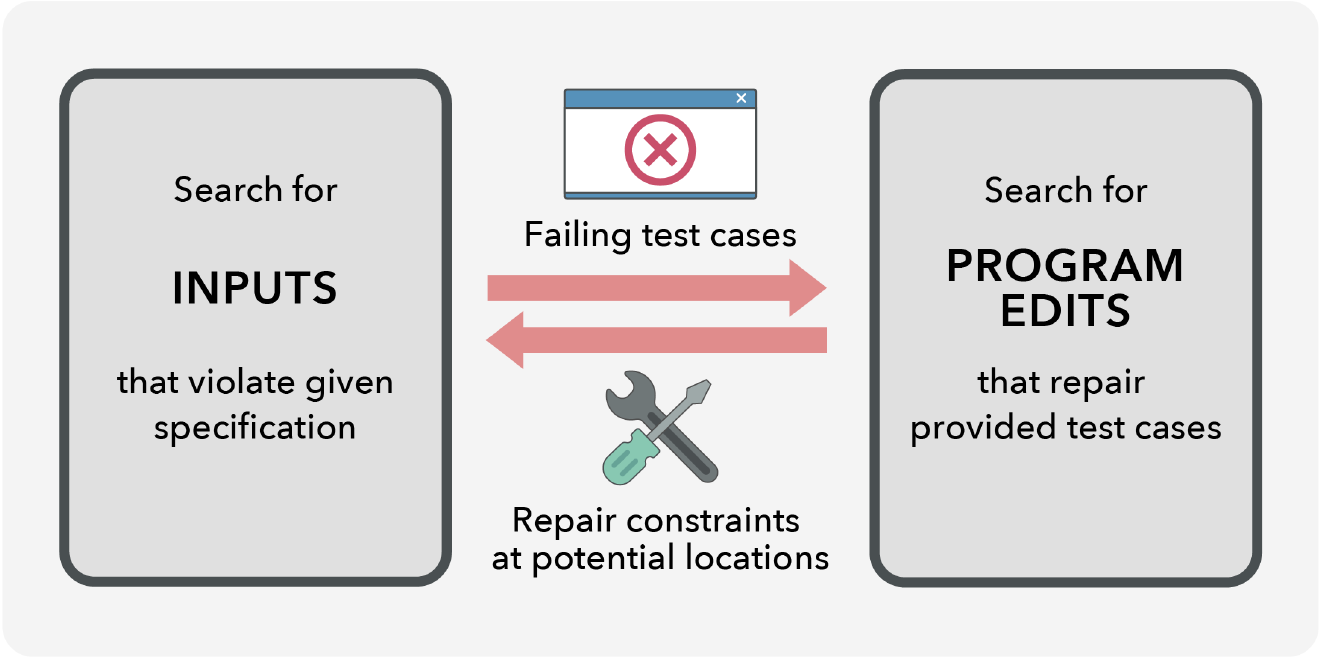}
    \caption{Fuzzing based co-exploration input space and program edits.}
    \label{fig:coexploration}
\end{figure}

CPR leverages symbolic/concolic execution to generate tests automatically. However, using symbolic execution requires working with a symbolic execution engine which is non-trivial. Hence, one can alternatively employ a (systematic) fuzzing technique such as~\cite{Boehme17}. Thus, finding inputs exposing vulnerabilities can be viewed as a biased random search over the space of program inputs. Similarly, the process of automated program repair has also been cast as a biased random search over the space of program edits. As a result, these searches can strengthen each other instead of running as separate processes. As illustrated in Figure~\ref{fig:coexploration}, the search for inputs can provide failing test cases that drive the search for program edits. The search for program edits can identify fix locations ({\em e.g.}, see \cite{asiaccs21} for preliminary work) and repair constraints derived from sanitizers, which further help to concentrate the search for inputs around these locations. The workflow (of the two searches strengthening each other) resembles an any-time patching/synthesis method that can be stopped anytime, where the correctness of the generated code gradually improves over time. More importantly, it closes an important gap in today's vulnerability discovery where software vulnerabilities are detected typically via fuzzing, and then the software remains exposed to these published vulnerabilities, which remain unfixed for long. We can view such as \textit{program protection} mechanism (fuzzing and patching together) as a new software process, where a program is subject to fuzzing and patching at the same time.

\subsection{Revisiting the challenges}

Despite the achievements in automated program repair, there are still challenges left and research work to do.
The \textbf{overfitting issue} (discussed in Chapter~\ref{sec:overfitting}) can be mitigated by the generation of additional test cases and patch ranking based on syntactic and semantic distance. However, since the developer's intention is rarely completely formalized, it remains a challenge to identify high-quality patches that are inspired by the provided specification but are general enough not to overfit. Related to this issue is the validation of plausible patches. While we can rule out patch candidates that violate a given specification, it is generally hard to determine a \textit{correct patch}. Even a given developer patch from a benchmark might not be the only way to repair it.
To help these issues, one way would be to perform additionally automated (or semi-automated) \textbf{specification inference}.  There has been significant advancements in this direction as mentioned in this article. At the same time, it will continue to be an area of interest in automated program repair

Another challenge remains in \textbf{complex repairs}. Many existing techniques focus on repairing errors, which can be repaired with a single-line fix. However, errors can also require the modification of large code chunks or edits at multiple locations. Again as discussed, there has been significant progress and repair tools like Angelix can produce multi-line fixes. Because of the inherent search space explosion involved in navigating/constructing multi-line fixes, this problem is likely to capture developer's attention in the future as well.

Repair techniques will also generally suffer from the \textbf{scalability} issue. Due to many possible combinations of fix locations, the identification of a plausible repair would require the exploration of a large search space which might be infeasible with current exploration and validation strategies and current patch space abstractions. As an implementation issue, the scalability challenge is particularly exacerbated due to the {\em need for recompilation}, when a search space of patch candidates is being traversed by a program repair tool. Solutions via bytecode mutation have been explored for Java programs \citep{PraPR}. These approaches avoid repairing at the source code level. However, to avoid recompilation in C program repair, one would need capabilities of binary interpretation and binary rewriting. This is because we will not have bytecode level intermediate representation to work with, for C program repair. 

In the end, the repair technology must be deployed where developers can easily integrate it into existing workflows like the continuous integration system. Existing industry deployments have already outlined a path for technology adoption. However, many questions remain to be answered, like how an efficient \textbf{interaction} between developers and repair tool can be established and how repair techniques can produce \textbf{trustworthy} patches.

\subsection{Future Outlook}

Automated program repair \citep{cacm19} is an emerging technology which seeks to reduce manual burden via automated fixing of errors and vulnerabilities. Apart from improving programmer productivity --- automated program repair technology has usage in reducing exposure of software systems to security vulnerabilities. Automated program repair techniques can also be used for programming education - where by repairing a student's programming attempt automated feedback can be given to students struggling to learn programming. 

In terms of technology, initial attempts at solving program repair were focused on using meta-heuristic searches \citep{GenProg}. By navigating an explicitly represented search space of program edits - such techniques can scale up to large programs but not to large search spaces. In other words, such techniques can work well on large programs which are "almost correct" - where the fix could be lifted from elsewhere in the program (or from past program versions). To scale up to large search spaces, we effectively need program repair techniques which can generate complex program edits such as multi-line edits. This is only possible if the search space of program edits is implicitly represented say as a symbolic repair constraint. Subsequently, fixes can be obtained by solving the repair constraint either via program synthesis (which involves a back-end constraint solving) or via an enumerative search over an implicitly defined search space. Semantic program repair techniques \citep{semfix} divide the task of program repair into such a repair constraint generation and patch synthesis steps. 

Along with the development of search-based and semantic program repair methods, researchers have studied the role of machine learning in program repair. One successful usage of machine learning in program repair is in ranking candidate patches \citep{prophet} after the candidates obtained by a technique such as enumerative search over a well-defined search space. Subsequently, a host of pattern-based and learning-based techniques have been proposed for program repair which steadily try to improve the quality of patches since the techniques want to ensure that the neural repair techniques can at least produce compilable programs \citep{monperrus19}.

Though deep learning based techniques have seen significant progress, patch quality has been one concern, as the learning techniques have no significant understanding of the program semantics, or even simple program properties such as program dependencies. Thus, instead of continually improving the quality of patches produced by deep learning methods - one could use learning techniques in an alternative fashion for automated program repair. Specifically, we note that language model based code generation \citep{codex}, as evidenced by tools like Github Copilot and AWS Codewhisperer, have recently gained traction. The recent release of Copilot only a few months ago includes an edit mode \citep{Codex-e} which can edit/insert text, apart from completing text. Such an enhanced language model based code companion engine can also serve as a program repair tool. As found in most recent results language model based code repair can even outperform pattern-based and learning based program repair tools in fixing automatically generated code (see the report in \cite{arxivMay22}). 

Looking forward, it is feasible to envision a role of language model based code generator/transformers for automated repair. While the exact role remains to seen and it is up to future research, we can already observe certain trends which may be worth commenting upon. If the focus shifts in the future on significant scaffoldings of code being automatically generated, automated program repair techniques could be employed to automatically improve the quality of automatically generated code. While this seems like a tall order at a first glance - we feel this can {\em beyond} engineering prompts for an automated code generator like Codex. In particular, program artifacts/properties can be integrated into language models. Alternatively, the various program candidates generated by a language model based generator can be treated by an augmented semantic program repair tool, to curate, extract and piece together patch ingredients - with the goal of generating complex patches via program analysis of automatically generated code.  This can be achieved in many possible ways, e.g., semantic analysis techniques can be used to find equivalence classes among partial code snippets obtained from various program candidates, and code from those equivalence classes can be pieced together into a complete program. We thus feel that the language model based code generation could provide newer opportunities (rather than threats) to existing (analysis based) program repair techniques.